\documentclass[12pt]{article}
\pdfoutput=1
\usepackage[textheight=9in,textwidth=6.5in]{geometry}
\usepackage{epsfig}
\usepackage{amsfonts}
\usepackage{amsmath}
\usepackage{amssymb}

\newcommand{\CO}{\mathcal{O}}

\newcommand{\be}{\begin{equation}}
\newcommand{\ee}{\end{equation}}
\newcommand{\ba}{\begin{eqnarray}}
\newcommand{\ea}{\end{eqnarray}}
\newcommand{\bea}{\begin{eqnarray}}
\newcommand{\eea}{\end{eqnarray}}
\newcommand{\bi}{\begin{itemize}}  
\newcommand{\ei}{\end{itemize}}

\newcommand{\nn}{\nonumber}
\newcommand{\aslash}[1]{\,\,{\raise.15ex\hbox{/}\mkern-12mu #1}}
\newcommand{\bslash}[1]{\,\,{\raise.15ex\hbox{/}\mkern-9mu #1}}
\newcommand{\cn}{{\mathcal N}}
\newcommand{\nt}{\tilde n}
\newcommand{\nb}{\bar n}

\def\>{\rangle}
\def\<{\langle}
\renewcommand{\bar}{\overline}
\renewcommand{\tilde}{\widetilde}
\renewcommand{\hat}{\widehat}

\newcommand\lrpar{\raise .8ex\hbox{$^\leftrightarrow$} \hspace{-9pt}
\partial}
\newcommand\lpar{\raise .8ex\hbox{$^\leftarrow$} \hspace{-9pt}
\partial}
\newcommand\rpar{\raise .8ex\hbox{$^\rightarrow$} \hspace{-9pt}
\partial}
\newcommand\lrd{\raise .8ex\hbox{$^\leftrightarrow$} \hspace{-9pt}
\nabla}

\newcommand{\gsim}{\lower.7ex\hbox{$\;\stackrel{\textstyle>}{\sim}\;$}}
\newcommand{\lsim}{\lower.7ex\hbox{$\;\stackrel{\textstyle<}{\sim}\;$}}
\usepackage{hyperref}

  \usepackage{color}

\begin{document}

\baselineskip=18pt

\setcounter{footnote}{0}
\setcounter{figure}{0}
\setcounter{table}{0}

\begin{titlepage}

\begin{center}
\vspace{1cm}

{\Large \bf  A  Natural Language for AdS/CFT Correlators}

\vspace{0.8cm}

{A. Liam Fitzpatrick$^1$, Jared Kaplan$^2$, Joao Penedones$^3$, Suvrat Raju$^4$, Balt C. van Rees$^5$}

\vspace{.5cm}

{\it $^1$ Department of Physics, Boston University, Boston, MA 02215, USA.}\\
{\it $^2$ SLAC National Accelerator Laboratory, 2575 Sand Hill, Menlo Park, CA 94025, USA.} \\
{\it $^3$ Perimeter Institute for Theoretical Physics, Waterloo, Ontario N2L 2Y5, Canada.}\\
{\it $^4$ Harish-Chandra Research Institute, Chatnag Marg, Jhunsi, Allahabad 211019, India.}\\
{\it $^5$ C. N. Yang Institute for Theoretical Physics, State University of New York, \\ Stony Brook, NY 11794-3840, USA.}\\

\end{center}

\vspace{1cm}

\begin{abstract}

We provide dramatic evidence that `Mellin space' is the natural home for correlation functions in CFTs with weakly coupled bulk duals.  In Mellin space, CFT correlators have poles corresponding to an OPE decomposition into `left' and `right' sub-correlators, in direct analogy with the factorization channels of scattering amplitudes. In the regime where these correlators can be computed by tree level Witten diagrams in AdS, we derive an explicit formula for the residues of Mellin amplitudes at the corresponding factorization poles, and we use the conformal Casimir to show that these amplitudes obey algebraic finite difference equations. 
By analyzing the recursive structure of our factorization formula we obtain simple diagrammatic rules for the construction of Mellin amplitudes corresponding to tree-level Witten diagrams in any bulk scalar theory.  We prove the diagrammatic rules using our finite difference equations.  
Finally, we show that our factorization formula and our diagrammatic rules morph into the flat space S-Matrix of the bulk theory, reproducing the usual Feynman rules, when we take the flat space limit of AdS/CFT. Throughout we emphasize a deep analogy with the properties of flat space scattering amplitudes in momentum space, which suggests that the Mellin amplitude may provide a holographic definition of the flat space S-Matrix.

\end{abstract}

\bigskip

\end{titlepage}

%%%%%%%%%%%%%%%%%%%%%%%%%%%%%%%%%%%%%%%%%%%%%%%%%%%%%%%%%%%%%%%
\setcounter{tocdepth}{1}
\tableofcontents

\section{Introduction}
Progress in physics often requires a formalism that makes both the symmetries and the dynamics manifest and simple.  
For example, recently we have seen dramatic progress in S-Matrix theory facilitated by the use of the spinor-helicity formalism, on-shell superspace, and twistor space. 
  We will argue that the Mellin representation  \cite{Mack, MackSummary, Penedones:2010ue} is the most natural framework for CFT correlation functions, especially in the large $N$ expansion.
The benefits of adopting `Mellin space' are structurally identical to the crucial, yet rather pedestrian progression from position to momentum space for correlation functions and scattering amplitudes in flat spacetime.  

Arguably the most important dynamical property of a conformal field theory is its obedience to the operator product expansion \cite{ Ferrara:1973vz, Ferrara:1974nf, Polyakov:1974gs,Sofia}, which says that for any two operators $\CO_1$ and $\CO_2$, we have 
\be
\CO_1(x) \CO_2(0) = 
\sum_{p} \sum_{m=0}^\infty C^{\mu_1\dots\mu_m}_{1 2 p, \nu_1\dots\nu_l}(x) \partial_{\mu_1}\dots  \partial_{\mu_m} \mathcal{O}_p^{\nu_1\dots\nu_l}(0)
\ee 
as an exact operator relation, for some coefficient functions $C_{1 2 p}$ that are kinematically determined up to an overall coefficient for each conformal primary field $\CO_p$.  
If we begin with an $n$-operator correlator and use the OPE to reduce it 
recursively by squeezing together $k$ and $n-k$ of the original operators, then we naturally end up with something akin to a factorization channel.   

The Mellin representation displays these factorization channels as poles.  We will argue that the residues at these poles are intimately related to lower point correlation functions, 
so that the Mellin amplitude inherits a recursive structure from the OPE.  We support this with a somewhat formal argument in section \ref{sect:FactorizationofCFT}   and then develop it extensively for large $N$ CFTs in section \ref{sect:FactorizationonAdSPropagators}. However, the presence of these poles follows simply from the definition \cite{Mack, MackSummary} of the conformally invariant Mellin amplitude $M_n(\delta_{ij})$ in terms of a CFT correlator via
\be
\langle \CO_1(x_1) \cdots \CO_n(x_n) \rangle 
= \int [d \delta] M_n(\delta_{ij}) \prod_{i < j}  (x_i - x_j)^{-2\delta_{ij}} \Gamma(\delta_{ij}) 
\ee
The integration variables $\delta_{ij}$ are the CFT analogue of the kinematic invariants $p_i \cdot p_j$ in scattering amplitudes, and we will explain the precise definition of the contour integral in the next section.  Terms in the OPE of the correlator involving an operator of dimension $\Delta$ will have definite power law dependence on the $x_i$ coordinates, and this specific power law can only be reproduced by the Mellin amplitude if $M_n(\delta_{ij})$ has poles on which an appropriate linear combination of the $\delta_{ij}$ variables can be localized.  

The power of the OPE is that, in principle, with knowledge of the spectrum of operators and the three point functions, we could recursively compute all of the correlation functions in any CFT.   Mellin space may play an interesting role in this program, but we will follow an easier route and study correlation functions in large $N$ CFTs, specifically those with an AdS dual.  The Mellin representation becomes far more powerful in CFTs with a large $N$ expansion and a small number of low-dimension operators \cite{Penedones:2010ue}, because in these theories multi-trace operators are equal to products of single trace operators, up to $1/N$ corrections.  This fact is in a sense built into the Mellin space formalism, as we will explain in section \ref{sect:MultitraceandMellin}, so that the contributions of multi-trace operators are automatically incorporated once single-trace exchanges are correctly reproduced.

The AdS/CFT correspondence \cite{Maldacena:1997re, Witten:1998qj,
Gubser:1998bc} has led to a long list of major insights into both
quantum gravity and gauge theory. In particular, it enabled the
computation of 4-point correlation functions in strongly coupled
conformal field theories using supergravity \cite{LiuTseytlin,Liu,D'Hoker:1998mz,Freedman:1998bj,Freedman:1998tz,D'Hoker,Arutyunov:2000py,Arutyunov:2002fh,Arutyunov:2003ae,Berdichevsky:2007xd,Uruchurtu:2008kp,Buchbinder:2010vw,Uruchurtu:2011wh}, see also
\cite{Dolan:2006ec} for an interesting related conjecture.  However,
progress on holographically computing correlation functions with more
than four external points has been relatively modest. (Some progress
was reported in \cite{Howtozintegrals}.)
 We hope to facilitate progress in this direction by deriving a precise factorization formula that makes it possible to recursively compute tree-level Witten diagrams for arbitrary bulk scalar field theories.  As a first demonstration of the power of our methods, we compute the 5-pt and 6-pt correlation function in $\phi_i \phi_j \phi_k$ theory in AdS.  For example, the Mellin representation of the unique 5-pt Witten diagram in $\phi^3$  theory with $\Delta = d = 4$ is
\be
M_5(\delta_{ij}) \propto
\left( \frac{1}{ \delta_a  \delta_b} +
 \frac{1}{ 3\delta_{a} (\delta_b - 1)} +
  \frac{1}{ 3 (\delta_a - 1) \delta_b } +
   \frac{5}{ 9(\delta_a - 1)( \delta_b - 1)}  \right)
\ee
where $\delta_a$ and $\delta_b$ are linear combinations of $\delta_{ij}$ that are analogous to the kinematic invariants that appear in familiar flat-space propagators, such as $(p_1 + p_2)^2$ and $(p_4+p_5)^2$.  

The fact that our factorization formula can be applied to any combination of factorization channels in any order strongly suggests that there must be an even simpler structure underlying the Mellin amplitudes.  In fact, we will derive a set of diagrammatic rules, the AdS/CFT equivalent of the momentum space Feynman rules, which permit a direct construction of the tree-level Mellin amplitudes in any scalar theory in terms of vertices and propagators.  For example, in section \ref{sec:FeynmanRules} we define propagators and vertices for  a general $\phi_i \phi_j \phi_k$ theory
\be
\frac{S_\Delta(m)}{\delta - m} , \ \ \ \ \
V_{\Delta_i \Delta_j \Delta_k} (m_i, m_j, m_k) \ee
which allow for a direct diagrammatic computation of AdS/CFT correlation functions.  Similar results should hold for vector and tensor theories, although aside from a brief discussion in Appendix \ref{sect:TensorFields} we leave these developments for future work.  

We emphasize that these rules are universal and diagrammatically local, so that the rule for one vertex in a given diagram does not depend on the rest of the diagram.  At the vertices we have `dimension conservation', which is analogous to momentum conservation in flat space and follows from conformal invariance.  The most familiar manifestation of `dimension conservation' is the fact that CFT two point functions vanish unless the two operators have the same dimension.

Although OPE factorization may be the physical principle behind the simplicity of AdS/CFT computations in Mellin space, this simplicity has another guise in the form of a functional equation, which we will derive in section \ref{sect:FunctionalEquation}.  An analogous equation was used in position space in \cite{Howtozintegrals} and it enabled major progress in the computability of AdS Witten diagrams. Furthermore, it was used by \cite{Dolan:2003hv} to find explicit expressions for conformal blocks with external scalar operators. The main idea is to use the fact that bulk to bulk propagators in AdS are Green's functions of the conformal Casimir.  One can use this Casimir to collapse bulk to bulk propagators into AdS delta functions, reducing AdS exchange diagrams to contact interactions.  In Mellin space, this observation becomes the AdS analogue of the very useful fact that $\nabla^2 \to -p^2$ when we Fourier transform to momentum space. Namely, when we apply the conformal Casimir to the Mellin representation, we find an enormous simplification, and a complicated differential equation becomes purely algebraic.

This functional equation has a variety of uses. We will use it to prove the validity of our recursive calculations and as a tool to obtain very general information about the analytic structure of Mellin amplitudes. It is also an interesting tool for computing the conformal block decomposition of various tensor structures, as we will describe in a forthcoming paper.

Mellin space has formal advantages because unlike CFT correlators in position space, Mellin amplitudes are simple meromorphic functions of their arguments. Again, we have a nice analogy with the S-Matrix in flat space, whose analyticity properties are very well-known and well-studied.  However, on the basis of the existence of a convergent OPE one may expect that Mellin amplitudes will always be meromorphic functions without branch cuts \cite{Mack}. In AdS this is reflected by the discrete spectra of quantum theories in AdS, so infinite sums replace the phase-space integrals that one encounters in flat space. The presence of poles and the absence of branch cuts was discussed in a 1-loop example in \cite{Penedones:2010ue}. 

We aim to show that Mellin space is a profoundly useful arena in which to study CFT correlation functions, but it may also illuminate the way in which the very large $N$ and $\lambda$ limit of AdS/CFT morphs into a holographic description of flat spacetime \cite{susskind, polchinski}, as several of us have discussed before \cite{GGP, JP, Penedones:2010ue, Katz, TakuyaFSL, Fitzpatrick:2011jn}. We obtain the flat space S-Matrix from AdS/CFT by studying high energy bulk states, which are dual to high dimension CFT operators, and so we should expect scattering amplitudes to be related to the large $\delta_{ij}$ behavior of the Mellin amplitude.  In fact, as was already argued in \cite{Penedones:2010ue}, we find that at large $\delta_{ij}$ the Mellin amplitude becomes the scattering amplitude of the AdS theory.  In section \ref{sect:FlatSpaceLimit} we show explicitly that our AdS/CFT factorization formula reduces to the usual factorization of tree-level amplitudes on their propagators when we take the flat space limit.  This can be viewed as a constructive proof that one can compute the complete tree-level flat space S-Matrix for scalar theories using only CFT correlators, addressing the issues of \cite{Gary:2009mi, GiddingsBulkLoc, Gary:2011kk} from a different angle.  We also show that the flat space limit works equally well for massless and massive particles.   
Therefore, we expect that any tree level scattering amplitude can be obtained as the flat space limit of the corresponding Mellin amplitude. 
The same should hold at loop level as well, as suggested by the 1-loop example studied in  \cite{Penedones:2010ue}. 
It is then natural to give a \emph{holographic} and \emph{non-perturbative} definition of gravitational scattering amplitudes as the flat space limit of the Mellin amplitudes of the dual CFT.

The outline of this paper is as follows. In section 2, we will use the operator product expansion to motivate the Mellin space approach to CFT correlation functions and discuss how the Mellin representation becomes especially useful in the large $N$ limit. We will then discuss the Mellin representation of a generic scalar Witten diagram and derive the aforementioned functional equation for the Mellin representation of the diagram. In section 3 we derive the factorization formula which allows us to split an arbitrary scalar tree-level Witten diagram into a `left' and a `right' piece. This factorization formula \emph{a priori} only gives us the poles (and residues) of the Mellin amplitude but we will claim that this is in fact the entire result. This claim is further motivated in section 4 by working out several  examples. In section 5 we provide an actual proof of our claim, by demonstrating that it is equivalent to the existence of specific Feynman rules in Mellin space and then showing that these Feynman rules satisfy the functional equation derived in section 2. In section 6 we verify that in the flat-space limit our Mellin amplitudes reproduce scattering amplitudes. We end in section 7 with a discussion. Three appendices discuss a more direct proof of the factorization formula of section 3, the extension to the exchange of bulk fields with spin and some technical developments.

While this project was being completed we became aware of the interesting work \cite{MPaulos}, which has some overlap with the present paper.

\section{Motivating Mellin Space}
In this section we will explain why Mellin space \cite{Mack} makes the physics of CFT correlation functions simple and transparent, in the same way that momentum space simplifies scattering amplitudes in flat spacetime. We will see that the key property of scattering amplitudes in momentum space is also present for CFT correlation functions in Mellin space: \emph{factorization} of correlation functions on propagator poles, with the residues given in terms of correlation functions with fewer operators. Furthermore, in CFTs with a large $N$ expansion, the Mellin representation of the dual Witten diagrams obeys a simple \emph{algebraic} equation. We will explain these two properties in sections \ref{sect:FactorizationofCFT} and \ref{sect:FunctionalEquation}, with a brief interlude to introduce some notation. Finally in section \ref{sect:MultitraceandMellin} we will review \cite{Penedones:2010ue} why the Mellin representation is particularly well suited to theories with a large $N$ expansion.

\subsection{Factorizaton of CFT Correlation Functions}
\label{sect:FactorizationofCFT}

First, let us motivate the Mellin transformation by studying one of the most basic properties of a CFT, namely the Operator Product Expansion (OPE).  

Consider a CFT correlation function of $n$ operators
\be
\label{eq:CFTcorr}
A_n(x_i) = \left\langle\prod_{i=1}^k \mathcal{O}_i\left(x_i\right)
\prod_{i=1+k}^n \mathcal{O}_i\left( x_i\right)\right\rangle
\ee
where we have divided the operators into two groups. (We will in this paper only consider correlation functions of scalar operators.) Upon recursively applying the OPE, we can write the product of operators as a sum
\be
\label{OPE}
\prod_{i=1}^k \mathcal{O}_i\left(x_i\right) = 
\sum_{p} \sum_{m=0}^\infty C^{\mu_1\dots\mu_m}_{p,\nu_1\dots\nu_l}(x_1,\dots,x_k) \partial_{\mu_1}\dots  \partial_{\mu_m} \mathcal{O}_p^{\nu_1\dots\nu_l}(x_k)
\ee
where $p$ labels primary operators.

The general idea behind the OPE is that we can expand in the distance between two operators in the limit that this distance is small.  Such an expansion is conceivable in any quantum field theory, but in a Euclidean CFT we expect the sum to have a finite radius of convergence because scale invariance implies that results for small separation should continue to hold (as long as there are no other operator insertions). We can apply the OPE to different sequential combinations of operators, and the equivalence of these different expansions, \emph{i.e.} crossing symmetry, provides a very powerful general constraint on CFTs \cite{Rattazzi:2008pe, Rychkov:2009ij}.

Consider now the action of a dilatation on the first $k$ operators, after using translation invariance to set $x_k=0$,
\be
\prod_{i=1}^k \mathcal{O}_i\left(e^{-\lambda}x_i\right) = 
\sum_{p} \sum_{m=0}^\infty e^{-\lambda (\Delta_p   +m) +\lambda \sum_i \Delta_i} C^{\mu_1\dots\mu_m}_{p,\nu_1\dots\nu_l}(x_1,\dots,0) \partial_{\mu_1}\dots  \partial_{\mu_m}
\mathcal{O}_p^{\nu_1\dots\nu_l}(0) \label{multipleOPE}
\ee 
where $\Delta_p$ is the dimension of the primary operator $\mathcal{O}_p^{\nu_1\dots\nu_l}(0)$.  Since this is an exact operator equation (as long as $\lambda$ is large enough to force the first $k-1$ points closer to $x_k = 0$ than any other point), we can substitute it into our original CFT correlation function to find
\be
\left\langle\prod_{i=1}^k \mathcal{O}_i\left(e^{-\lambda}x_i\right)
\prod_{i=1+k}^n \mathcal{O}_i\left( x_i\right)\right\rangle=
\sum_{p} \sum_{m=0}^\infty e^{-\lambda (\Delta_p +m) +\lambda \sum_i \Delta_i}
F_{p,m}(x_1,\dots,x_n) \label{OPEscaling}
\ee
where 
\be
F_{p,m}(x_1,\dots,x_n)=
C^{\mu_1\dots\mu_m}_{p,\nu_1\dots\nu_l}(x_1,\dots,0)
\left\langle  \partial_{\mu_1}\dots  \partial_{\mu_m}
\mathcal{O}_p^{\nu_1\dots\nu_l}(0)
\prod_{i=1+k}^n \mathcal{O}_i\left( x_i\right)\right\rangle
\ee
Our question: in what variables is the structure of the OPE manifest?

A natural answer is to use the variables that are conjugate to the dilatation parameter $\lambda$.  In these variables the CFT correlator will have a pole with residue given in terms of lower-point correlators. To implement this philosophy, one introduces the Mellin representation
\be
\label{mellindefn}
A_n(x_i) = \int [d \delta] M_n(\delta_{ij}) \prod_{i < j}^n  (x_i - x_j)^{-2\delta_{ij}} \Gamma(\delta_{ij})
\ee
where the parameters $\delta_{ij}$ are symmetric in $ij$, but $\delta_{ii} = 0$, and they are constrained to give the correct behavior under conformal transformations. This means that
\be
\sum_{j} \delta_{ij} = \Delta_i
\label{eq:deltaconstr}
\ee
Taking into account these constraints, the symbol $[d\delta]$ in \eqref{mellindefn} denotes an integral over a subset of precisely $n(n-3)/2$ of the $\delta_{ij}$ which are independent of each other, normalized as
\be
\int [d \delta] = \int \frac{d \delta_{12}}{2\pi i} \frac{d\delta_{13}}{2\pi i} \ldots
\ee
The contour of integration for each of the independent $\delta_{ij}$ runs parallel to the imaginary axis.  
An extremely useful analogy that will pervade what follows is to think of the $\delta_{ij}$ as kinematic invariants $p_i \cdot p_j$ in an $n$-particle scattering amplitude, and to think of the $\Delta_i$ as the masses of these $n$ particles.  Then the constraint eq. \ref{eq:deltaconstr} follows simply from the requirement of momentum conservation $\sum_j p_j=0$ and the on-shell conditions $p_i^2 = -\Delta_i$ \cite{Mack}.  We will discuss below why it is especially natural for theories with a large $N$ expansion to include the $\Gamma(\delta_{ij})$ factor in the definition of the Mellin amplitude.

Now if we rescale the $x_i \to e^{-\lambda} x_i$ for $i \leq k$ as above and consider the large $\lambda$ limit of the Mellin representation, we find
\begin{align}  \label{LRDecompJoao}
\int  [d \delta]\,
M_n(\delta_{ij})
e^{2\lambda   \sum_{i<j}^k \delta_{ij}}
\prod_{ i<j}^n \Gamma(\delta_{ij})
\prod_{ i<j}^k   (x_{ij}^2)^{-\delta_{ij}} 
\prod_{ i\le k < j}^n  \left(x_{j}^2-e^{-\lambda}2x_i\cdot x_j +e^{-2\lambda} x_i^2\right)^{- \delta_{ij}} 
\prod_{k<i<j}^n   (x_{ij}^2)^{-\delta_{ij}} \nn
\end{align}
To match the leading behaviour at large $\lambda$ between the Mellin amplitude and our OPE result, 
we consider the expansion
\be
\prod_{ i\le k < j}^n  \left(x_{j}^2-e^{-\lambda}2x_i\cdot x_j +e^{-2\lambda} x_i^2\right)^{- \delta_{ij}} =
\sum_{q=0}^\infty e^{-q \lambda  } Q_q(x_1,\dots,x_n)
\ee
where $Q_q$ is a polynomial of degree $q$ in $x_i$ with $i=1,\dots,k$. Therefore, the contribution of a spin $l$ operator to the OPE, comes from the $q=l$ term in this expansion.
Matching the $e^{-\lambda}$ scaling with 
 (\ref{OPEscaling}), we conclude that the Mellin amplitude must have  poles at
\be
\label{mellinpolesgeneral}
\sum_{i=1}^k \Delta_i    -  2 \sum_{i<j}^k \delta_{ij} =  \tau_p  + m
\ee
for all non-negative integers $m$.
Here, we have introduced $\tau_p = \Delta_p - l_p$, the twist of the operator $\CO_p^{\nu_1\dots\nu_l}$. 
 Notice that the left hand side is the precise analog of the flat space kinematic invariant $-(p_1 +\ldots + p_k)^2$. Corresponding poles arise explicitly when we consider Witten diagrams in AdS/CFT, and a major goal in what follows will be to give a precise and computationally useful formula for the residues of these poles.  

But let us first give an intuitive explanation for why these residues should be intimately related to lower point correlation functions. 
The residue corresponding to a specific OPE channel is most conveniently written by introducing for every primary field $\CO_p$ a corresponding \emph{shadow} field $\tilde \CO_p$\footnote{Very roughly speaking, one introduces shadow fields in order to write the operator {\bf 1} as a sum of \emph{primary} operators acting on the vacuum, $\sum_p \CO_p |0 \rangle \langle 0 | \tilde \CO_p$.  Shadows are necessary to ensure that the correlator transforms correctly under dilatations; their necessity is analogous to the fact that on a certain very formal sense, the bra and ket in-states $\langle p |$ and $| p \rangle$ have opposite energy.}, defined such that:
\be
\langle \CO_p(x) \tilde \CO_{p'} (y) \rangle = \delta^d(x-y) \delta_{p,p'}
\ee

Clearly, if $\CO_p$ has scaling dimension $\Delta_p$, the shadow field must have scaling dimension $d - \Delta_p$. An intuitive way to write the shadow field is via the convolution:
\be
\tilde \CO_p(x) = \int d^d y \frac{\CO_p(y)}{(x-y)^{ 2 (d- \Delta_p ) }}
\ee
but formally this integral is divergent and needs regularization. 

 Using the OPE, we find that at least schematically
\be
\label{Aalternative}
A_n(x_i) \sim \sum_{p} \int d^d y \left\langle\prod_{i=1}^k \mathcal{O}_i\left(x_i\right) 
 \CO_p (y)  \right\rangle \left\langle  \tilde \CO_p (y)
\prod_{i=1+k}^n \mathcal{O}_i\left( x_i\right)\right\rangle
\ee
This equation is however only formal as the integral over $y$ of the
insertion point of $\CO_p(y)$ implies that we may destroy the
convergence of the OPE of the other operators. Nevertheless, it can be
used to offer a reasonable CFT intuition of the OPE in Mellin space.
In particular, if we were to substitute the Mellin transform of the
two correlation functions on the right-hand side of
\eqref{Aalternative}, the resulting Mellin transform of $A_n$ has
poles precisely at \eqref{mellinpolesgeneral}.  These poles isolate specific terms in the sum,
and have residues which are given in terms of the product of Mellin transforms of the lower point
correlators. 
In section  \ref{sect:AdSCFTFactorization} we will see an explicit and  precise confirmation of this rough OPE intuition in the case of Witten diagrams.

\subsection{Conformally Covariant Notation}

We will be discussing CFT correlation functions, so it is natural to use variables \cite{Dirac,Weinberg:2010fx} that are acted on linearly by the Euclidean conformal group $SO(1,d+1)$.  
If we begin with $(d+2)$-dimensional Minkowski spacetime, then the conformal generators will simply be 
\be
J^{AB} = X^A \frac{\partial}{\partial X_B} - X^B \frac{\partial}{\partial X_A}
\label{ConformalGenerators}
\ee
so that conformally invariant functions can be constructed out of the covariant inner products $(X_i \cdot X_j)$.  We can view Euclidean AdS$_{d+1}$ as the hyperboloid
\be
X^2=-R^2\ ,\ \ \ \ \ \ \ \ \ \  X^0 > 0\ ,\ \ \ \ \ \ \ \ \ \ \ X \in \mathbb{M}^{d+2}\ ,
\ee
embedded in this $(d+2)$-dimensional Minkowski spacetime.  
We set $R=1$ in what follows. 
Furthermore, we can think of the conformal boundary of AdS as the space of null rays
\be
P^2=0\ ,\ \ \ \ \ \ \ \ \ \  P\sim \lambda P\ \ (\lambda \in \mathbb{R})\ ,\ \ \ \ \ \ \ \ \ \ \ P \in \mathbb{M}^{d+2}\ . 
\ee
Then, the correlations functions of primary scalar operators of  the dual CFT are encoded into $SO(1,d+1)$ invariant functions of the external points $P_i$, transforming homogeneously with weights $-\Delta_i$.  We will work extensively with this formalism in what follows; for more thorough discussions, see \cite{Weinberg:2010fx}.

Using the standard AdS/CFT prescription, CFT correlators can be computed in terms of Witten diagrams, which are bulk Feynman diagrams that connect to propagators ending on the boundary of AdS. The external legs in such diagrams represent AdS bulk to boundary propagators, which in this notation are simply given by
\be
G_{\partial B}(P,X) = \frac{\mathcal{C}_\Delta}{(-2 P \cdot X)^{\Delta}} \ ,
\ee
where
\be
\mathcal{C}_\Delta = \frac{\Gamma(\Delta)}{2\pi^\frac{d}{2} \Gamma\left(\Delta-\frac{d}{2}+1\right)}= \frac{\Gamma(h+c)}{2\pi^h \Gamma\left(c+1\right)}\ .
\ee
Here, we have used the notation
\be
\Delta=h+c\ ,\ \ \ \ \ \ \ \ \ \ \ \ \ \ d=2h\ .
\ee
This normalization was obtained by taking the limit of the bulk to bulk propagator; it differs by a factor of $(2\Delta - d)^{-1}$ from the normalization of \cite{Freedman:1998tz}. 

Using again the results of \cite{Freedman:1998tz}, this implies that in our conventions the two-point functions are normalized as
\be
\langle \mathcal{O}_\Delta(P_1)  \mathcal{O}_\Delta(P_2) \rangle = \frac{\mathcal{C}_\Delta}
{(-2P_1 \cdot P_2)^\Delta}\ . \label{AdSnormalization}
\ee

Finally, to recover the usual expressions in physical $\mathbb{M}^d$ or $\mathbb{R}^d$  we choose the light cone section
\be
P=(P^+,P^-,P^\mu)=(1,x^2,x^\mu)\ ,
\ee
where $\mu=0,1,\dots,d-1$. Then $P_{12}\equiv -2P_1 \cdot P_2=(x_1-x_2)^2$.  We will use $P_i$ and $x_i$ variables interchangeably in what follows.

\subsection{The Functional Equation, or AdS/CFT Turns Algebraic}
\label{sect:FunctionalEquation}

Let us begin by giving a brief and tortured-looking review of why momentum space drastically simplifies tree-level computations of S-Matrix elements in flat space, using $\phi^3$ theory as a simple example.  A directly analogous procedure will lead to a beautiful functional equation for the Mellin representation of correlators in AdS/CFT.

Consider an $n$-pt correlation function for $\phi^3$ theory in flat space.  Isolating the class of position-space diagrams with a propagator connecting particles 1 and 2 to the rest, we have
\be
C(x_i) 
= \int d^4 x d^4 y \frac{1}{4\pi^2(x_1 - x)^2}\frac{1}{ 4\pi^2(x_2 - x)^2 } \frac{1}{4\pi^2(x-y)^2} F(y,x_3,...,x_n)
\ee
Thus if we act on this correlator with $(\nabla_{x_1} + \nabla_{x_2})^2$, we can exchange this operator inside the integral for $\nabla_x^2$, which collapses the propagator $\frac{1}{4\pi^2(x-y)^2}$ to a delta function.  Now if we consider the Fourier transform $\tilde C(p_i)$ of the correlator, this means that
\be
(p_1 + p_2)^2 \tilde C(p_i) = C_0
\ee
where $C_0$ is the same set of Feynman diagrams with the dependence on particles 1 and 2 reduced to a contact interaction.  The transition to momentum space therefore allows us to solve for general tree level correlation functions without doing any integrals.

\begin{figure}
 \centering
    \includegraphics[width=0.98\textwidth]{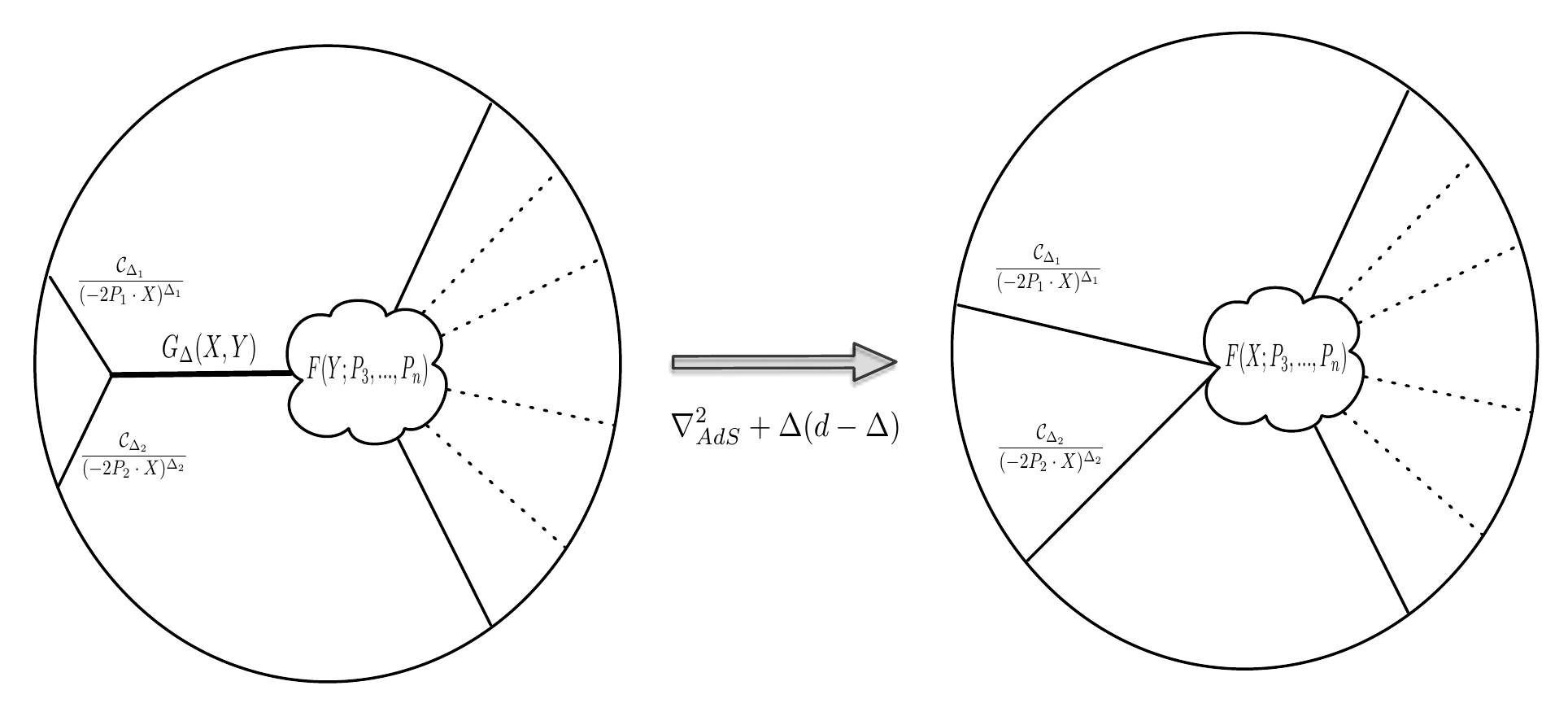}
      \caption{\small By acting with the conformal Casimir on a Witten diagram with a bulk to bulk propagator, we collapse the propagator into a delta function.  We derive the functional equation by looking at this process in Mellin space.}
      \label{fig:FunctionalEquation}
\end{figure}

Now let us perform the analogous steps for correlation functions in AdS, with the Mellin representation playing the role of the momentum space amplitude.  This procedure has been used in position space by \cite{Howtozintegrals}.  Consider a Witten diagram where particles $1$ and $2$ are connected to the rest of the diagram by a $\phi^3$ vertex and a bulk-bulk propagator as shown in figure \ref{fig:FunctionalEquation},
\be
A (P_1,P_2, ..., P_n)=\int_{\rm AdS} dX dY 
\frac{\mathcal{C}_{\Delta_1}}{(-2P_1\cdot X)^{\Delta_1}}
\frac{\mathcal{C}_{\Delta_2}}{(-2P_2\cdot X)^{\Delta_2}}
G_\Delta(X,Y) F(Y, P_3,...,P_n)
\ee
where $G_\Delta(X,Y)$ is the bulk to bulk propagator for a field with dimension $\Delta$.  The equivalent of the box operator in flat space is the Casimir of the conformal group, which is just the sum of the squares of the generators.  These take an especially simple form in terms of the $P_i$ variables, as we saw above in equation (\ref{ConformalGenerators}).  
The Casimir for the first two particles is
\be
\frac{1}{2}(J_1 + J_2)^2 = 2 P_1 \cdot P_2   \frac{\partial}{\partial P_1} \cdot  \frac{\partial}{\partial P_2} - 2 P_1^A P_2^B  \frac{\partial}{\partial P_1^B}  \frac{\partial}{\partial P_2^A} + \sum_{i=1}^2 P_i^A P_i^B \frac{\partial}{\partial P_i^A} \frac{\partial}{\partial P_i^B} - (d-1) P_i \cdot \frac{\partial}{\partial P_i}
\ee
and when it acts on the correlator, inside the integral it is equivalent to $\frac{1}{2} J_X^2 =  -\nabla_{\rm AdS}^2$.  Since the bulk to bulk propagator is the Green's function of this operator, 
\be
\left[ \nabla_{\rm AdS}^2 -\Delta(\Delta-d) \right]G_\Delta(X,Y) = -\delta(X,Y)\ ,
\ee
it collapses $G_\Delta(X,Y)$ into a delta function.  This gives an equation
\be
\left[ \frac{1}{2}(J_1 + J_2)^2 - \Delta (d - \Delta) \right] A = A_0
\ee
where in $A_0$ the propagator has been collapsed into a contact interaction. In \cite{Howtozintegrals}, this was used to convert Witten diagrams with bulk to bulk propagators to contact interactions. 

In Mellin space, this equation takes a remarkably simple form. 
When the conformal Casimir of particles 1 and 2 acts on the product $\prod_{i < j} (P_{ij})^{-\delta_{ij}}$ in the definition \eqref{mellindefn} of the Mellin amplitude, where $P_{ij} = -2 P_i \cdot P_j$, we find
\be
\left[ (\delta_{LR} - \Delta)(d - \Delta - \delta_{LR}) + \sum_{i \neq j \geq 3} 2 \delta_{1i} \delta_{2j} \left(1 - \frac{P_{1j} P_{2i}}{P_{1i} P_{2j}} + \frac{P_{12} P_{ij}}{P_{1i} P_{2j}} \right) \right] \prod_{i < j} (P_{ij})^{-\delta_{ij}}
\ee
where $\delta_{LR} = \Delta_1 + \Delta_2 - 2 \delta_{12}$ is the natural analog of the momentum space variable $-(p_1+p_2)^2$; later on we will see that the Mellin amplitude has poles in this $\delta_{LR}$.  This expression can be simplified by noting that multiplication by the kinematic invariants $P_{ij}$ is equivalent to shifting the $\delta_{ij}$, so that for example
\be
\frac{P_{12}P_{34}}{P_{13}P_{24}} A(P_1,\dots,P_n) =
 \int [d\delta] \left(  \frac{\delta_{12}\delta_{34}M(\delta_{12}+1,\delta_{13}-1,\delta_{34}+1,\delta_{24}-1
,\dots)}{(\delta_{13}-1)(\delta_{24}-1)} \right)\prod_{i<j}^n \Gamma(\delta_{ij})
 P_{ij} ^{-\delta_{ij}}\nn
\ee
This allows us to write the functional equation
\be
(\delta_{LR} - \Delta)(d - \Delta - \delta_{LR})M + \sum_{i \neq j \geq 3} 2 \left( \delta_{1i} \delta_{2j} M - \delta_{1j} \delta_{2i} M_{1i,2j}^{1j,2i} + \delta_{12} \delta_{ij} M_{1i,2j}^{12,ij} \right) = M_0
\ee
where we define
\be
M^{12,ij}_{1i,2j} =M(\delta_{12}+1,\delta_{ij}+1,\delta_{1i}-1,\delta_{2j}-1,\dots) 
\ee
and analogously for the other indexed $M$s.   This gives a purely algebraic equation for any Mellin amplitude $M$ with a propagator connecting particles $1$ and $2$ to the rest of the diagram.  Because of the finite differences this equation is more intricate than the analogous equation in momentum space, but it will be extremely useful later on for proving general results. 
In particular, in section \ref{sec:FeynmanRules}  we will derive a set of Feynman rules for Mellin amplitudes and use the functional equation to prove that they correctly compute Mellin space Witten diagrams.  For completeness, the general functional equation corresponding to any propagator is
\be
\label{GeneralFunctionalEquation}
M_0 = (\delta_{LR} - \Delta)(d - \Delta - \delta_{LR})M + \sum_{ab \leq k <  ij}  \left( \delta_{ai} \delta_{bj} M - \delta_{aj} \delta_{bi} M_{ai,bj}^{aj,bi} + \delta_{ab} \delta_{ij} M_{ai,bj}^{ab,ij} \right)
\ee
where the propagator separates the first $k$ from the last $n-k$ operators.

We should also point out that the functional equation is useful beyond
its application to Witten diagrams. To see this, return to the OPE 
equation \eqref{OPE}. Acting with the conformal Casimir on both sides, we see that whenever the operator $O_p$ on the right belongs to a conformal 
representation of lowest weight $\Delta_p$ and spin $l_p$, we find:
\begin{equation}
{1 \over 2} ( \sum_{i=1}^k J_i)^2 \prod_{i=1}^k \mathcal{O}_i\left(x_i\right) = -
\sum_{p} g(\Delta_p,l_p) \sum_{m=0}^\infty C^{\mu_1\dots\mu_m}_{p,\nu_1\dots\nu_l}(x_1,\dots,x_k) \partial_{\mu_1}\dots  \partial_{\mu_m}  \mathcal{O}_p^{\nu_1\dots\nu_l}(x_k),
\end{equation}
where the conformal Casimir $g(\Delta,l) = \Delta (\Delta - d) + l (l + d - 2)$. 
So, the contribution of a primary $O_p$ and {\em all its descendants} 
to a correlation function
can be packaged into a single solution of the homogeneous functional equation. For the four point function, this solution is determined entirely by the kinematics.  It was used by Dolan and Osborn to find explicit and simple expressions for these contributions, which are the familiar conformal blocks \cite{Dolan:2000ut, Dolan:2003hv}. In Mellin space, conformal blocks look even simpler, as we will describe in a forthcoming paper where we will perform a general analysis of the solutions of the functional equation. 

%\BvR{We can also write down a more general functional equation associated to any intermediate bulk-bulk propagator in a Witten diagram. It takes the following form..}
%It can be solved directly in many cases if we assume that the Mellin amplitude is given by a sum over integer spaced poles in the linear combinations of $\delta_{ij}$, such as our $\delta_{LR}$ above, that appear in propagators.

\subsection{The Mellin Representation at Large $N$}
\label{sect:MultitraceandMellin}
We have motivated the Mellin amplitude as a natural representation of CFT correlation functions which makes the structure of the OPE manifest and obeys algebraic equations %at large $N$, \emph{i.e.} 
whenever there is a local bulk description available.  In this section we will explain why we have included the factors of $\Gamma(\delta_{ij})$ in the definition of the amplitude.  %the relation between multi-trace operators and the product of $\Gamma(\delta_{ij})$ factors in the Mellin amplitude, which shows that the Mellin representation is especially well-suited for large $N$ theories. 

%Let us discuss whether the Mellin amplitude is the unique object with these properties, and why it is natural to include the factors of $\Gamma(\delta_{ij})$ in its definition.

Consider a large $N$ CFT with single-trace primary operators $\CO_1$ and $\CO_2$ of dimension $\Delta_1$ and $\Delta_2$. When we investigate their OPE, we will always have contributions from operators such as $\CO_1 \partial^{2n} \CO_2$ with dimension $\Delta_1 + \Delta_2 + 2n + \CO(\frac{1}{N})$. The product of $\Gamma$ functions in our definition of the Mellin amplitude then guarantees that the full Mellin integrand has poles whenever $\delta_{ij} = -n_{ij}$ for integers $n_{ij}$.  In our simple example, the poles at $\delta_{12} = -n_{12}$ produce residues of the integral that combine $\CO_1$ and $\CO_2$ into an operator of dimension $\Delta_1 + \Delta_2 + 2n_{12}$, exactly as desired.  

In general, we can combine $k$ operators in the same way.  Taking the first $k$ for convenience, we find poles at $\delta_{ij} = -n_{ij}$ for all $i,j \leq k$.  This set of coincident poles produces the multi-trace operator
\be
\partial^{\sum^k_j n_{1j}} \CO_1 \partial^{\sum_j^k n_{2j}}  \CO_2 \ldots \partial^{\sum_j^k n_{kj}}  \CO_k
\ee
where the derivatives are contracted so that between each $i$ and $j$ there are $n_{ij}$ contractions.  The residue at this pole is related to the correlation function of this operator with the other $n-k$ operators via our factorization formula, giving a sort of LSZ prescription for extracting the correlation functions of many composite operators.  
%The presence of the $\Gamma(\delta_{ij})$ factors in the Mellin representation guarantees the presence of these composite operators.

Notice that the above argument is very specific to CFTs with a perturbative description.  In more general CFTs the analogous `multi-trace' operators have finite anomalous dimensions and the poles of the $\Gamma(\delta_{ij})$ factors are not at the correct location to account for the multi-trace contributions to the OPE. In those cases $M(\delta_{ij})$ must not only have additional poles accounting for the multi-trace operators but also zeroes to cancel off the poles from the $\Gamma(\delta_{ij})$ factors \cite{Mack}. In our case the $\Gamma(\delta_{ij})$ already give poles at the right location and we do not need to worry about zeroes in the Mellin amplitude.  Perhaps a different definition of $M(\delta_{ij})$ would be appropriate for such theories.

Of course multi-trace operators do gain anomalous dimensions beyond the leading order in $1/N$, which can be read off from the leading connected higher-point correlation functions. The anomalous dimensions of the multi-trace operators can be obtained from the Mellin amplitudes. To see how this works, note that a contour integral around a double pole gives
\be
\oint  \frac{d \delta}{2\pi i} \frac{x^{- \delta}}{(\delta - \Delta)^2}  = -x^{-\Delta} \log x\,.
\ee
Logarithms are just the perturbative manifestation of the anomalous dimensions, because $x^\gamma \approx 1+\gamma \log x$.  So anomalous dimensions can be extracted by studying the double and higher poles of the Mellin amplitude.  As a concrete example, the 4-pt amplitude in $g (\phi \chi)^2$ theory has factors of $\Gamma(\delta_{ij})$ in the Mellin integrand.  So in the OPE channel combining $\CO_\phi$ and $\CO_\chi$ as operators $1$ and $2$, we have $\delta_{12} = \delta_{34}$ due to the constraints (recall that this is analogous to $(p_1 + p_2)^2 = (p_3 + p_4)^2$).  This leads to double poles in the Mellin integrand due to $\Gamma(\delta_{12}) \Gamma(\delta_{34})$, which tell us that the operator $\CO_\phi \CO_\chi$ has received an anomalous dimension proportional to $g$.  However, if we look at the double-trace operator $\CO_\phi \CO_\phi$ in a theory where $\Delta_\phi - \Delta_\chi$ is non-integral, then there are only single poles, since $\delta_{12} = \delta_{34} - \Delta_\phi + \Delta_\chi$.  So the operator $\CO_\phi^2$ does not receive an anomalous dimension in this theory, as expected.  One can go on to consider more complicated examples, although it is necessary to disentangle the contributions from primaries and descendants.

The above discussion leads to the general expectation that the inclusion of the $\Gamma(\delta_{ij})$ factors in the defining equation (\ref{mellindefn}) should lead to simple Mellin amplitudes for Witten diagrams. This effect is most extreme when we consider the Mellin amplitude of an $n$-point scalar contact interaction in AdS, which is simply a constant \cite{Penedones:2010ue}, 
\be
\label{ContactInteraction}
\int_{AdS} d^{d+1} X \prod_{i=1}^n \frac{\mathcal{C}_{\Delta_i} }{(-2 P_i \cdot X)^{\Delta_i}} =
 \frac{\pi^h }{2}\Gamma\left(\frac{\sum_{i=1}^n \Delta_i - d}{2}\right)
\prod_{i=1}^n \frac{\mathcal{C}_{\Delta_i}}{\Gamma(\Delta_i)}
 \int [d \delta] \prod_{i<j}^n \Gamma(\delta_{ij}) P_{ij}^{-\delta_{ij}}\,. 
\ee
This is analogous to the fact that an insertion of a $\phi^n$ vertex is a constant in the momentum space Feynman rules for scattering amplitudes.

Since this formula will be one of the main ingredients in the remainder of the paper, let us quickly review its derivation, which is essentially a slight generalization of a result by Symanzik \cite{Symanzik}. One first writes the bulk-boundary propagators as
\be
\frac{1 }{(-2 P_i \cdot X)^{\Delta_i}} = \frac{1}{\Gamma(\Delta_i)} \int_0^\infty \frac{dt_i}{t_i} t_i^{\Delta_i} e^{- 2 t_i P_i \cdot X}
\ee
In this representation, it is easy to integrate over the bulk coordinate $X$ to obtain an exponential of $\sum t_i t_j P_{ij}$.  Then the key is to represent some of the terms in the exponential using the standard Mellin identity
\be
e^{-z} = \int \frac{ds}{2\pi i} \ \! \Gamma(s) z^{-s}
\ee
and evaluate the $t_i$ integrals, leading to the Mellin representation with the constraints that we have discussed.  This is how our standard representation for the Mellin amplitude arises in the case of the simplest AdS amplitude.

%%% Local Variables: 
%%% mode: latex
%%% TeX-master: "MellinFactorization"
%%% End: 

\section{Factorization of AdS/CFT Correlators}
\label{sect:AdSCFTFactorization}

In this section we will show how to recursively compute the tree-level Mellin amplitude for any theory of scalar fields, including any number of derivative couplings.  As modest examples, in section \ref{sec:examples}, we will compute the $5$-pt and $6$-pt functions in a theory with 3-pt contact interactions among scalar fields with arbitrary masses, dual to operators with dimensions $\Delta_i$. 

These computational advances are made possible by a factorization formula that we will derive shortly, which says that a Witten diagram with a propagator that divides the amplitude into a left and right piece (see Fig. \ref{fig:factorizationbubu}) will have simple poles at
\be
\delta_{LR} =\Delta +2m\ , \ \ \ \ \mathrm{where} \ \ \ \  \delta_{LR} = \sum_{i=1}^k \Delta_i - 2\sum_{i < j \leq k} \delta_{ij}
\ee
is the direct analogue of the familiar intermediate propagator variable from flat space scattering amplitudes, $-(p_1 + \cdots + p_k)^2$.  Here $\Delta$ is the dimension and twist of the exchanged operator; since we are dealing with bulk scalar fields these are identical.  The residues at these poles are
\be
-4 \pi^h \frac{\Gamma(\Delta-h+1)  m!}{(\Delta-h+1)_m} \,L_m(\delta_{ij})\,R_m(\delta_{ij})
\label{MainFormula}
\ee
where
\begin{align}
L_m (\delta_{ij})&=
\sum_{\sum n_{ij}=m} M^L_{k+1} (\delta_{ij} + n_{ij}) \prod_{i < j}^k \frac{  (\delta_{ij})_{n_{ij}}}{n_{ij}!} \label{definitionofLm}\\
R_m(\delta_{ij}) &= 
\sum_{ \sum n_{ij}=m} 
M_{n-k+1}^R(\delta_{ij}+n_{ij}) 
\prod_{k<i<j}^n \frac{  (\delta_{ij})_{n_{ij}} }{n_{ij}!}  
\end{align}
which depend only on the lower point diagrams or Mellin amplitudes, $M_{k+1}^L$ and $M^R_{n+1-k}$.  We remind the reader that the Pochhammer symbol $(x)_n = \Gamma(x+n)/\Gamma(x)$. The derivation is given in section \ref{sect:FactorizationonAdSPropagators}. The basic idea is to re-write the Mellin representation of a Witten diagram with a particular propagator in terms of the amplitudes to the left and right of this propagator.  Then we massage the propagator into a form such that the entire Witten diagram is written in terms of lower-point Mellin amplitudes.  This will allow us to identify all of the poles and residues in the appropriate $\delta_{LR}$ variable.  

Our factorization formula can be applied to any factorization channel.   The equivalence of all the different possible recursive applications of the formula seems to be a very strong constraint on the form of Mellin amplitudes.  This suggests that there may exist a single set of diagrammatic rules whereby an arbitrary Mellin amplitude can be constructed.  In fact, in section \ref{sec:FeynmanRules} we will see that such a set of rules exists, and we will prove that it gives the same results as the factorization formula. 

Finally, in section \ref{sec:FeynmanRules} we will show that our diagrammatic rules and our factorization formula satisfy the functional equation from section \ref{sect:FunctionalEquation}, and in appendix \ref{sect:MellinDeterminedByPoles} we complete the proof that our formulas are identical to the Mellin representation of the Witten diagram.   Thus in any scalar theory, one can compute all tree-level Witten diagrams either recursively, or by explicit construction using the diagrammatic rules.

\subsection{Factorization on AdS Propagators}
\label{sect:FactorizationonAdSPropagators}

\subsubsection{An Integral Formula}

\begin{figure}
 \centering
    \includegraphics[width=0.9\textwidth]{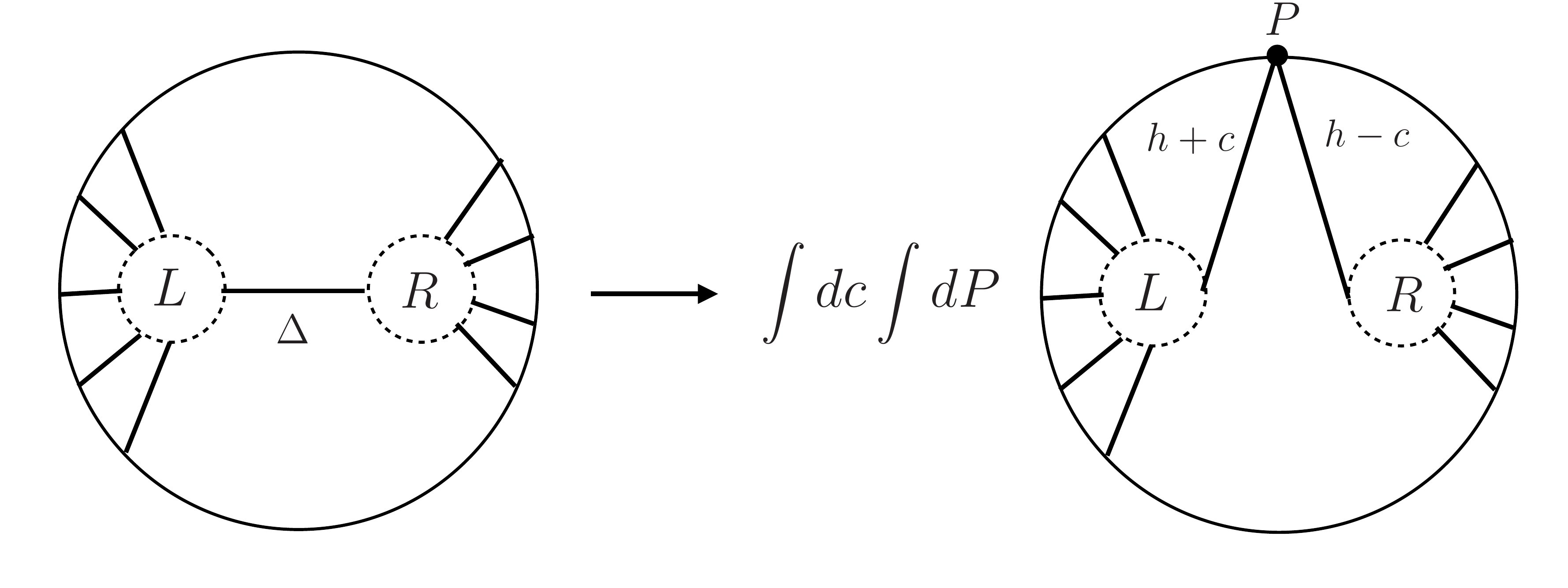}
      \caption{A pictorial representation of the derivation of the factorization formula.}
      \label{fig:factorizationbubu}
\end{figure}

Our starting point is the the following formula for the bulk-to-bulk scalar propagator \cite{Penedones:2010ue}
\be
G_{BB}(X,Y)=\int_{-i\infty}^{i\infty} \frac{dc}{2\pi i} \frac{2 c^2}{c^2-(\Delta-h)^2}
\int dP \frac{\mathcal{C}_{h+c}}{(-2P\cdot X)^{h+c}} \frac{\mathcal{C}_{h-c}}{(-2P\cdot Y)^{h-c}}
\ee
where we recall that $h=d/2$.  Notice the appearance of two bulk-to-boundary propagators in the integrand.
Beginning with an $n$-point Witten diagram, we will use this representation for a specific internal propagator. This propagator will break the $n$-point Witten diagram into two Witten diagrams integrated over a common boundary point,
\be
A_n(P_i)=\int_{-i\infty}^{i\infty} \frac{dc}{2\pi i} \frac{2c^2}{c^2-(\Delta-h)^2}
\int dP A_{k+1}^L (P_i,P)A_{n-k+1}^R (P_i,P) \label{factorAn}
\ee
We represented equation \eqref{factorAn} pictorially in Fig. \ref{fig:factorizationbubu}. The left (L) amplitude has $k+1$ external legs with weights $\Delta_i$ at point $P_i$ for $i=1,2,\dots,k$ and weight $h+c$ at point $P$. The right (R) amplitude has $n-k+1$ external legs with weights $\Delta_i$ at point $P_i$ for $i=k+1,k+2,\dots,n$ and weight $h-c$ at point $P$.  In other words, the operator with coordinate $P$ on $A^R$ is the shadow of the operator with coordinate $P$ on $A^L$. 
Now we write the left and right amplitudes in the Mellin representation
\begin{align}
\label{Aleftright}
&A_{k+1}^L (P_i,P)=   \int  [d \tilde{\delta}]_L[dl]_L\,
M_{k+1}^L(\tilde{\delta}_{ij},l_i)
\prod_{i<j}^k \Gamma(\tilde{\delta}_{ij}) (-2P_i\cdot P_j)^{-\tilde{\delta}_{ij}}
\prod_{i=1}^k \Gamma(l_i) (-2P_i\cdot P)^{-l_i}  
\\
&A_{n-k+1}^R (P_i,P)=  \int  [d \tilde{\delta}]_R [dl]_R\,
M_{n-k+1}^R(\tilde{\delta}_{ij},l_i)
\prod_{k<i<j}^n \Gamma(\tilde{\delta}_{ij}) (-2P_i\cdot P_j)^{-\tilde{\delta}_{ij}}
\prod_{i=k+1}^n \Gamma(l_i) (-2P_i\cdot P)^{-l_i} \nonumber 
\end{align}
 The integration measure $[d \tilde{\delta}]_L[dl]_L$ is constrained by
\be
\begin{split}
\sum_{i=1}^k l_i &= h+c \\
l_i+ \sum_{j\neq i}^k \tilde{\delta}_{ij} &=\Delta_i \qquad (i=1,2,\dots,k)
\end{split}
\ee
and analogously for $[d \tilde{\delta}]_R[dl]_R$ we have the constraints
\be
\begin{split}
\sum_{i=k+1}^n l_i &= h-c \\
l_i+ \sum_{k<j\neq i}^n \tilde{\delta}_{ij} &=\Delta_i \qquad  (i=k+1,\dots,n)\,.
\end{split}
\label{constraints'}
\ee
%  \be
% \sum_{i=1}^k l_i=h+c\ ,\ \ \ \ \sum_{i=k+1}^n l_i=h-c \ , \ \ \ \ \ \left\{ \begin{array}{ll}  l_i+ \sum_{j\neq i}^k \tilde{\delta}_{ij} =\Delta_i\ ,  &  \  (i=1,2,\dots,k) \ , \\  l_i+ \sum_{k<j\neq i}^n \tilde{\delta}_{ij} =\Delta_i \ , & \ (i=k+1,\dots,n) \ \end{array} \right\}
% \label{constraints'}.
%  \ee
 The equations in \eqref{Aleftright} may appear complex, but they encode very simple information, namely that $A^L$ is a CFT correlator between the $k$ operators on the left of the propagator and a new operator with dimension $h+c$, and equivalently for $A^R$.  The constraint equations are identical to what we would find from momentum conservation if we introduced fictitious $p_i$ with $\delta_{ij} = p_i \cdot p_j$ and $p_i^2 = -\Delta_i$.
    
Inserting now the Mellin representations \eqref{Aleftright} in (\ref{factorAn}) we find
\begin{align}
&A_n(P_i)=\int_{-i\infty}^{i\infty} \frac{dc}{2\pi i} \frac{2c^2}{c^2-(\Delta-h)^2} 
 \int  [d \tilde{\delta}]_L[dl]_L\,
M_{k+1}^L(\tilde{\delta}_{ij},l_i)
\prod_{i<j}^k \Gamma(\tilde{\delta}_{ij}) (-2P_i\cdot P_j)^{-\tilde{\delta}_{ij}}
\nonumber \\&
\int  [d \tilde{\delta}]_R [dl]_R\,
M_{n-k+1}^R(\tilde{\delta}_{ij},l_i)
\prod_{k<i<j}^n \Gamma(\tilde{\delta}_{ij}) (-2P_i\cdot P_j)^{-\tilde{\delta}_{ij}}
\int dP \prod_{i=1}^n \Gamma(l_i) (-2P_i\cdot P)^{-l_i}
\end{align}
Notice that the contour integral over $c$ requires knowledge of the Mellin amplitudes $M^L_{k+1}$ and $M^R_{n-k+1}$ for general (complex) external scaling dimensions $\Delta_i$. As we will see explicitly in all the examples below, for general Witten diagrams we obtain a Mellin amplitude which is an analytic function of the $\Delta_i$ and so this is not problematic.

%As we saw in the case of the fundamental equation (\ref{ContactInteraction}), %\JP{Not quite. Improve this reference after correcting in section 2.}
%which is very similar to the Mellin representation of an AdS contact interaction, 
The last integral was studied by Symanzik \cite{Symanzik} and has a simple Mellin representation
\be
\int d^d P \prod_{i=1}^n \Gamma(l_i) (-2P_i\cdot P)^{-l_i}
=\pi^h \int  [d \delta]\, 
\prod_{i<j}^n \Gamma( \delta_{ij}) P_{ij}^{-\delta_{ij}} 
\ee
where the measure $[d \delta]$ is constrained by $\sum_{j\neq i}^n \delta_{ij} = l_i$.  We have reduced the dependence on all external kinematic invariants to the Mellin form!  This means that we can shift the integration variables $\delta_{ij} \to \delta_{ij} -\tilde{\delta}_{ij}$ to obtain the following expression for the Mellin amplitude of the original $n$-point diagram
\be
M_n(\delta_{ij})= \int_{-i\infty}^{i\infty} \frac{dc}{2\pi i} \frac{2\pi^h c^2}{c^2-(\Delta-h)^2}\  L \times R 
\label{factorMellin}
\ee
where
\begin{align}
L&=\int  [d \tilde{\delta}]_L[dl]_L\,
M_{k+1}^L(\tilde{\delta}_{ij},l_i)
\prod_{i<j}^k \frac{\Gamma(\tilde{\delta}_{ij}) \Gamma(\delta_{ij}-\tilde{\delta}_{ij})}{\Gamma(\delta_{ij})},\label{Lintegrals}\\
R&=\int  [d \tilde{\delta}]_R[dl]_R\,
M_{n-k+1}^R(\tilde{\delta}_{ij},l_i)
\prod_{k<i<j}^n \frac{\Gamma(\tilde{\delta}_{ij}) \Gamma(\delta_{ij}-\tilde{\delta}_{ij})}{\Gamma(\delta_{ij})}. \nn
\end{align}
It is crucial that the factor $L$  depends only on $\delta_{ij}$ with $1\le i,j \le k$ and the $R$ factor  depends only on $\delta_{ij}$ with $k< i,j \le n$.
% Also note the similarity of these expressions to the shadow transformation derived in the pervious section.
Using the constraints (\ref{constraints'}) we can solve for and eliminate the $l_i$ variables, which are the last vestige of the spacetime version of the internal propagator.  This leaves the single constraint  
\be
2\sum_{  i<j}^k \tilde{\delta}_{ij}+h\pm c -\sum_{i=1}^k  \Delta_i=0 \label{singleconstraintI}
\ee
which expresses the dimension $h\pm c$ of the internally propagating operator in terms of the other integration variables in $L$ and $R$, respectively.

Let us end the presentation of the factorization formula with some comments regarding the relation with the OPE decomposition as presented in equation \eqref{Aalternative}. At first sight, the above factorization is very similar to the factorization appearing in equation \eqref{Aalternative}. The difference lies in the additional integral over $c$ in the evaluation of the Witten diagram, which in the analogous CFT computation is localized at $c = \Delta - h$. Although we will not prove this statement here, we claim that this additional integral is eventually responsible for correctly taking into account the contributions of multi-trace operators. (At the level of the four-point function this follows from equation (38) of \cite{Penedones:2010ue}, where the integral over $c$ is explicitly responsible for the multi-trace poles in $\delta_{12}$ and $\delta_{34}$.) Furthermore, because the choice of contour for the $c$-integral breaks the symmetry between the field and the shadow field our expressions can be made completely exact and are not of the formal form as in equation \eqref{Aalternative}. It would be interesting to make the relation between the multi-trace contributions and the $c$ integral precise, for which an analogous discussion of the conformal partial wave decomposition in \cite{Mack} should also be very useful.

\subsubsection{Identifying the Poles}

The advantage of equation (\ref{factorMellin}) is that we can use it to determine the residues of the poles of the $n$-point Mellin amplitude in the factorization variable
\be
\label{defndeltaLR}
\delta_{LR}=\sum_{i=1}^k\sum_{j=k+1}^n \delta_{ij} 
= \sum_{i=1}^k \left(\Delta_i-\sum_{j\neq i}^k \delta_{ij}\right) , 
\ee
Notice that the equivalence between the two expressions on the right-hand side follows from the constraints \eqref{eq:deltaconstr}. As we discussed above, $\delta_{LR}$ is the direct analogue of the kinematic invariant that vanishes at the factorization channel in a scattering amplitude, such as $-(p_1 + \cdots + p_k)^2$ in an $n$-pt scattering amplitude.

Contour integrals have singularities when a pair of poles in the integrand collides and squeezes the contour of integration between them.  We have a many-dimensional contour integral, so identifying such occurrences would naively be a daunting proposition, but our task is greatly simplified by the structure of the integrand, its contour, and the fact that we are focusing on $\delta_{LR}$.  In particular, since $\delta_{LR}$ involves all of the left or all of the right $\delta_{ij}$, we can obtain a pole in this variable only if we use the poles from all the $\Gamma(\delta_{ij} - \tilde \delta_{ij})$ in (\ref{Lintegrals}), which form a semi-infinite sequence towards positive real infinity in the $\tilde \delta_{ij}$ integration variables.  We obtain a pole in $\delta_{LR}$ when these singularities collide with the `propagator singularity' at $c = \pm (\Delta - h)$, where $c$ is a function of the $\tilde \delta_{ij}$ from the constraints.  

Before studying the general case, let us analyze the case where $M^L$ and $M^R$ are independent of $\delta_{ij}$ so we can perform the integrals in \eqref{Lintegrals} explicitly. This case corresponds to a Witten diagram with a single bulk-to-bulk propagator connecting a $(k+1)$-vertex to a $(n-k+1)$-vertex. To compute the integrals in \eqref{Lintegrals}, one starts by eliminating $\delta_{12}$ using the constraint \eqref{singleconstraintI}. Then, all other $\delta_{ij}$ are independent integration variables and, for constant $M^L$, the integrals can be computed iteratively using Barnes' lemma:
\be
\frac{1}{2\pi i} \int_{-i\infty}^{i\infty} \Gamma(a+ s)\Gamma(b+s)\Gamma(c-s)\Gamma(d-s)ds = \frac{\Gamma(a+c)\Gamma(a+d)\Gamma(b+c)\Gamma(b+d)}{\Gamma(a+b+c+d)}\ .
\ee
This gives
\be
\label{Lexactcontact}
L =  \frac{\Gamma\left(\frac{\sum_{i=1}^k \Delta_i - h - c}{2}\right) 
\Gamma\left(\frac{h + c - \delta_{LR}}{2}\right)}{\Gamma\left(\frac{\sum_{i=1}^k \Delta_i - \delta_{LR}}{2}\right)}M^L\ ,
\ee
which has poles at $c=\delta_{LR}-h-2m$, for $m=0,1,2,\dots$, with residue
\be
2\frac{(-1)^m}{m!} \frac{\Gamma(\sum_{i< j}^k \delta_{ij} + m)}{\Gamma(\sum_{i< j}^k \delta_{ij})}M^L\ .
\ee
In this expression, $M^L$ depends on the integration variable $c$. It is convenient to
make this dependence explicit. Using \eqref{ContactInteraction} we find that
\be
M^L(c)= \frac{\Gamma\left(\frac{\sum_{i=1}^k \Delta_i - h + c}{2}\right) }{4\Gamma(c+1)}
\prod_{i=1}^k \frac{\mathcal{C}_{\Delta_i}}{\Gamma(\Delta_i)} =
\frac{\Gamma(\Delta-h+1) \Gamma\left(\frac{\sum_{i=1}^k \Delta_i - h + c}{2}\right)
}{\Gamma(1+c)\Gamma\left(\frac{\sum_{i=1}^k \Delta_i +\Delta- 2h}{2}\right)}M^L_{k+1}
\ee
where $M^L_{k+1}$ stands for the left Mellin amplitude with physical external dimensions $\Delta_i$ and $\Delta$. Similarly, the right part of the diagram gives
\be
\label{Rexactcontact}
R=  \frac{\Gamma\left(\frac{\sum_{i>k}^n \Delta_i - h + c}{2}\right) 
\Gamma\left(\frac{h - c - \delta_{LR}}{2}\right)}{\Gamma\left(\frac{\sum_{i>k}^n \Delta_i - \delta_{LR}}{2}\right)}
\frac{\Gamma(\Delta-h+1) \Gamma\left(\frac{\sum_{i>k}^n \Delta_i - h - c}{2}\right)
}{\Gamma(1-c)\Gamma\left(\frac{\sum_{i>k}^n \Delta_i +\Delta-2h}{2}\right)}M^R_{n-k+1}\ ,
\ee
which has poles at $-c=\delta_{LR}-h-2m$ for $m=0,1,2,\dots$.
%\be
%2\frac{(-1)^m}{m!} \frac{\Gamma(\sum_{k<i< j}^n \delta_{ij} + m)}{\Gamma(\sum_{k<i< j}^n \delta_{ij})}M^R\ .
%\ee
The poles in $\delta_{LR}$ of the $n$-point Mellin amplitude arise from pinching the $c$ integration contour in (\ref{factorMellin}) between two poles of the integrand. 
\begin{figure} 
  \centering
    \includegraphics[width=0.6\textwidth]{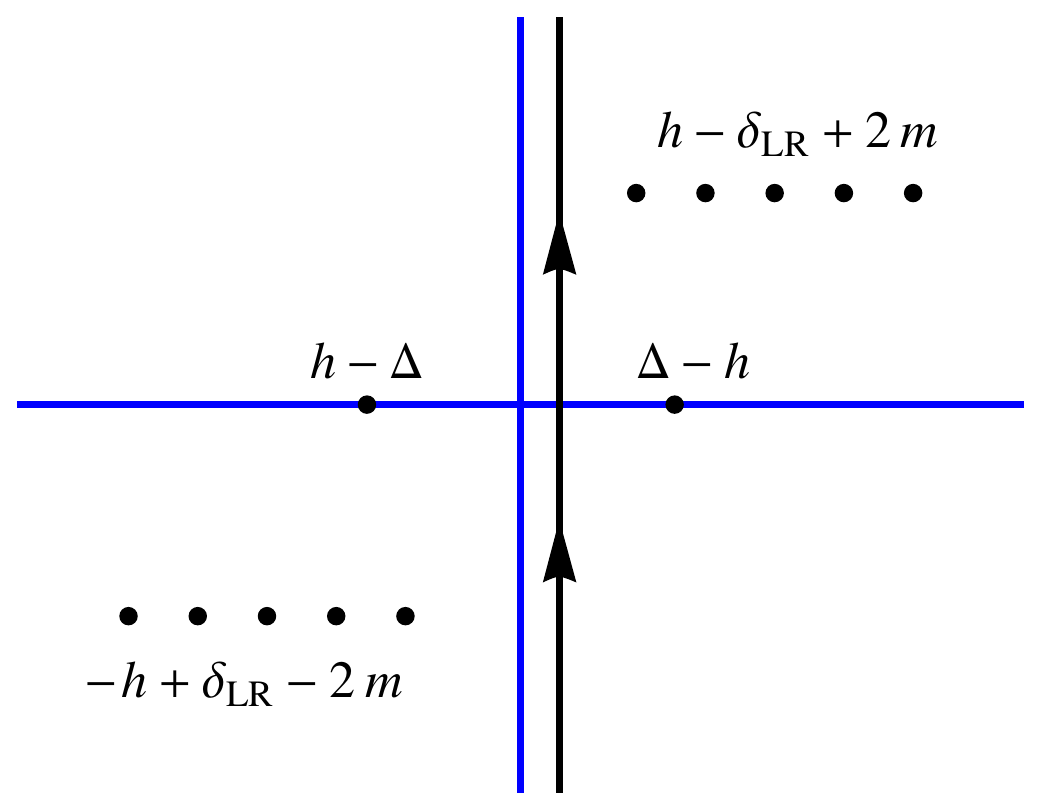} 
      \caption{Structure of poles of the integrand of the factorization equation \eqref{factorMellin} in the $c$ complex plane. When $\delta_{LR}=\Delta+2m$, the integration contour is pinched between poles at two places. }
      \label{fig:poles}
\end{figure}
From Fig. \ref{fig:poles}, it is clear that, for $\delta_{LR}=\Delta+2m$, the contour will be pinched in two places, at $c=\Delta-h$ and at $c=h-\Delta$. The contribution to the residue of $M_n$ at $\delta_{LR}=\Delta+2m$ from the pinching at $c=\Delta-h$ involves the product of the  residue of $L$ times $R$ evaluated at the pole. This gives a residue of the form 
\begin{align}
&
\pi^h(\Delta-h) \ 2\frac{(-1)^m}{m!} \left[ \frac{\Gamma(\sum_{i< j}^k \delta_{ij} + m)}{\Gamma(\sum_{i< j}^k \delta_{ij})}M^L_{k+1} \ R|_{c=\Delta-h}\right]_{\delta_{LR}=\Delta+2m} \nonumber \\
=& -\frac{2\pi^h \Gamma(\Delta-h+1)}{m! (\Delta-h+1)_m } M^L_{k+1} M^R_{n-k+1}  
\left[  \Big(\sum_{i< j}^k \delta_{ij}\Big)_m \,\Big(\sum_{k<i< j}^n \delta_{ij} \Big)_m\right] _{\delta_{LR} = \Delta + 2m} \label{leftrightresidue}
\end{align}
which is symmetric between left and right. The other contribution is equal and just doubles this residue. In \eqref{leftrightresidue}, we are always evaluating the residue at the pole $\delta_{LR} = \Delta + 2m$, which using \eqref{defndeltaLR} implies a constraint for the $\delta_{ij}$ appearing in \eqref{leftrightresidue}. Notice however that those $\delta_{ij}$, being the arguments of the full Mellin amplitude \eqref{factorMellin}, are in principle not subject to this extra constraint and we therefore added the square brackets and the explicit $\delta_{LR} = \Delta + 2m$ to indicate this extra constraint. %One way to implement the constraint would be by solving for one of the $\delta_{ij}$ in \eqref{leftrightresidue}, for example $\delta_{12}$, and express it in terms of the other $\delta_{ij}$ and $\Delta + 2m$. This however unnecessarily breaks the manifest permutation symmetries of the above expression and this is why we opted for the notation above.
%\JP{I think this example helps the reader but if you think it makes the text too heavy we can move it to an appendix. In any case, it is important for us because it explains the missing factor of 4.}  

Let us now consider the general case.  First consider the contribution from the collision of poles of $L$ with the pole at $c=\Delta-h$ in (\ref{factorMellin}). The relevant poles from the left $\tilde \delta_{ij}$ arise from the pinching of
\be
\Gamma(\delta_{ij} - \tilde \delta_{ij}) \ \ \ \mathrm{and} \ \ \ \frac{2\pi^h c^2}{c^2-(\Delta-h)^2}
\ee
where we remind the reader that $c$ is linked to the $\tilde \delta_{ij}$ by constraints.  This happens when the $\delta_{ij}$ variables are such that
\be
\tilde \delta_{ij} = \delta_{ij} + n_{ij} % \qquad \qquad (ij) \neq (12)
\ee
This gives the following contribution to the residue of $M_n$ at the pole $\delta_{LR}=\Delta+2m$, 
\begin{align}
 2\pi^h (\Delta-h) \left[\sum_{\sum n_{ij} = m}   M^L_{k+1} (\delta_{ij} + n_{ij}) \prod_{i < j}^k \frac{(-1)^{n_{ij}}}{n_{ij}!} \frac{\Gamma(\delta_{ij} + n_{ij})}{\Gamma(\delta_{ij})} \times R \right]_{\delta_{LR} = \Delta + 2m}
\label{eq:MellinLeftI}
\end{align}
where it is important that the sum of the $n_{ij}$ are constrained to equal $m$. Notice that evaluating the $\delta_{ij}$ at the pole $\delta_{LR} = \Delta + 2m$ also ensures that the arguments of $M^L_{k+1}$ always satisfy the appropriate constraints (which follows from \eqref{defndeltaLR}) where $M^L_{k+1}$ is well-defined. This is almost our factorization formula eq. (\ref{MainFormula}), except that here the arguments in $R$ are constrained by eq. (\ref{singleconstraintI}) with $c=\Delta-h$.  Thus, it has the appearance of a lower-point Mellin amplitude where one of the legs has been replaced with a shadow field, exactly as we should expect from our discussion in section \ref{sect:FactorizationofCFT}.  In Appendix \ref{app:shadow}, we prove an identity relating the Mellin amplitude $\tilde{M}$ with a  shadow field replacement to the original Mellin amplitude:
\begin{align}
&\left[\int [d\tilde{\delta}] \tilde{M}_{n-k+1}(\tilde{\delta}_{ij}) \prod_{k<i<j}^n \frac{
 \Gamma(\tilde{\delta}_{ij})\Gamma(\delta_{ij}-\tilde{\delta}_{ij})}{\Gamma(\delta_{ij})} \right]_{\delta_{LR} = \Delta + 2m} \nn \\
 =& 
 \left[- \frac{\Gamma(\Delta-h) (-1)^m m!}{(\Delta-h+1)_m}
\sum_{ \sum n_{ij}=m} 
M_{n-k+1}(\delta_{ij}+n_{ij}) 
\prod_{k<i<j}^n \frac{  (\delta_{ij})_{n_{ij}} }{n_{ij}!} \right]_{\delta_{LR} = \Delta + 2m} 
\label{eq:genshadow}
\end{align}
Notice that the arguments of $M^R_{n-k+1}$ again satisfy the required constraints.
%where the $\delta_{ij}$ satisfy the constraints for a $n$-point Mellin amplitude with weights $$\Delta+2m,\ \Delta_{k+1},\ \dots,\ \Delta_n\ .$$
Inserting this identity into equation (\ref{eq:MellinLeftI}), we obtain half of our factorization formula \eqref{MainFormula}. The other half comes from the collision of poles in $R$ with the pole at $c=h-\Delta$ in (\ref{factorMellin}).

\subsubsection{The Complete Factorization Formula and Its Interpretation}
We have shown that any Witten diagram will have a Mellin representation with the above poles and residues in the $\delta_{LR}$ channel. If the Mellin amplitude vanishes for large $\delta_{LR}$ then it would be completely determined by its poles and residues, and we would be able to write:
\be
M =  \sum_{m=0}^\infty   \frac{Res(m)}{\delta_{LR} - \Delta - 2m}  
\label{factorizationansatzmt}
\ee
with 
\be
Res(m) = -\frac{ 4\pi^h \Gamma^2(\Delta-h+1)  m!}{\Gamma(\Delta-h+1+m)} \left[ L_m(\delta_{ij})  R_m(\delta_{ij}) \right]_{\delta_{LR} = \Delta + 2m}
\label{factorizationansatzresmt}
\ee
where $L_m$ and $R_m$ are given in \eqref{definitionofLm}.  In fact, we will see in all examples that a stronger statement is true.  Our formula is equivalent to its projection onto all of its poles, not just the specific $\delta_{LR}$ singularity in the factorization formula, so that all of the explicit Pochhammer symbols $(\delta_{ij})_{n_{ij}}$ can be evaluated at poles.  If $M$ vanishes as any propagator goes to infinity, then this follows from the simple fact from complex analysis that 
\be
\sum_i \frac{f_i(z)}{z - a_i} = \sum_i \frac{f_i(a_i)}{z - a_i}
\ee
when the sum vanishes as $z \to \infty$.  In what follows, when we refer to our factorization formula we will almost always be referring to equation (\ref{factorizationansatzresmt}) with all $\delta_{ij}$ in the numerator projected onto poles, because it is this pole-projected formula that we will be able to prove.

In the remainder of this paper we provide strong evidence that the Mellin amplitude is in fact completely determined by its poles and therefore \eqref{factorizationansatzmt} \emph{is} the full answer. This we will do as follows. We will first show that our factorization formula implies a set of diagrammatic rules for the computation of Mellin amplitudes, and then we will show that these rules satisfy the functional equation from section \ref{sect:FunctionalEquation}. Assuming that Witten diagrams are polynomially bounded at large $\delta_{ij}$, as we discuss in appendix \ref{sect:MellinDeterminedByPoles}, this leads to a proof of the factorization formula \eqref{factorizationansatzmt} as well as our diagrammatic rules.  Additionally, we provide a more direct proof of \eqref{factorizationansatzmt} and a few other technical details in appendix \ref{sect:MellinDeterminedByPoles}.

\subsection{Adding Derivative Interactions}

The results above generalize to scalar theories with arbitrary derivative interactions, due to a beautiful interplay between the interaction vertices and our factorization formula.  In fact, we will see that all dependence on the $\delta_{ij}$ from derivatives passes through our factorization formula and simply leads to an overall factor multiplying the Mellin amplitude and the shift of a few constant factors.

As shown in \cite{Penedones:2010ue}, the Mellin amplitude for a contact interaction with an arbitrary number of derivatives is a polynomial. 
%\JP{The original statement was too strong} 
A convenient basis for these polynomials is
\be
M_k = g \prod_{i<j}^k (\delta_{ij})_{a_{ij}}
\ee
where the $a_{ij}$ are integers related to the number of derivatives coupling field $i$ to field $j$. (Notice however that the derivation below goes through for non-integral $a_{ij}$ as well.)  If we plug this amplitude into  our factorization formula, we find
\bea
R_m &=&  \sum_{\sum n_{ij} =m} \prod_{i<j} \frac{(\delta_{ij})_{n_{ij}}}{n_{ij}!} \times g \prod_{ij} (\delta_{ij} + n_{ij})_{a_{ij}} \\
&=& g \prod_{ij} (\delta_{ij})_{a_{ij}} \times \sum_{\sum n_{ij} =m} \prod_{i<j} \frac{(\delta_{ij} + a_{ij})_{n_{ij}}}{n_{ij}!} 
\eea
and equivalently for $L_m$.  In other words, $M_k$ has passed through the Pochhammer symbols from the factorization formula, and so we can evaluate it at the original $\delta_{ij}$ and not at $\delta_{ij} + n_{ij}$.  The only effect of the derivative interaction is to shift the Pochhammer symbols in the factorization formula, but since these $\delta_{ij}$ are naturally evaluated at poles, this only shifts certain constants in the residues at those poles.  

This effect persists under recursion, so Mellin amplitudes with many derivative interactions are simply given by the equivalent amplitude without derivatives with some constant shifts and an overall factor from all of the various polynomial Mellin amplitudes from the derivative interactions.  In particular, this means that if Mellin amplitudes without derivative interactions are entirely determined by their poles, then our factorization formula applies to all scalar theories in AdS.  Another way of saying this is that when we add derivative interactions, the `skeleton diagrams' with only the propagators are basically just `dressed' by a polynomial coming from the derivatives at vertices.

\section{Sample Computations}
\label{sec:examples}
In this section we will demonstrate the power of our formalism by computing the 5-pt and 6-pt amplitudes in a scalar field theory with 3-pt interaction vertices. Notice that, as will become clear below, using the factorization formula it is even easier to compute amplitudes in theories with general $\nabla^a \phi^b$ vertices, since the greatest complication arises from having many bulk to bulk propagators.

\begin{figure} 
  \centering
    \includegraphics[width=0.4\textwidth]{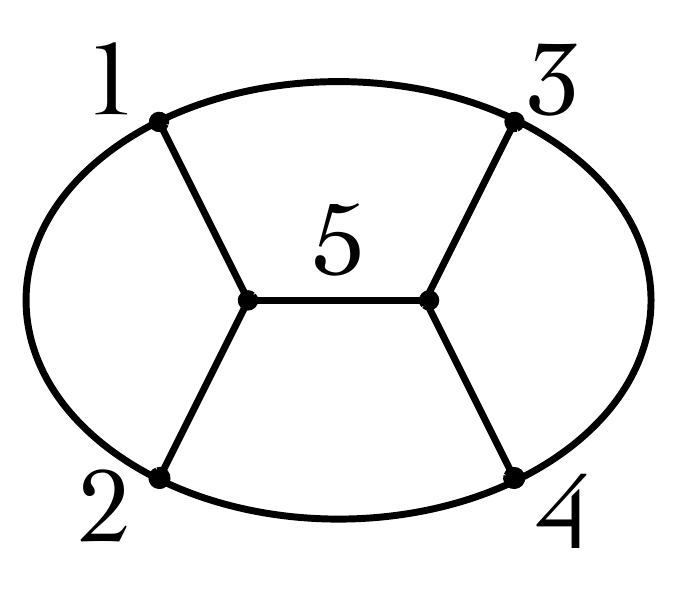}
    \includegraphics[width=0.45\textwidth]{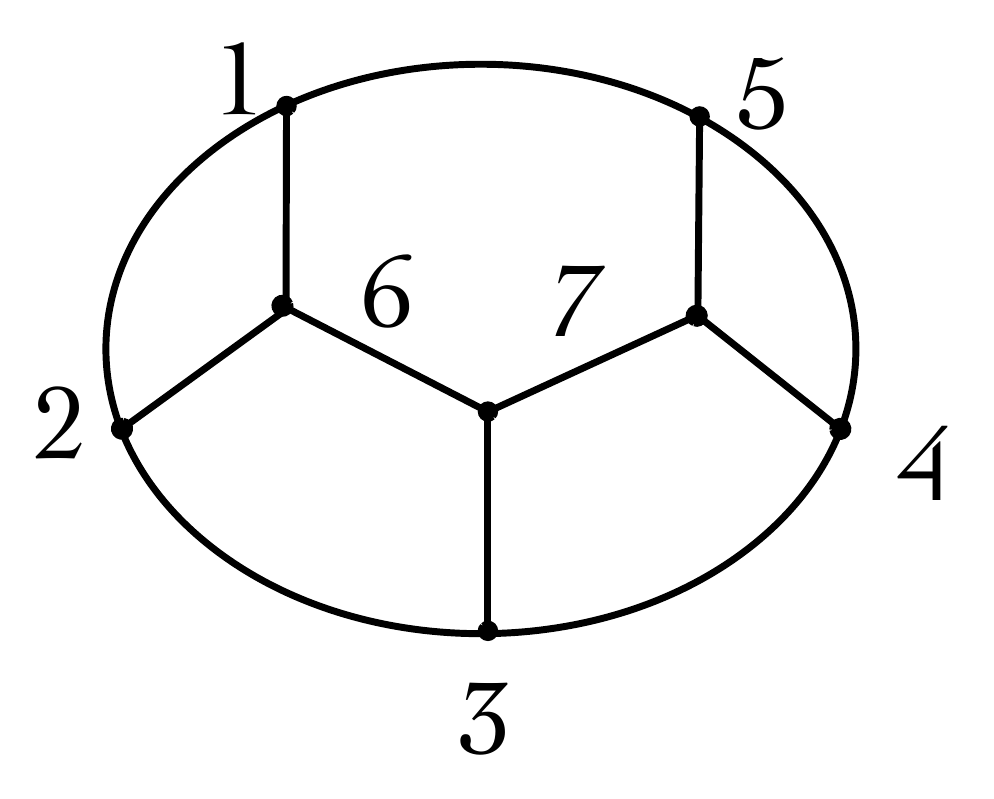}
      \caption{Four-point and five-point Witten diagrams in cubic scalar theory.}
      \label{4ptFigure}
\end{figure}

Before moving on to a non-trivial computation, let us see how our formalism works in the simplest case, that of the 4-pt function. Suppose specifically that we have the bulk interaction vertices $\lambda \phi_1 \phi_2 \phi_5$ and $\lambda \phi_3 \phi_4 \phi_5$, and we want the Mellin amplitude for $\langle \CO_1 \CO_2 \CO_3 \CO_4 \rangle$ from $\phi_5$ exchange, as shown in Fig. \ref{4ptFigure}.  
Applying equation (\ref{factorizationansatzmt}), we find the Mellin amplitude is
\ba
M_4(\delta_{ij}) &=& \sum_m \frac{1}{\delta_{LR} - \Delta_5-2m} 
\frac{-4 \pi^h \Gamma(\Delta_5 -h+1) m!}{(\Delta_5 -h+1)_m}
\left[ \left( \lambda_{125} \frac{(\delta_{12})_m}{m!} \right)\left( \lambda_{345} \frac{(\delta_{34})_m }{m!} \right) \right]_{\delta_{LR} = \Delta+2m} \nn\\
&=&  \sum_m \frac{1}{\delta_5 -m} 
\frac{-2 \pi^h \Gamma(\Delta_5 -h+1) m!}{(\Delta_5 -h+1)_m}
\left( \lambda_{125} \frac{1}{m!(\Delta_{12,5})_{-m}} \right)\left( \lambda_{345} \frac{1}{m!(\Delta_{34,5})_{-m}} \right)  ,
\ea
where $\Delta_{ij,k} \equiv \frac{\Delta_i + \Delta_j - \Delta_k}{2}$ and $\lambda_{ijk}$ is the 3-pt Mellin amplitude for a contact Witten diagram with external dimensions $\Delta_i$, $\Delta_j$ and $\Delta_k$.
In the second line, we have used the fact that $2\delta_{12} = -\delta_{LR} +
\Delta_1 + \Delta_2, 2 \delta_{34} =-\delta_{LR} + \Delta_3 + \Delta_4$, and
the identity $(a-m)_m = \frac{1}{(a)_{-m}}$.  
We have also introduced the notation
\be
2\delta_5 \equiv \delta_{LR} - \Delta_5 .
\ee
This will be convenient in amplitudes with multiple propagators because
in those cases there will be poles in many different specific linear combinations 
of the $\delta_{ij}$.  In this respect, the $\delta_i$'s are analogous to the combinations
$(p_1+\cdots+ p_k)^2+M^2$ in a flat-space diagram;  in fact, in terms of the equivalent fictitious Mellin ``momenta'' $p_i$,
$2\delta_5 = -(p_1+p_2)^2 - \Delta_5 = -(p_3+p_4)^2-\Delta_5$.
The major difference is that, while flat-space amplitudes have a single pole for each propagator, Mellin amplitudes have the full tower of poles
in eq. (\ref{factorizationansatzmt}):
\ba
\textrm{S-Matrix: } \frac{1}{\left( \sum p_i \right)^2+M^2} &\longleftrightarrow&
\textrm{ Mellin: } \frac{1}{\delta - m} \frac{-2\pi^h \Gamma(\Delta-h+1)m!}{(\Delta-h+1)_m}, \ \ m=0,1,\dots
\ea

Note also that the vertices, while more complicated than the simple flat-space factor $\lambda$, do not actually introduce additional $\delta_{ij}$-dependence, despite the naive appearance of equation (\ref{factorizationansatzresmt}), because
$\delta_{12}$ and $\delta_{34}$ are completely fixed by the constraint
$\delta_{LR}= \Delta_5 + 2m$.  So, continuing the comparison of the four-point function with that of flat space, the vertices differ by
\ba
\textrm{S-Matrix: } \lambda &\longleftrightarrow&
\textrm{ Mellin: } \frac{ \lambda_{ijk}}{m!(\Delta_{ij,k})_{-m}} , \ \ m=0,1, \dots
\ea
We will next turn to the evaluation of some higher-point amplitudes, where we will see that much of this structure continues to hold.  Thus although in section \ref{sect:AdSCFTFactorization} we were only able to prove that our factorization formula reproduced the correct poles and residues, in this and all other examples that we have computed, miraculous identities and cancellations seem to guarantee that this is the entire result. In other words, even the naive version of our formula \eqref{factorizationansatzmt}, with in particular the explicit Pochhammer factors $(\delta_{ij})_{n_{ij}}$ in equation \eqref{factorizationansatzresmt}, actually does vanish at infinity.

The components of the 4-point calculation will be so ubiquitous that it is useful to introduce notation for the factors that accompany the propagator, and for the vertices:
\be
S_\Delta(m) =  -\frac{2 \pi^h \Gamma(\Delta-h+1)m!}{(\Delta-h+1)_m}, \ \ \ \ \
V_{\Delta_i \Delta_j \Delta_k} (m) =\frac{\lambda_{ijk}}{m!(\Delta_{ij,k})_{-m}}.
\label{eq:prop}
\ee
This is our first hint of the Mellin space diagrammatic, which we will develop in section \ref{sec:FeynmanRules}.  Note that the vertex is not symmetric, because only the leg $k$ is `off-shell'.  We will obtain a general expression for this vertex with all legs `off-shell' in equation (\ref{3ptVertexExample}).

\subsection{5-pt Amplitude}

The next simplest amplitude is the five-point diagram shown in Fig.
\ref{4ptFigure}.  We can choose to apply our factorization formula,
equation (\ref{factorizationansatzmt}), on either propagator.  In this example, the diagram
is symmetric so the computation is identical either way, so let us decompose
on the internal line 7.  Then, the left Mellin amplitude
is the four-point diagram from the previous subsection, which
depends only on the  $\delta_{ij}$ through the combination 
\be
2\delta_6 = -(p_1+p_2)^2 -\Delta_6 = -2\delta_{12} + \Delta_1 + \Delta_2 -\Delta_6,
\ee
which is essentially the ``momentum'' flowing through line 6. Applying
(\ref{factorizationansatzmt}) with $2\delta_7 \equiv \delta_{LR}-\Delta_7$  thus gives
\ba
M_5(\delta_6,\delta_7) &=&
  \sum_m \frac{1}{\delta_7-m}  \frac{-2 \pi^h \Gamma(\Delta_7-h+1)m!}{(\Delta_7-h+1)_m} \nn\\
&& \times \left[ \left( \sum_{\sum n_{ij} = m} M_4(\delta_6-n_{12}) 
\frac{(\delta_{12})_{n_{12}}}{n_{12}!}
\frac{(\delta_{13})_{n_{13}}}{n_{13}!}
\frac{(\delta_{23})_{n_{23}}}{n_{23}!} \right) 
\left(\lambda_{457} \frac{(\delta_{45})_m}{m!} \right) \right]_{\delta_7=m}.
\label{eq:5pt1}
\ea
The fact that this entire amplitude can be written as just a function
of $\delta_6$ and $\delta_7$ is not yet manifest, since naively the
Pochhammer symbols contribute dependence on $\delta_{12},\delta_{13}$ and
$\delta_{23}$ separately.  However, $M_4$ depends only on $\delta_6= -\delta_{12} + \Delta_{12,6}$, so we can first do the sum over $n_{13}$ and $n_{23}$
with their sum $\tilde{n}=n_{13}+n_{23}$ fixed.  This is aided by the general identity\footnote{
This identity is easily proven by first noting that
$\sum_{n=0}^\infty \frac{z^n (\delta)_n }{n!} = (1-z)^{-\delta}$,
and then matching powers of $z$ in the product $\prod_{i<j} (1+z)^{-\delta_{ij}} = (1+z)^{-\sum_{i< j} \delta_{ij}}$.
}
\be
 \frac{\Gamma(\sum_{i< j} \delta_{ij} + \tilde{n})}{\tilde{n}!} = \sum_{\sum n_{ij} = \tilde{n}} \prod_{i < j}  \frac{\Gamma(\delta_{ij} + n_{ij})}{n!}  
\label{eq:pochsumid} .
\ee
Therefore, only the combinations $\delta_{12}$ and $\delta_{13}+\delta_{23}$
actually appear in the 5-point Mellin amplitude.  But, it is easy
to see that these combinations are completely fixed by $\delta_6$ and 
$\delta_7$!  The most immediate way to see this is by considering the
``momenta'' flowing through the diagram: $(p_1+p_2)$ flows through propagator
6, and $(p_1+p_2+p_3)=-(p_4+p_5)$ flows through propagator 7, so we have that
\ba
&& \delta_{45} = -\delta_7 + \Delta_{45,7}, \ \ \ \delta_{12} = -\delta_6 + \Delta_{12,6}, \ \ \  \delta_{12}+\delta_{13}+ \delta_{23} = -\delta_6 + \Delta_{123,7}.
\ea
Now, let us substitute our expansion for the four-point Mellin amplitude in 
eq. (\ref{eq:5pt1}):
\begin{align}
&M_5(\delta_6, \delta_7) =
 \sum_m \frac{1}{\delta_7-m}  \frac{-2 \pi^h \Gamma(\Delta_7-h+1)m!}{(\Delta_7-h+1)_m} \left( \frac{ \lambda_{457}}{m!(\Delta_{45,7})_{-m}} \right) \\&
\sum_{n_{12}+\tilde{n}=m} \sum_k \frac{V_{\Delta_1 \Delta_2 \Delta_6}(k-n_{12}) S_{\Delta_6}(k-n_{12})
V_{\Delta_3 \Delta_7 \Delta_6}(k-n_{12}) }{\delta_6-k} 
\frac{(\Delta_{12,6}-\delta_6)_{n_{12}}}{n_{12}!} \frac{(\Delta_{36,7} +\delta_6-m)_{\tilde{n}} }{\tilde{n}!} \nn
\end{align}
We have checked numerically that this is a valid, explicit and symmetric formula for the five-point function,
but we can simplify it by `projecting it onto its poles', \emph{i.e.} replace $\delta_6 \to k$ in the numerator.  In fact, in section \ref{sec:FeynmanRules} we will explicitly prove that our factorization formula holds \emph{after} this simplifying projection is performed, although, somewhat miraculously, in all examples this step has not actually been necessary.

The projection eliminates the $\delta_{ij}$ dependence from the Pochhammer symbols in the numerator, and we find
\ba
M_5(\delta_6, \delta_7) &=& \sum_{m,k} \frac{V_{\Delta_1 \Delta_2 \Delta_6}(k)
S_{\Delta_6}(k) V_{\Delta_3 \Delta_6 \Delta_7}(k,m) S_{\Delta_7}(m)
V_{\Delta_4 \Delta_5 \Delta_7}(m) }{(\delta_6-k)(\delta_7-m)} ,
\label{eq:5pt}
\ea
where we have defined a generalization of the vertex function
to include two indices:\footnote{Arriving at this expression for $V_{\Delta_3 \Delta_6 \Delta_7}(k,m)$ requires the use of a hypergeometric 
transformation identity.}
\ba
V_{\Delta_3 \Delta_6 \Delta_7}(k,m) &\equiv&
\frac{\lambda_{367}}{k!(\Delta_{37,6})_{-k} m! (\Delta_{36,7})_{-m} } 
{}_3F_2\left( {-k,-m,\frac{\Delta_3+\Delta_6+\Delta_7}{2} -h \atop \Delta_{367}-m ,  \Delta_{376} -k};1\right) .
\ea
This vertex is manifestly symmetric in $(\Delta_6,k) \leftrightarrow
(\Delta_7,m)$, as is required by reflection symmetry of the diagram.
Furthermore, one may easily see that it reduces to our earlier
vertex function $V_{\Delta_i \Delta_j \Delta_k}(m)$ when one of the indices
is set to zero:
\be
V_{\Delta_i \Delta_j \Delta_k}(0,m) = V_{\Delta_i \Delta_j \Delta_k}(m).
\ee

\subsection{6-pt Amplitude}

\begin{figure} 
  \centering
    \includegraphics[width=0.45\textwidth]{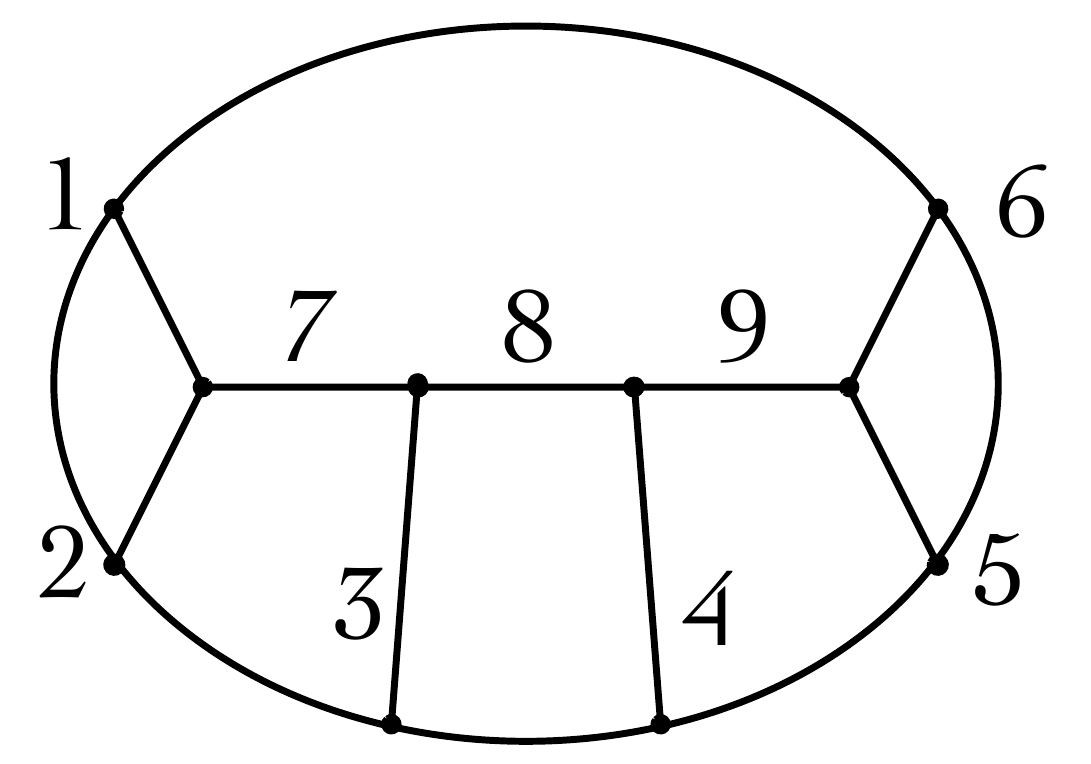} \includegraphics[width=0.45\textwidth]{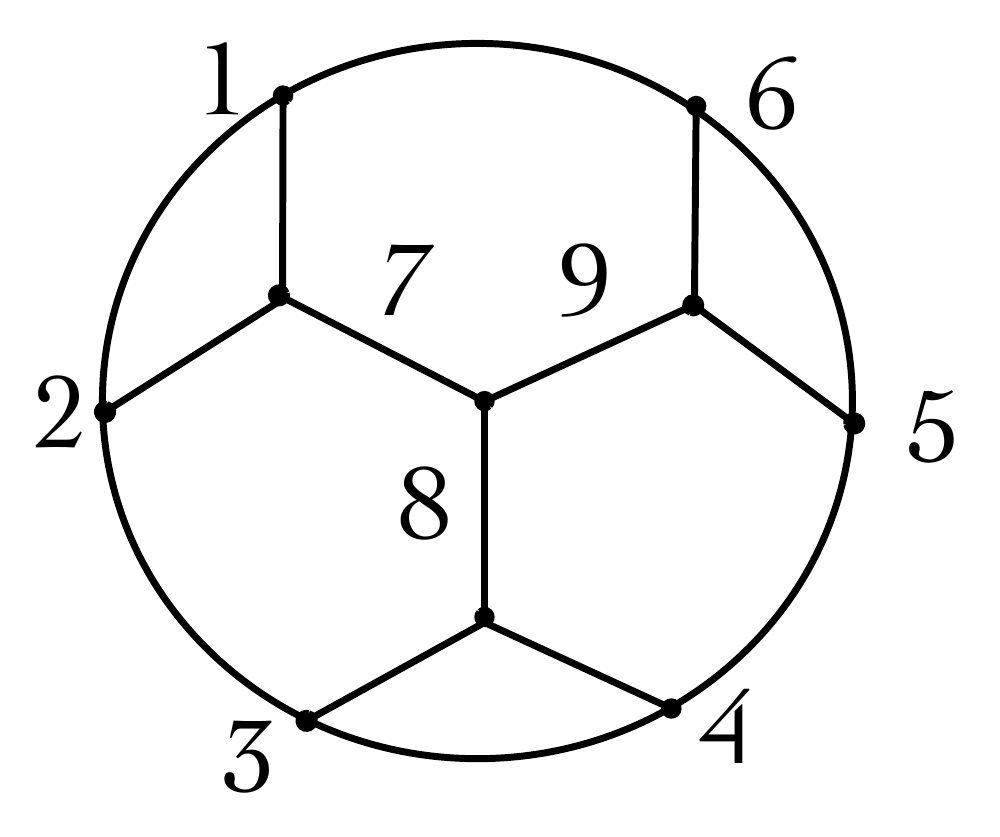}
      \caption{Left (Right): Six-point linear (star) 
Witten diagram in cubic scalar theory.}
      \label{6ptFigure}
\end{figure}

For the next example, consider the ``star'' 6-point diagram
in Fig. \ref{6ptFigure}.  We will apply the factorization formula to
line 9, so the left amplitude is now the 5-point Mellin amplitude
$M_5(\delta_7,\delta_8)$ from the previous subsection.
We will see that the final six-point amplitude depends only on the combinations
$\delta_7,\delta_8, $ and $\delta_9$, which satisfy
\ba
&& \delta_{12} =-\delta_7 + \Delta_{12,7} , \ \ \ \ 
\delta_{34} = - \delta_8 + \Delta_{34,8} , \ \ \ \ \delta_{56} = -\delta_9
 + \Delta_{56,9} \nn\\
&& \delta_{12} + \delta_{13} + \delta_{14} + \delta_{23} + \delta_{24} + \delta_{34} = - \delta_9 + \Delta_{1234,9}.
\ea
These relations are most intuitively understood by noting that 
the ``momentum'' flowing through lines 7,8, and 9
are $(p_1+p_2),(p_3+p_4)$, and $-(p_1+p_2+p_3+p_4)=(p_5+p_6)$, 
respectively.

In evaluating (\ref{MainFormula}), we can again take advantage of the fact that $M_5$ depends only on
$\delta_7$ and $\delta_8$, rather than all possible $\delta_{ij}$'s,
to immediately sum over $n_{13},n_{23},n_{14},n_{24}$, subject
to the constraint $n_{13}+n_{23}+n_{14}+n_{24}=\tilde{n}$.
The 6-point Mellin amplitude can then be written
\begin{align}
&M_6 = \sum_m \frac{S_{\Delta_9}(m)}{\delta_9-m}  \left( \frac{ \lambda_{569}}{m! (\Delta_{56,9})_{-m}} \right) \\
&
  \left( \sum_{n_{12}+n_{34}+\tilde{n}=m} M_5(\delta_7-n_{12},\delta_8-n_{34})
  \frac{(\Delta_{12,7}-\delta_7)_{n_{12}}}{n_{12}!}
\frac{(\Delta_{34,8}-\delta_8)_{n_{34}}}{n_{34}!}
\frac{(\Delta_{78,9}+\delta_7 + \delta_8 -m)_{\tilde{n}}}{\tilde{n}!}
\right) .
\nn
\end{align}
Upon substituting the expression (\ref{eq:5pt}) for $M_5$, we obtain
an explicit expression for $M_6$.  However, as before, we can simplify
further by evaluating $\delta_7$ and $\delta_8$ at the residues of
all poles This step simplifies the calculation, and it means that we are explicitly using the version of our factorization formula that we wrote down in section \ref{sect:AdSCFTFactorization}. This diagram is particularly interesting because it is the
lowest-point amplitude that contains a cubic vertex that connects
three internal or `off-shell' propagators.  As such, it is the simplest diagram that
one should calculate in order to obtain the generalization
of $V_{\Delta_i \Delta_j \Delta_k}$ to three different, non-zero indices.
A short computation shows that the Mellin amplitude $M_6$ for this `star' Witten diagram is
\ba
\label{eq:6pt}
M_6(\delta_7, \delta_8,\delta_9) &=&
 \sum_{l,k,m} \frac{V_{\Delta_1 \Delta_2 \Delta_7}(l) S_{\Delta_7}(l)
V_{\Delta_3 \Delta_4 \Delta_8}(k) S_{\Delta_8}(k)
V_{\Delta_5 \Delta_6 \Delta_9}(m) S_{\Delta_9}(m)
}{(\delta_7-l)(\delta_8-k)(\delta_9-m)}V_{\Delta_7 \Delta_8 \Delta_9}(l,k,m)  \nn 
\ea
where
\ba
\label{3ptVertexExample}
&&  V_{\Delta_7 \Delta_8 \Delta_9} (l,k,m) =
\frac{\lambda_{789}}{l! k! m! (\Delta_{78,9}+l+k)_{-m} (\Delta_{89,7})_{-l} (\Delta_{79,8})_{-k} }  
\label{eq:3ptvertex}\\
 &&  \times
\sum_{n_{12}, n_{34}, n =0}^\infty \left( \frac{(-l)_{n_{12} + n} (-k)_{n_{34}+n} (-m)_{n_{12} +n_{34}} }{n_{12}! n_{34}! n!} 
\right. \nn\\
&& \ \ \ \ \ \ \ \left. 
\frac{(h-\Delta_7 -l)_{n_{12}} (h- \Delta_8-k)_{n_{34}} \left( \frac{\Delta_7 + \Delta_8 + \Delta_9 -d}{2} \right)_n }
{(1-\Delta_{78,9}-l-k)_{n_{12} + n_{34}} (\Delta_{89,7}-l)_{n_{12} + n} (\Delta_{79,8}-k)_{n_{34} +n} } \right) . \nn
\ea
Although it is not obvious from the way this expression is written, it is easy
to verify numerically that it is symmetric under interchange of any
two $(\Delta,i)$ dimension-index pairs, e.g. $(\Delta_7,l) \leftrightarrow
(\Delta_9,m)$.  Moreover, when any of the indices vanishes, it reduces to
the vertex functions we have already encountered in the 5-point amplitude:
\be
V_{\Delta_i \Delta_j \Delta_k}(0,k,m) = V_{\Delta_i \Delta_j \Delta_k}(k,m).
\ee
So far, the above formula for $V_{\Delta_i \Delta_j \Delta_k}(l,k,m)$ is
simply another way of packaging the 6-point amplitude, and one could
reasonable expect that a new vertex function would appear at each vertex
as one considered higher and higher $n$-point amplitudes.  
Surprisingly, this turns out not to be the case: the functions
$S_\Delta(m)$ and $V_{\Delta_i \Delta_j \Delta_k}(l,k,m)$ are all that is
needed in order to write down the most general $n$-point amplitude
in $\phi^3$ theory.  These remarkable diagrammatic rules for Mellin amplitudes are the subject of 
section \ref{sec:FeynmanRules}, to which we now turn.

\section{Mellin Space Diagrammatic Rules }
\label{sec:FeynmanRules}

In section \ref{sect:FactorizationonAdSPropagators} we derived a factorization formula for Mellin amplitudes that can be applied to any factorization channel, and in section \ref{sec:examples} we used this formula to recursively compute several examples.  The fact that our factorization formula gives equivalent results when the recursive steps are applied in different orders suggests that there exist universal diagrammatic rules that allow for the construction of any Mellin amplitude.  The purpose of this section is to derive and prove these rules.  We also obtain a practical benefit, because the computation of complicated diagrams becomes standard and straightforward.

\subsection{The Diagrammatic Rules and Factorization}

To begin, we will demonstrate that there is a special case where Mellin amplitudes may be computed with essentially exactly the usual flat space procedure.  This case is the $\lambda \phi^3$ theory, in any number of dimensions up to 6,\footnote{Above 6 dimensions, $\Delta=2$ is not allowed in a unitary CFT.} when the CFT operator dual to $\phi$ has dimension $\Delta=2$.  The reason that the usual kind of diagrammatic rules automatically apply to this case is that every time we use (\ref{MainFormula}) to add on a cubic vertex to a lower-point diagram, we encounter the factor (in, say, $R$):
\be
R_m= \frac{M_3}{m! (1)_{-m}}
\ee
where $M_3$ is the 3-pt Mellin amplitude.
Thus, the sum over $m$ always truncates at $m=0$, and therefore all the sums over the $n_{ij}$'s do as well.  Since all amplitudes in this theory can be built up by adding on three-point vertices, there are never any sums to do at all!  For an arbitrary diagram, the rules for the Mellin amplitude are simply to include factors of:
\begin{enumerate}

\item
$\frac{S_2(0)}{\delta_i} $ for each propagator, (where $\delta_i$ is the appropriate linear combination of $\delta_{ij}$'s),

\item
$M_3$ for each vertex.

\end{enumerate}

So, in this special case, the calculation of Mellin amplitudes is identical to the calculation of the corresponding S-matrix elements in flat space, as long as we replace the kinematic invariants $p_i \cdot p_j \to \delta_{ij}$.  For instance, we can write the five-point diagram from figure \ref{4ptFigure} in terms of the `Mellin momenta' as
\be
M_5(p_1, \dots, p_5) = M_3 \frac{-2S_2(0)}{(p_1+p_2)^2 + \Delta_6} 
M_3 \frac{-2S_2(0)}{(p_4+p_5)^2 + \Delta_7}M_3.
\ee

For general $\Delta$, however, the sum on $m$ will not truncate at $m=0$, and each propagator will be associated with multiple poles: 
\bea
M&=& \sum_{\{ m_a \} } \frac{ M(m_1, \dots,m_s)}{(\delta_1- m_1) \dots(\delta_s - m_s)}, 
\label{eq:Mpoles}
\eea
where 
\be
 \delta_i = -\frac{\Delta_i + K_i^2}{2}
\ee
 is the square of the total `momentum' flowing through the $i$-th propagator.
  Thus, the best that we could hope for generally is to have diagrammatic rules for 
the residues $M(m_1, \dots, m_s)$. We have already seen in our examples in the
previous section in equations (\ref{eq:5pt}) and (\ref{eq:6pt}) what form such rules might take. Thus we will optimistically guess that more generally, one can calculate $M(m_1, \dots, m_s)$ according to the following rules:
\begin{enumerate}

\item
$\frac{S_{\Delta_a}(m_a)}{\delta_a - m_a}$ for each propagator, where $S_{\Delta_a}(m_a) = \frac{-2 \pi^h \Gamma(\Delta_a-h+1) m_a!}{(\Delta_a-h+1)_{m_a}} $

\item
$ V_{\Delta_i \Delta_j \Delta_k}(m_i,m_j,m_k)$ for each vertex.

\end{enumerate}

\begin{figure} 
  \centering
    \includegraphics[width=0.4\textwidth]{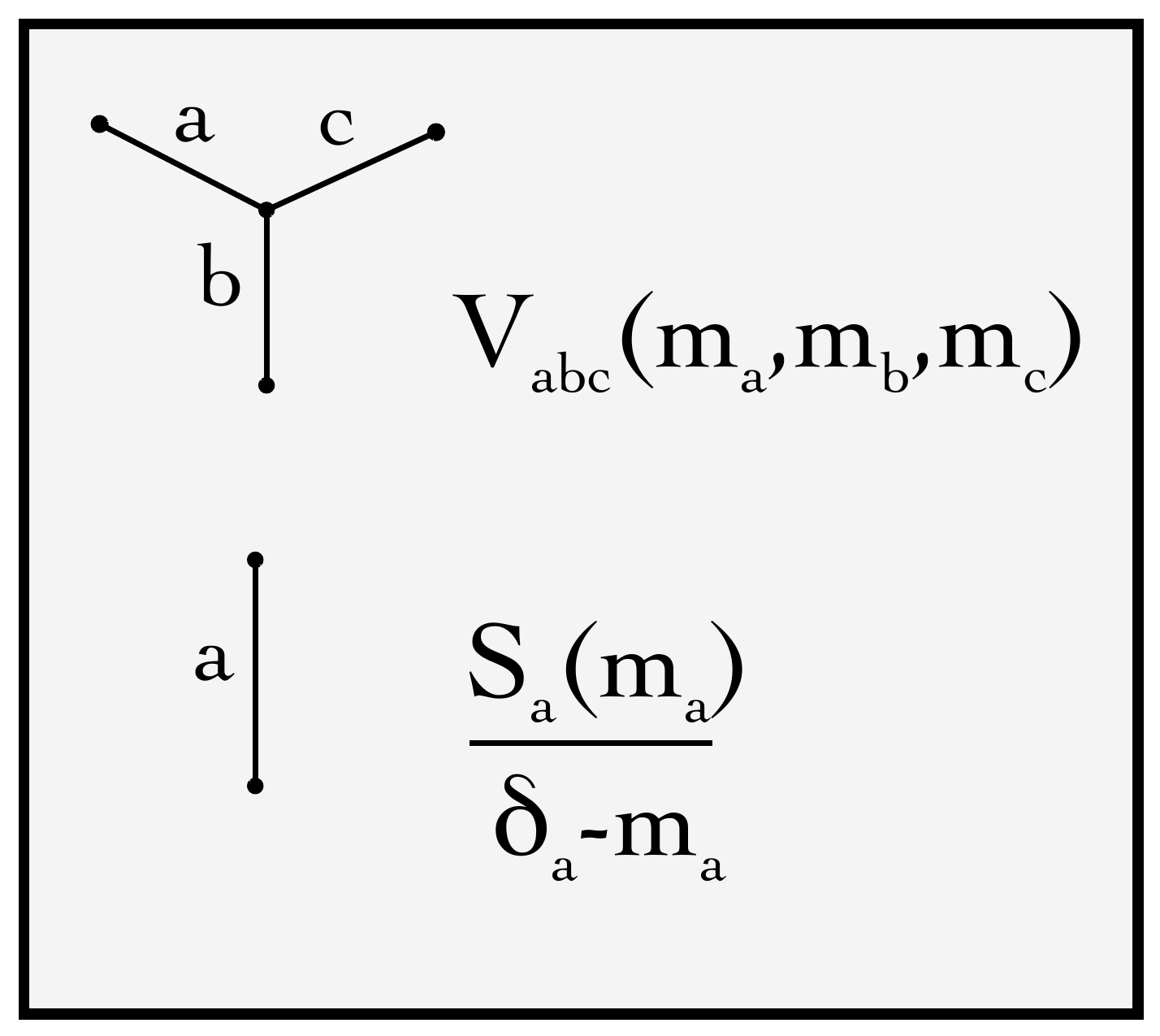}
      \caption{Diagrammatic Rules. The expression for $V_{abc}$ is given in eq. (\ref{eq:3ptvertex}), and that for $S_a$ is given in eq. (\ref{eq:prop}). }
      \label{fig:rules}
\end{figure}

These rules are depicted schematically in Fig. (\ref{fig:rules}).\footnote{All external lines are taken to 
have index $m = 0$. 
To save space, we will abbreviate $S_{\Delta_a}(m)$ to
$S_a(m)$, and similarly for $V$.

}  We will show below that this very simple procedure is exactly correct for higher-point diagrams as well!  On the one hand, this is rather surprising from the point of view of the factorization formula (\ref{MainFormula}), where naively any diagrammatic vertex factors like $V_{\Delta_a \Delta_b \Delta_c}$ would have to depend on all the indices of all internal lines in the diagram.  On the other hand, the existence of such a set of rules is very natural in that it automatically explains why the factorization formula gives the same answer when applied to any propagator in diagram,  a very strong consistency condition.

 The first step is to simplify the Pochhammer symbols.  Since all dependence
 on the $\delta_{ij}$s in $M_L$ and $M_R$ is through the propagator
 variables, we can always use eq. (\ref{eq:pochsumid}) like we did 
 in the case of the five-point function in section \ref{sec:examples}.  There we grouped all
 the $\delta_{ij}$s into terms that depend on only the propagator variables, the $\delta_i$s.  However, each $\delta_{ij}$ appears in a Pochhammer symbol exactly once, so each $\delta_{ij}$ can only appear in a single linear combination of the $\delta_i$.  For instance, in the five-point function, we had to use the latter two identities in eq. (\ref{eq:pochsumid}) to write $\delta_{12}$ and $\delta_{23}+\delta_{13}$ as
 \be
  \delta_{12} = -\delta_6+\Delta_{12,6},
\ \ \ \ \ \ \ \ \
 \delta_{13} + \delta_{23} = - \delta_7+\delta_6+\Delta_{36,7}.
 \ee
 
\begin{figure}
  \centering
    \includegraphics[width=0.75\textwidth]{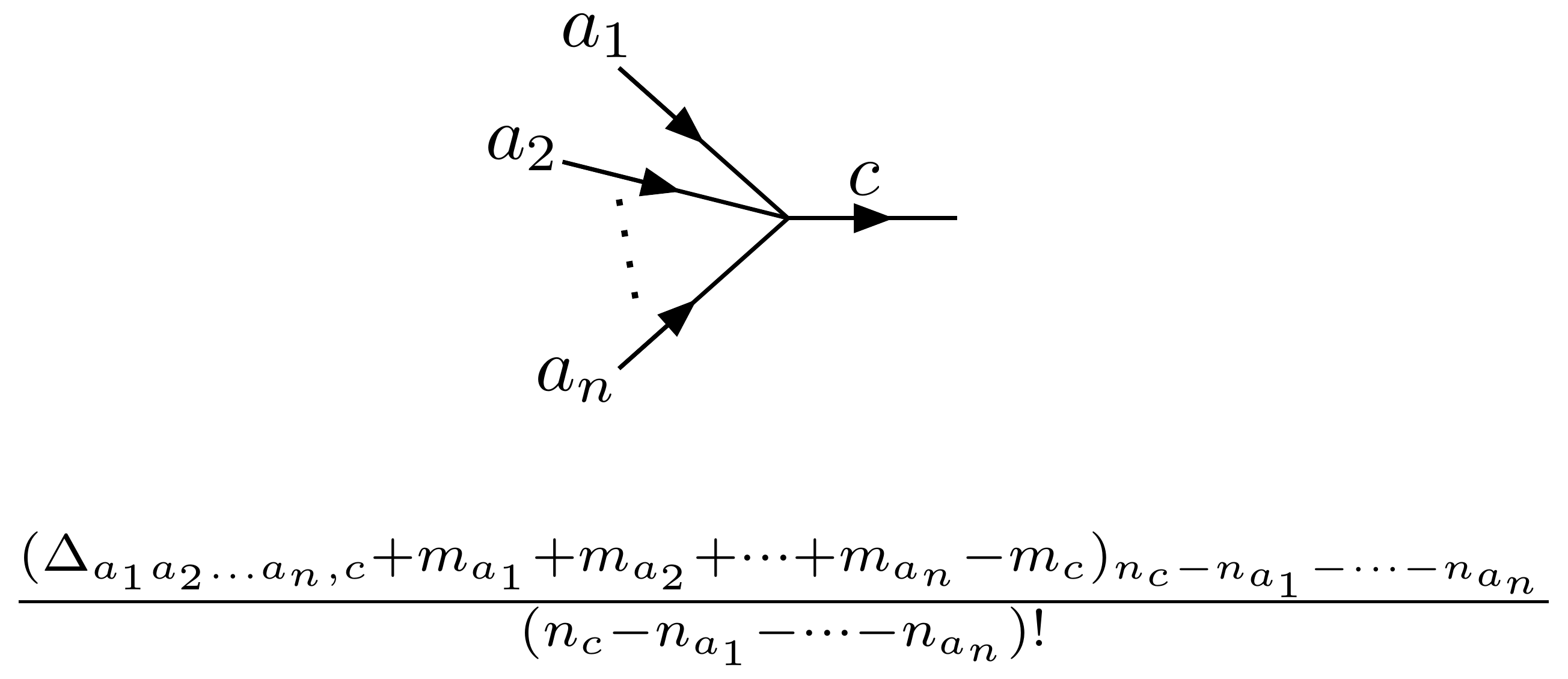}
      \caption{\small When evaluating the factorization formula on its poles, products of Pochhammer symbols reduce generally to a simple expression associated with each vertex.}
      \label{fig:pochhammerresidue}
\end{figure}

 In general, this regrouping can always be performed, so that for every vertex we have a Pochhammer symbol of the form
 \be
 \frac{(\Delta_{ab,c} +\delta_a + \delta_b - \delta_c)_{n_c-n_a-n_b}}{(n_c-n_a-n_b)! },
 \ee
as depicted in Fig. \ref{fig:pochhammerresidue}. Here, $a$ and $b$ are the two propagators leading into the vertex and $c$ is the propagator leading out, as we work from the external lines inward toward the $\delta_{LR}$ propagator.  For completeness, we note that this identity generalizes to arbitrary n-pt interactions, with the $a$ and $b$ indices replaced by a sum over all the propagators flowing into the vertex, towards the propagator on which we are factorizing.  Each $\delta_{ij}$ shows up in exactly one such Pochhammer, because 
 starting with external vertices $i$ and $j$ and flowing inward through the diagram, there is always a unique vertex where the two meet. The sum is then reduced from a sum over all the $n_{ij}$'s to a sum over $n_i$'s, one for each propagator, defined by the condition that $\delta_{ij}\rightarrow \delta_{ij} + n_{ij}$ is equivalent to $\delta_i \rightarrow \delta_i - n_i$.\footnote{To be a little more explicit, if $p_i = (k_1 + k_2 + \dots + k_s)$, where $k_i$ are all external ``momenta'', then $n_i = \sum_{i<j}^s n_{ij}$. } An important nicety of defining the $n_i$'s this way is that the poles are always shifted by 
 \be
 \frac{1}{\delta_i- m_i} \rightarrow \frac{1}{\delta_i - m_i -n_i}
 \ee
 As we will soon see, this makes it possible for us to absorb all of the $n_i$s via a re-definition of the sums.

To demonstrate how the Feynman rules follow from the factorization formula, let us factorize a Witten diagram along a certain `central' propagator and assume that the associated $M_L$ and $M_R$ have been computed from the diagrammatic rules.  We will prove that if we compute use the factorization formula \eqref{factorizationansatzmt} to compute $M_n$ from $M_L$ and $M_R$, then the result will be equal to what we would have found had we computed it entirely from the diagrammatic rules.

It is helpful to illustrate our arguments first with an example, so let us consider the eight-point diagram below, as it will exhibit all the features we will use for the general proof. Factorizing on the ``e'' propagator, $M_L$ is a seven-point diagram and $M_R$ is a three-point diagram.  Assuming that we have applied the diagrammatic rules to $M_L$ and $M_R$, we obtain for $M_8$
\begin{center}
    \includegraphics[width=0.5\textwidth]{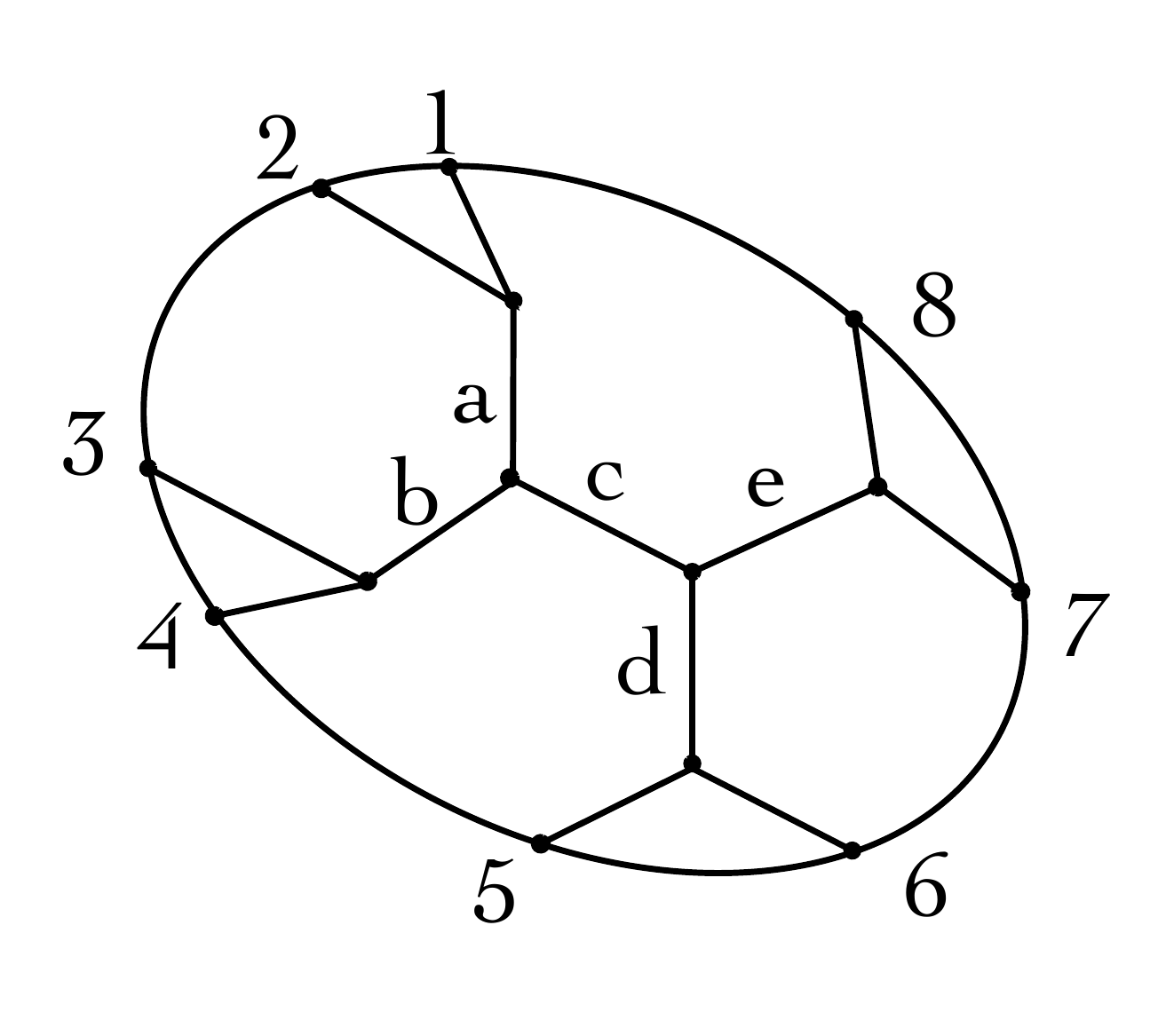}
\end{center}
\ba
M_8 &=& \sum_{m_e} \frac{S_e(m)}{\delta_e - m_e}  \nn\\
&& \left[ \sum_{ n_a,n_b \atop n_c+n_d + n_e =m_e} \left( \sum_{m_a,m_b, \atop m_c,m_d} \frac{V_{12a}(m_a)S_a(m_a) V_{34b}(m_b)S_b(m_b) V_{abc}(m_a,m_b,m_c)V_{56d}(m_d)S_d(m_d)V_{cde}(m_e)}{(\delta_a - m_a-n_a)(\delta_b-m_b-n_b)(\delta_c-m_c-n_c)(\delta_d-m_d-n_d)} \right. \right.  \nn\\
&& \left. \left. \frac{(\Delta_{12,a} - \delta_a)_{n_a}}{n_a!} \frac{(\Delta_{34,b} - \delta_b)_{n_b} }{n_b!} \frac{(\Delta_{ab,c} + \delta_a + \delta_b - \delta_c)_{n_c-n_a-n_b}}{(n_c-n_a-n_b)!} \frac{(\Delta_{56,d} - \delta_d)_{n_d}}{n_d!}
\right. \right. \nn\\
&& \left. \left. \frac{(\Delta_{cd,e}+\delta_c +\delta_d-m_e)_{n_e}}{n_e!} \right) \right] \left[ \frac{(\Delta_{78,e}-m_e)_{m_e}}{m_e!} \right] 
\label{eq:fig8ex}
\ea
We can eliminate $n_e$ through $n_c+n_d+n_e=m_e$, and then the sum on $n_i$'s is unrestricted.  To simplify further, we first redefine $m_i \rightarrow m_i - n_i$ in the sums on $m_i$ in order to shift the poles back to $m_i$, and then as usual we evaluate all the $\delta_i$'s in the numerator on the poles (i.e. $\delta_i \rightarrow m_i$).  

Now we want to show that $M_8$ also satisfies the diagrammatic rules.  We will first show that the correct $V$ and $S$ factors are associated with the residues of the poles in $\delta_a$ and $ \delta_b$.  This follows from the following identity, which we have verified numerically:
\ba
&& \sum^{m_a,m_b}_{n_a, n_b=0} \left( V_{12a}(m_a - n_a)S_a(m_a-n_a) \frac{(\Delta_{12,a} -m_a)_{n_a}}{n_a!} \right) \left( V_{34,b}(m_b-n_b)S_b(m_b-n_b) \frac{(\Delta_{34,b} - m_b)_{n_b}}{n_b!} \right) \nn\\
&& \times 
V_{ab,c}(m_a-n_a,m_b-n_b, m_c-n_c) \frac{(\Delta_{ab,c} +m_a + m_b-m_c)_{n_c - n_a -n_b}}{(n_c -n_a - n_b)!}\label{eq:rulesidentity} \\
&& \ \ \ \ \ \ \ = V_{12a}(m_a)S_a(m_a)V_{34b}(m_b)S_b(m_b) \frac{V_{abc}(m_a,m_b,m_c)}{V_{abc}(m_c)} V_{abc}(m_c-n_c) \frac{(\Delta_{ab,c} -m_c)_{n_c}}{n_c!} \nn
\ea
In words, this identity implies that after summing over $n_a$ and $n_b$ 
in (\ref{eq:fig8ex}), all the factors associated with $m_a$ and $m_b$ become exactly what they should be according to the diagrammatic rules.  Furthermore, the factors associated with $m_c$ are $\frac{1}{V_{abc}(m_c)}$ times exactly what we would have started with if we had considered the diagram with external lines 1, 2, 3 and 4 stripped off. We have already seen in the previous section that a five-point function satisfies the rules; repeating that analysis here, the factor $V_{abc}(m_c)$ of the five-point diagram cancels the $\frac{1}{V_{abc}(m_c)}$ in the identity above to give exactly the correct result. 

\begin{figure}
  \centering
    \includegraphics[width=0.95\textwidth]{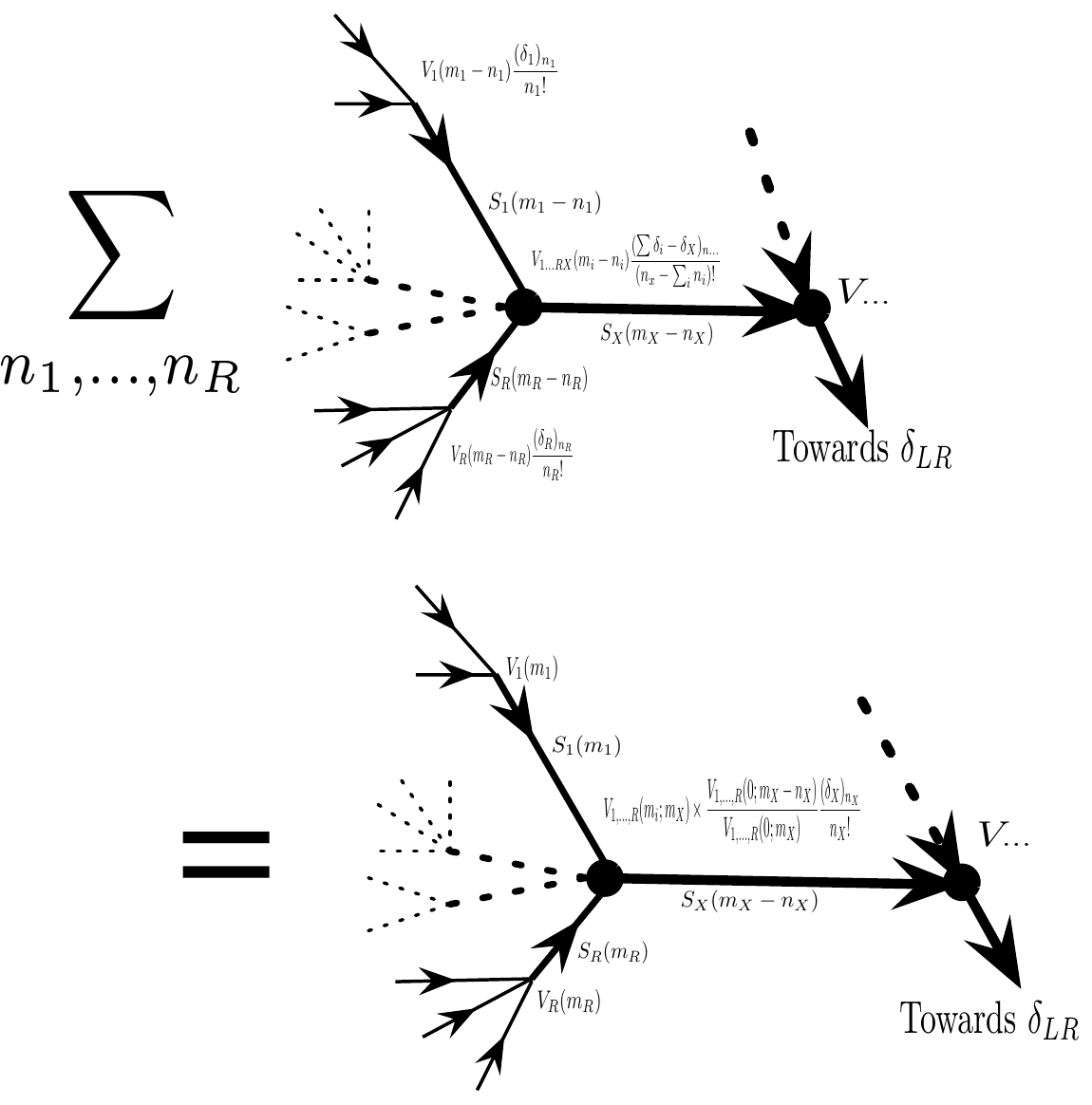}
      \caption{\small This is the general diagrammatic identity which is necessary to prove that our factorization formula and our diagrammatic rules are identical.  The Pochhammer factors localize at vertices in the diagram, and then the sum over $n_{ij}$ from the factorization formula can be performed locally on the diagram, working from the external legs inward towards the factorization or `$\delta_{LR}$' propagator. Applying this identity to the sums turns the factorization formula into the diagrammatic rules.  The precise form of the Pochhammer symbols is schematic, refer to equation (\ref{eq:GeneralDiagrammaticIdentity}) for the full form.}
      \label{fig:DiagrammaticIdentity}
\end{figure}

If we had started with an even larger diagram, we would just apply eq. (\ref{eq:rulesidentity}) repeatedly, reducing at each stage to the needed factors times a reduced diagram. To be a bit more explicit, consider a general diagram $M_n$, factorized along a `central' bulk to bulk propagator with a vertex with dimension $\Delta_e$, taking the form
\ba
M_n &=& \sum_m \frac{S_e(m_e)}{\delta_e - m_e} \left[ \sum_{\{ n_i \}} \sum_{\{ m_i \}} \left( \prod_{\rm prop} \frac{S_i(m_i-n_i)}{\delta_i - m_i} \right) 
\frac{(\Delta_{yz,e} + m_y + m_z -m_e)_{m_e - n_y - n_z}}{(m_e - n_y - n_z)!} \right. \\
&& \times \left. \left( \prod_{\rm vert} V_{ijk}(m_i - n_i, m_j -n_j, m_k -n_k) \frac{( \Delta_{ij,k} + m_i +m_j-m_k)_{n_k - n_i -n_j}}{(n_k - n_i - n_j )!} \right) \right] \times R_m , \nn
\ea
where $y,z$ are the lines connected to $e$ on the left. The first product inside the brackets is over propagators, and the second is over all vertices. 
The identity (\ref{eq:rulesidentity}) now says that, working ``outwards in'' according to figure \ref{fig:DiagrammaticIdentity},
 the sums on the $n_i$'s exactly turn the Pochhammer symbols, shifted vertex factors, and shifted propagator factors associated with $m_i$ into the appropriate final factors, times factors that allow us to consider a reduced diagram with the $i$ legs stripped off.  Thus, the poles of the factorization formula (\ref{factorizationansatzresmt}) are exactly given by the diagrammatic rules we have presented. 

A crucial aspect of the identity in equation \eqref{eq:rulesidentity} is that while it may seem complex, it is localized to a particular vertex in the diagram.  It says that if we work from the outside of a tree diagram inward, toward the central propagator with an associated $\delta_{LR}$, then the sums over the $n_i$ variables simplify at each vertex independently.   Without this property, it is hard to imagine that these diagrammatic rules could consistently reproduce our factorization formula.

The relative simplicity of this localization allows us to immediately write down and check the necessary identities for any scalar theory.  In a theory with both 3-pt $\phi_i \phi_j \phi_k$ couplings and an $(R+1)$-pt coupling, we would find a generalized identity
\ba
\label{eq:GeneralDiagrammaticIdentity}
&& \prod_{i=1}^R  \sum^{m_{a_{i3}}}_{n_{a_{i3}}=0} \left( V_{a_{i1} a_{i2} a_{i3} }(m_{a_{i3}} - n_{i3}) S_{a_{i3}}(m_{a_{i3}}-n_{a_{i3}}) \frac{(\Delta_{{a_{i1}}{a_{i2}},{a_{i3}}} -m_{a_{i3}})_{n_{a_{i3}}}}{n_{a_{i3}}!} \right)  \nn\\
&& \times 
V_{a_{13},...,a_{R3} X}(m_{a_{i3}}-n_{a_{i3}}; m_X - n_X) \frac{ \left( \Delta_{a_{13}...a_{R3}, X} + \sum_i^R m_{a_{i3}} - m_X \right)_{n_X - \sum_i^R n_{a_{i3}}}}{\left( n_X -\sum_i^R n_{a_{i3}} \right)!} \\
&& = 
\left[  \prod_{i=1}^R  \left( V_{a_{i1} a_{i2} a_{i3}}(m_{a_{i3}}) S_{a_{i3}}(m_{a_{i3}}) \right) \right]  
V_{a_{13},...,a_{R3} X}(m_{a_{i3}}; m_X) 
\nn \\ && \times \frac{V_{a_{13},...,a_{R3} X}(0; m_X-n_X)}{V_{a_{13},...,a_{R3} X}(0; m_X)}  
\frac{(\Delta_{a_{13}...a_{R3} X} -m_X)_{n_X}}{n_X!}
\nn
\ea
The first line of this identity is a product of all the vertices, propagators, and associated Pochhammers that lead into our $(R+1)$-pt vertex, the second line is the $(R+1)$-pt vertex and its associated Pochhammer.  The third line is a product of simplified vertices with the $n$ dependence eliminated, so that they take precisely the form that the diagrammatic rules dictate. In the final line we have an `on-shell' $m_i = 0$ version of the $(R+1)$-pt vertex, which will feed in naturally to the next vertex, leading towards the factorization propagator $\delta_{LR}$.  We have checked numerically that this identity holds in the case of theories with 3-pt and 4-pt interactions, and we have written out the general 4-pt vertex in appendix \ref{sec:diagid}.  In theories with many different interaction vertices, there are equivalent identities involving all combinations of the vertices.

\subsection{The Diagrammatic Rules Satisfy the Functional Equation}

Next, we will show that the diagrammatic rules we have just presented satisfy the functional equation. As we discussed in section 2, the general functional equation takes the form
\be
M_0 = (\delta_{LR} - \Delta)(d - \Delta - \delta_{LR})M + \sum_{ab \leq k <  ij}  \left( \delta_{ai} \delta_{bj} M - \delta_{aj} \delta_{bi} M_{ai,bj}^{aj,bi} + \delta_{ab} \delta_{ij} M_{ai,bj}^{ab,ij} \right)
\ee
where we recall that the shifted amplitudes are $M_{ai,bj}^{ab,ij} = M(\delta_{ab}+1, \delta_{ij}+1, \delta_{ai}-1, \delta_{bj}-1,...)$, and we have chosen a specific propagator $\delta_{LR}$, which divides the Mellin amplitude into an $M^L_k$ and $M^R_{n-k+1}$.  A feature of our diagrammatic rules is that they only explicitly depend on the left and right $\delta_{ij}$, and not those with $i \leq k < j$.  Now the shifts such as $M_{ai,bj}^{aj,bi}$ in the functional equation only involve L-R combinations, so when we plug in our factorization formula, we find that $M_{ai,bj}^{aj,bi} = M$.  
This means that the first and second terms in the sum in equation \ref{GeneralFunctionalEquation} cancel, so the functional equation reduces to
\be
M_0 = (\delta_{LR}-\Delta)(d-\Delta-\delta_{LR}) M + \sum_{ab\le k <ij} \delta_{ab} \delta_{ij} M^{ab,ij}_{ai,bj} .
\label{eq:funceqred}
\ee
Now, since the large $\delta_{ij}$ behavior of the diagrammatic rules are easy to read off, we can immediately see that the term linear in $\delta_{LR}$ in the RHS of this equation cancels.  The constant term is by definition independent of $\delta_{LR}$, so this will have to be shown to match $M_0$ on the LHS. Our strategy will be to first show that the RHS is independent of $\delta_{LR}$. But, any possible 
dependence on $\delta_{LR}$ in the RHS  is clearly in the part that falls like $\delta_{ij}^{-1}$. So, if we just want to prove that the RHS is independent of $\delta_{LR}$, then all expressions at any step can be evaluated on their poles. Afterwards, we will show that
the remaining $\delta_{LR}$-independent piece matches $M_0$.\footnote{
We will not prove here that the overall numerical coefficient matches that in $M_0$, although we will obtain the correct parametric dependence on the couplings.  }

First, note the effect of the shifts in $M^{ab,ij}_{ai,bj}$.  By assumption, $ab$ and $ij$ are on opposite sides of the $\delta_{LR}$ propagator, so $\delta_{LR}\rightarrow \delta_{LR}-2$.  The other propagators are shifted as follows.  Consider a propagator $\delta_i$ on the left part of the diagram.  If $a$ and $b$ are both to the left of this propagator, then $\delta_i\rightarrow \delta_i-1$ in $M^{ab,ij}_{ai,bj}$, otherwise it has no change.  A symmetric statement holds for propagators on the right part of the diagram. 

At this point, we can make effective use of the results from the previous subsection.  Specifically, let us write the Mellin amplitude as
\be
M  = \sum_{m_e} M_L(\delta_{ab},m_e) \frac{S_e(m_e)}{\delta_e - m_e} M_R(\delta_{ij}, m_e),
\ee
and rewrite the second term on the RHS of (\ref{eq:funceqred}) as
\ba
\sum_{ab\le k <ij} \delta_{ab} \delta_{ij} M^{ab,ij}_{ai,bj} &=& 4 \sum_{m_e} \frac{S_e(m_e)}{\delta_e-m_e-1} 
\left( \sum_{\sum n_{ab}=1} M_L(\delta_{ab}+n_{ab}, m_e) \prod_{a<b\le k} \frac{(\delta_{ab})_{n_{ab}}}{n_{ab}!}  \right) \nn\\
&& \times \left(  \sum_{\sum n_{ij} =1} M_R(\delta_{ij}+n_{ij}, m_e) \prod_{k<i<j} \frac{(\delta_{ij})_{n_{ij}}}{n_{ij}!}  \right) .
\label{eq:secondRHSsimp}
\ea
where we have re-written the $\delta_{ab} \delta_{ij}$ factors as Pochhammer symbols. Now, we can further simplify the terms in brackets using the result from the previous subsection that
\ba
M_L(\delta_{ab}+n_{ab}, m_e) &=&
 \sum_{n_{ab}' = m_e}\left( \prod_{a < b \le k} \frac{(\delta_{ab}+n_{ab})_{n_{ab}'}}{n_{ab}'!} \right) M_L(\delta_{ab}+n_{ab}+n'_{ab}) ,
\ea
when all $\delta_{ab}$'s are evaluated on the poles. 
Substituting this into eq. (\ref{eq:secondRHSsimp})\footnote{It is necessary here to use the fact that 
\ba
\sum_{\sum n_{ab} =1} \sum_{\sum n_{ab}' =m_e} 
\prod_{a<b \le k} \frac{(\delta_{ab})_{n_{ab}}}{n_{ab}!} \frac{(\delta_{ab} + n_{ab}')_{n_{ab}'}}{n_{ab}'} &=& \sum_{a<b\le k} \sum_{\sum n_{ab}'' = m_e + 1} \prod_{a<b \le k} \frac{(\delta_{ab})_{n_{ab}''}}{n_{ab}''} n_{ab}'' \nn\\
  &=& (m_e+1)  \sum_{\sum n_{ab}'' = m_e + 1} \prod_{a<b \le k} \frac{(\delta_{ab})_{n_{ab}''}}{n_{ab}''} .
\ea
}, 
we find that
\ba
\left. \sum_{ab\le k <ij} \delta_{ab} \delta_{ij} M^{ab,ij}_{ai,bj} \right|_{\rm poles} &=&   
4 \sum_{m_e} \frac{(m_e+1)^2 S_e(m_e)}{\delta_e-m_e-1} 
\left( M_L(\delta_{ab}+n_{ab}, m_e+1) \prod_{a<b\le k} \frac{(\delta_{ab})_{n_{ab}}}{n_{ab}!}  \right)\nn\\
&&\left.  \times \left(  M_R(\delta_{ij}+n_{ij}, m_e+1) \prod_{k<i<j} \frac{(\delta_{ij})_{n_{ij}}}{n_{ij}!}  \right)  \right|_{\rm poles} \nn\\
&=& \left. m_e (m_e + \Delta_e - h) M \right|_{\rm poles},
\ea
where in going to the last line we have shifted the summation index $m_e \rightarrow m_e -1$ and used the fact that $\frac{S_e(m_e-1)}{S_e(m_e)} = \frac{\Delta_e +m_e-h}{m_e}$. Now, since $(\delta_{LR} -\Delta_e)(d-\Delta_e - \delta_{LR}) = 
- 4 \delta_e (\delta_e + \Delta_e -h)$, the poles in the above expression clearly cancel those in the first term in eq. (\ref{eq:funceqred})!

We still need to show that the left-over $\delta_{LR}$-independent piece, call it $M_{red}$, actually has the correct dependence on the remaining $\delta_{ij}$'s to match $M_0$.  To show that we produce the correct $M_0$, we can simply substitute $M_{red}$ into the functional equation for any of the other legs, and keep doing this until we have acted on every propagator in the diagram.  After the last propagator is removed this way, the resulting $M_0$ will be that for a contact interaction, which is simply independent of all the $\delta_{ij}$'s.  Thus, if we can prove that this procedure reduces the Mellin amplitude constructed from the diagrammatic rules to just a constant, then we will be done.\footnote{We will address quite generally the issue of homogeneous solutions in appendix \ref{sect:MellinDeterminedByPoles}.} 

However, this follows immediately if we use a nice property of the relevant conformal Casimirs.  The RHS of the functional equation (\ref{GeneralFunctionalEquation}) is the action of the Casimir $(\sum_{i \in L} J_i)^2$, where $L$ denotes the set of all external lines to the left (with respect to an a priori chosen vertex at the far right of the diagram) of the propagator being acted upon. Because of the tree structure of the diagrams, for any two propagators with corresponding $L$, $L'$, either $L$ and $L'$ will be disjoint or one of them will be a subset of another. In the former case, it is obvious that the two Casimirs commute. A short computation shows that they commute in the latter case as well:
\ba
\left[ \left( \sum_{i \in L}J_i\right)^2 , \left( \sum_{i \in L' \subset L} J_i\right)^2 \right] 
= 2 \left(\sum_{ i \in L-L'} J_i \right) \cdot \left[  \left(\sum_{ i \in L'} J_i \right),  \left(\sum_{ i \in L'} J_i \right)^2 \right] = 0.
\ea
Therefore, all the Casimirs associated with propagators commute with each other. Let us act on $M$ with the Casimirs corresponding to all the propagators in the diagram.  Since the Casimirs commute, we can act with any one of them first, so  the result must be independent of all the $\delta_{ij}$'s,
which completes the proof.

\section{The Flat Space Limit of AdS/CFT and the S-Matrix}
\label{sect:FlatSpaceLimit}

One can obtain a holographic description of the flat space S-Matrix \cite{susskind, polchinski, GGP, JP, Penedones:2010ue, Katz, TakuyaFSL, Fitzpatrick:2011jn} by taking a limit of the AdS/CFT correspondence in which the curvature of AdS goes to zero.  This fact has received more attention recently, and in particular in \cite{Penedones:2010ue} one of us argued for an intimate connection between the Mellin representation of AdS/CFT correlation functions and the flat space S-Matrix.  

Roughly speaking, the claim is that if one interprets the $\delta_{ij}$ as the kinematic invariants $p_i \cdot p_j$ of a scattering amplitude, then in the large $\delta_{ij}$ limit the Mellin amplitude will reproduce the flat space S-Matrix.  This result has a very simple physical justification.  Time translations in global AdS are generated by the dilatation operator of the CFT, so the dimension of an operator in the CFT directly translates into the energy of a corresponding bulk state.  Since the AdS radius $R$ is a dimensionful quantity, what we really mean by the flat space limit is $ER \to \infty$ for all bulk energies $E$.  We have seen repeatedly that the Mellin space coordinates $\delta_{ij}$ are related to the dimensions of operators in the CFT, so it is natural to expect that the large $\delta_{ij}$ limit of a Mellin amplitude is related to the physics of bulk states with large $ER$, which compute the flat space S-Matrix.  We will now show how this works quantitatively.

\subsection{The Flat Space Limit of AdS/CFT Factorization}

At a  computational and pictorial level, we know that Witten diagrams describe scattering events in AdS spacetime.  
In \cite{Penedones:2010ue}, an explicit formula was given relating the large $\delta_{ij}$ behavior of a Mellin amplitude to the corresponding flat space scattering amplitude.
With the present normalizations, this formula reads
\be
M(\delta_{ij}) \approx \frac{\pi^h}{2}\prod_{i=1}^n \frac{\mathcal{C}_{\Delta_i}}{\Gamma(\Delta_i)} 
\ \int_0^\infty d\beta \,\beta^{\frac{1}{2}\sum \Delta_i -h-1} e^{-\beta}\,
T(p_i\cdot p_j= 2\beta \delta_{ij})\ ,\ \ \ \ \ \ \ \ 
\delta_{ij} \gg 1\ .
\label{FSlimit}
\ee
where the flat space S-Matrix is $S = 1 + i(2\pi)^{d+1}\delta\left(\sum p_i\right) T$.  

The transformation in (\ref{FSlimit}) may appear complicated but its effect is relatively simple.  To understand why something like (\ref{FSlimit}) is necessary, first consider the case of massless scalars scattering through a single contact interaction with some fixed number of derivatives.  In this case, the only dimensionful parameter is the coupling constant, which is an overall factor in the amplitude $T(p_i \cdot p_j)$. Thus,  $\beta$ in (\ref{FSlimit}) just factors out by dimensional analysis, and the Mellin amplitude is simply proportional to the flat-space scattering amplitude.  The only effect of the transformation is to modify the overall coefficient. The reason that a tranformation is needed for more general theories is that this overall coefficient depends on the dimension of the interaction.  It was shown in \cite{Penedones:2010ue} that for all scalar contact interactions, (\ref{FSlimit}) produces exactly the correct overall coefficient in order to match the Mellin amplitude.  This is a very general check, since one may typically consider an arbitrary scattering amplitude as a linear combination of such interactions below some cut-off.  So, roughly what (\ref{FSlimit}) says is that to get the flat-space S-matrix from $M(\delta_{ij})$, one simply performs a series expansion at large $\delta_{ij}$ and goes through the series term by term, altering the coefficients by hand in a way that depends only on the power of the $\delta_{ij}$'s.

Now, the goal of this section is to show that the flat space limit of the factorization formula 
\begin{align}
M(\delta_{ij})=-\sum_{m=0}^\infty \frac{4\pi^h \Gamma(\Delta-h+1)m!}{(\Delta-h+1)_m} 
\frac{L_m (\delta_{ij})R_m(\delta_{ij}) }{\delta_{LR}-\Delta-2m}
\label{FactorAnsatz}
\end{align}
reduces to the usual factorization of scattering amplitudes.

We recall that $L_m$ is given by 
\be
L_m(\delta_{ij})=\left[\sum_{\sum n_{ij}=m} M^L(\delta_{ij}+n_{ij}) \prod_{i<j} \frac{(\delta_{ij})_{n_{ij}}}{n_{ij}!} \right]_{\delta_{LR}=\Delta+2m}
\ee
and similarly for $R_m$. 
We start by considering the limit of $\delta_{ij} \gg 1$ with fixed internal and external dimensions $\Delta$ and $\Delta_i$. In this case, the flat space limit will give rise to  scattering amplitudes of massless particles because the mass squared of the bulk fields is of the order of the AdS curvature.
This turns equation \eqref{FactorAnsatz} into
\begin{align}
M(\delta_{ij})\approx\frac{4\pi^h \Gamma(\Delta-h+1)}{-\delta_{LR}}\sum_{m=0}^\infty \frac{m!}{(\Delta-h+1)_m} 
L_m(\delta_{ij}) R_m (\delta_{ij})
\label{FSLfactor}
\end{align}
We see that the multiple poles of the Mellin amplitude at $\delta_{LR}=\Delta+2m$ all contribute to the pole of the scattering amplitude at $(\sum^k_{i=1} p_i )^2=0$.
In order to show that the sum over all contributions reproduces the correct residue of the scattering amplitude, we need to understand the large $\delta_{ij}$ limit of $L_m$.  It turns out that at large $\delta_{ij}$, 
$L_m$ simplifies significantly,
\begin{align} 
L_m (\delta_{ij})&\approx  
\frac{ 1}{m!} \left. \left(\frac{\partial}{\partial t}\right)^m
t^{\frac{1}{2}(\sum \Delta_{i} -\Delta)-1}
 M_L(t \delta_{ij}) \right|_{t=1}\ ,\ \ \ \ \ \ \ \ \ \ 
 \delta_{ij} \gg 1\ . \label{LmML}
\end{align}
We will now prove this relation.
To that end, consider the example
\be
M^L(\delta_{ij}) =\prod_{i<j}^k (\delta_{ij})_{a_{ij}}\ . \label{basicform}
\ee
 This set of functions is very broad and can be used as a basis.
Thus, to prove (\ref{LmML}) in general, it will be sufficient to prove it for
(\ref{basicform}).  
This is a convenient basis to use because, for this type of left Mellin amplitude, one can perform the sum in $L_m(\delta_{ij})$ explicitly,
\begin{align}
L_m(\delta_{ij})&=\left[\sum_{\sum n_{ij}=m} 
\prod_{i<j}^k \frac{(\delta_{ij}+a_{ij})_{n_{ij}}}{n_{ij}!}(\delta_{ij})_{a_{ij}}\right]_{\delta_{LR}=\Delta+2m} \nn \\
&= \left[
\frac{(\sum \delta_{ij}+\sum a_{ij})_{m}}{m!} \prod_{i<j}^k (\delta_{ij})_{a_{ij}} \right]_{\delta_{LR}=\Delta+2m}\nn \\
&=
\frac{\left(\frac{1}{2}(\sum \Delta_{i} -\Delta)+\sum a_{ij}-m\right)_{m}}{m!} 
 \prod_{i<j}^k (\delta_{ij})_{a_{ij}} 
\end{align}
where, in the second line, we used the identity \eqref{eq:pochsumid} and, in the last line, we evaluated $\sum_{i<j}^k \delta_{ij} = (\sum^k_{i=1}\Delta_i -\delta_{LR})/2$
on the pole $\delta_{LR}=\Delta+2m$.
For large $\delta_{ij}$, we have
\begin{align}
M^L(\delta_{ij}) \approx\prod_{i<j}^k (\delta_{ij})^{a_{ij}} \ ,\ \ \ \ \ \ \ \ 
L_m (\delta_{ij})\approx  
\frac{\left(\frac{1}{2}(\sum \Delta_{i} -\Delta)+\sum a_{ij}-m\right)_{m}}{m!} 
 \prod_{i<j}^k (\delta_{ij})^{a_{ij}} \ ,
\end{align}
in perfect agreement with  (\ref{LmML}).

Let us now substitute this simplified form into (\ref{FSLfactor}) and
invoke the flat space limit formula (\ref{FSlimit}) for $M^L$ and $M^R$.  We obtain the large $\delta_{ij}$ limit of the Mellin amplitude in terms of the left and right scattering amplitudes:
\begin{align}
M(\delta_{ij})
\approx& -
\frac{\pi^h}{4\delta_{LR} } 
\prod_{i=1}^n 
\frac{\mathcal{C}_{\Delta_i}}{\Gamma(\Delta_i)} 
\int_0^\infty d\beta_L \,\beta_L^{\frac{1}{2}(\Delta+\sum_{i=1}^k \Delta_i) -h-1}
 T^L(p_i\cdot p_j= 2\beta_L \delta_{ij})\nn\\&
 \int_0^\infty d\beta_R \,\beta_R^{\frac{1}{2}(\Delta+\sum_{i>k}^n \Delta_i) -h-1} 
T^R(p_i\cdot p_j= 2\beta_R \delta_{ij})\label{Mlargedelta}\\&
\sum_{m=0}^\infty \frac{1}{m!} 
\left. \left(\frac{\partial}{\partial t_L}\frac{\partial}{\partial t_R}\right)^m
\frac{(t_Lt_R)^{h -\Delta-1}e^{-\beta_L/t_L-\beta_R/t_R}}{\Gamma(\Delta-h+1+m)}
\right|_{t_L=t_R=1}, \nn
\end{align}
where we have rescaled the integration variables $\beta_L, \beta_R$. 
This looks like a rather complicated expression but it simplifies dramatically due to the following identity
\begin{align}
\sum_{m=0}^\infty \frac{1}{m!} 
\left. \left(\frac{\partial}{\partial t_L}\frac{\partial}{\partial t_R}\right)^m
\frac{(t_Lt_R)^{h -\Delta-1}e^{-\beta_L/t_L-\beta_R/t_R}}{\Gamma(\Delta-h+1+m)}
\right|_{t_L=t_R=1} 
=\beta_L^{h-\Delta} e^{-\beta_L} \, \delta(\beta_L-\beta_R)\ ,
\label{identitybetaLR}
\end{align}
which we prove in Appendix \ref{sect:FlatSpaceLimitIdentities}.
Finally,  the large $\delta_{ij}$ behaviour of the Mellin amplitude simplifies to 
\begin{align}
M(\delta_{ij})\approx
\frac{\pi^h}{2 } 
\prod_{i=1}^n 
\frac{\mathcal{C}_{\Delta_i}}{\Gamma(\Delta_i)} \int_0^\infty d\beta \,\beta^{\frac{1}{2}\sum \Delta_i -h-1} e^{-\beta}
\frac{T^L(p_i\cdot p_j= 2\beta \delta_{ij})  
T^R(p_i\cdot p_j= 2\beta \delta_{ij}) }{-2 \beta \delta_{LR}}\ ,
\end{align}
in agreement with the factorization of scattering amplitudes for massless scalars,
\be
T(p_i\cdot p_j) = \frac{T^L(p_i\cdot p_j)  
T^R(p_i\cdot p_j) }{(\sum_{i=1}^k p_i )^2}\ .
\ee

\subsubsection{Massive Propagators in Flat Space}
\label{subsect:MassiveFlatSpaceLimit}

The mass of a scalar field in AdS$_{d+1}$ is $\Delta(\Delta-d)/R^2$, where $\Delta$ is the conformal dimension of the dual operator and $R$ is the AdS radius of curvature.
Thus, in order to keep a finite mass in the flat space limit $R\to \infty$, one must scale the dimensions $\Delta \to \infty$ of the scalar operators.

More precisely, one considers the limit of large $\Delta$ and $\delta_{ij}$ with fixed ratio $\delta_{ij}/\Delta^2$.
Let us then return to (\ref{FactorAnsatz}) with (\ref{LmML}) and (\ref{FSlimit}) applied to $M^L$ and $M^R$, and study this limit
\begin{align}
M(\delta_{ij})
\approx& -
\frac{\pi^h}{4 } 
\prod_{i=1}^n 
\frac{\mathcal{C}_{\Delta_i}}{\Gamma(\Delta_i)} \int_0^\infty d\beta_L \,\beta_L^{\frac{1}{2}\sum_L \Delta_i -h-1} T^L(p_i\cdot p_j= 2\beta_L \delta_{ij})\nn \\&
 \int_0^\infty d\beta_R \,\beta_R^{\frac{1}{2}\sum_R \Delta_i -h-1} 
T^R(p_i\cdot p_j= 2\beta_R \delta_{ij})\nn  \\&
\sum_{m=0}^\infty \left. \left(\frac{\partial}{\partial t_L}\frac{\partial}{\partial t_R}\right)^m\frac{(\beta_L\beta_R)^{\frac{\Delta}{2}} 
(t_Lt_R)^{h -\Delta-1}e^{-\beta_L/t_L-\beta_R/t_R}
}{(\delta_{LR}-\Delta-2m)\Gamma(\Delta-h+1+m)m!} 
\right|_{t_L=t_R=1}
\label{eq:MlargedeltaII}
\end{align}
In appendix \ref{sect:FlatSpaceLimitIdentities}, we prove that the limit of the last line is
\be
\delta(\beta_L - \beta_R) e^{-\beta_L} \beta_L^h \frac{2\beta_L}{2\beta_L \delta_{LR} -\Delta^2}\ .
\ee

Thus, the limit of (\ref{eq:MlargedeltaII}) is
\begin{align}
M(\delta_{ij})\approx
\frac{\pi^h}{2} 
\prod_{i=1}^n 
\frac{\mathcal{C}_{\Delta_i}}{\Gamma(\Delta_i)} \int_0^\infty d\beta \,\beta^{\frac{1}{2}\sum \Delta_i -h-1} e^{-\beta}
\frac{T^L(p_i\cdot p_j= 2\beta \delta_{ij})  
T^R(p_i\cdot p_j= 2\beta \delta_{ij}) }{-2 \beta \delta_{LR}+\Delta^2}
\label{massivefactor}
\end{align}
as expected from the factorization of scattering amplitudes
\be
T(p_i\cdot p_j) = \frac{T^L(p_i\cdot p_j)  
T^R(p_i\cdot p_j) }{(\sum_{i=1}^k p_i )^2 +M^2}\ ,
\ee
where we identified $\Delta^2$ with the mass squared of the exchanged particle.

It is worth noting that the poles of the Mellin amplitude at $\delta_{LR}=\Delta+2m$ turned into the pole at $\delta_{LR}=\Delta^2/(2\beta)$ in the integrand of \eqref{massivefactor}. The reason is that, in the flat space limit, the infinite sum over poles is dominated by the poles with $m$ of order $\Delta^2 \gg 1$.
More precisely, in the limit of large $\delta_{ij}$ with fixed $\delta_{ij}/\Delta^2$, the infinite sequence of poles of the Mellin amplitude gives rise to a branch cut along the positive real axis of $\delta_{LR}/\Delta^2$.

\subsection{Diagrammatic Rules}

In the last section, we have shown that the flat space limit of the AdS factorization formula reduces to the usual factorization of scattering amplitudes.
In this section, we will show that the flat space limit of the AdS Feynman rules proposed in section \ref{sec:FeynmanRules} gives the usual Feynman rules for scattering amplitudes.

The flat space limit corresponds to the large $\delta_{ij}$ behaviour of Mellin amplitudes.  
If we take this limit with fixed $\Delta$'s, then equation (\ref{eq:Mpoles}) simplifies to 
\be
M(\delta_{ij})\approx 
\Big( \sum_{\{m_a\}} M(m_1,\dots,m_s)\Big)
\prod_{a=1}^s \frac{1}{\delta_a} \ ,
\ee
where the index $a=1,\dots, s$ labels the internal propagators of the Witten diagram.
The function $M(m_1,\dots,m_s)$ is computed using the diagrammatic rules of  section \ref{sec:FeynmanRules}.
This large $\delta_{ij}$ behaviour of the Mellin amplitudes should be compared with the prediction from the flat space limit formula (\ref{FSlimit}).
Inserting the appropriate scattering amplitude for massless scalars in (\ref{FSlimit}), one obtains
\be
M(\delta_{ij}) \approx  
\frac{\pi^h}{2}\left( \prod_{i=1}^n \frac{\mathcal{C}_{\Delta_i}}{\Gamma(\Delta_i)} \right)
\Gamma\left(\frac{1}{2}\sum_{i=1}^n \Delta_i -h-s\right)
\prod_{a=1}^s \frac{-1}{4\delta_a} \ ,
\ee
where we have set to 1 the coupling constants associated to each interaction vertex.
We conclude that the dependence on the kinematic variables $\delta_{ij}$ is the correct one. To finish the proof we just need to show that the overall normalization also agrees, i.e. we must show that
\be
\sum_{\{m_a\}} M(m_1,\dots,m_s)=
\frac{\pi^h}{2 }\left( \prod_{i=1}^n \frac{\mathcal{C}_{\Delta_i}}{\Gamma(\Delta_i)} \right)
\Gamma\left(\frac{1}{2}\sum_{i=1}^n \Delta_i -h-s\right) \frac{(-1)^s}{4^s}\ .
\label{FRnorma}
\ee

We will perform the sum over $\{m_a\}$ recursively, starting from the bulk propagators closer to the external legs of the tree level Witten diagram.

The first case to consider is a part of the Witten diagram that connects one bulk propagator ($\Delta_8,m_8$) to 3 external legs ($\Delta_1,\Delta_2,\Delta_3$) like in the left diagram in Fig. \ref{6ptFigure}. 
In this case, we need to compute the sum 
\begin{align}
&\sum_{m_7=0}^\infty V_{\Delta_1 \Delta_2 \Delta_7}(0,0,m_7) S_{\Delta_7}(m_7) V_{\Delta_3 \Delta_7 \Delta_8}(0,m_7,m_8)
\label{FRsum1}\\
=&
\frac{   -2\pi^h\lambda_{127}\lambda_{378}
\Gamma^2(\Delta_7-h+1)
 \Gamma\left(\frac{\Delta_1+\Delta_2+\Delta_3+\Delta_8}{2}-h-1\right)
 }{m_8!
 (\Delta_{123,8}-1)_{-m_8} \Gamma\left(\frac{\Delta_1+\Delta_2+\Delta_7}{2}-h\right)
 \Gamma\left(\frac{\Delta_3+\Delta_7+\Delta_8}{2}-h\right)
 }\ .\nn
\end{align}
It is then easy to use this result twice to compute the sum
\begin{align}
&\sum_{m_7,m_8=0}^\infty V_{\Delta_1 \Delta_2 \Delta_7}(0,0,m_7) 
S_{\Delta_7}(m_7) V_{\Delta_3 \Delta_7 \Delta_8}(0,m_7,m_8)
S_{\Delta_8}(m_8) V_{\Delta_4 \Delta_8 \Delta_9}(0,m_8,m_9)\nn \\
=&
\frac{  4 \pi^{2h}\lambda_{127}\lambda_{378}\lambda_{489}
\Gamma^2(\Delta_7-h+1)\Gamma^2(\Delta_8-h+1)
 \Gamma\left(\frac{\Delta_1+\Delta_2+\Delta_3+\Delta_4+\Delta_9}{2}-h-2\right)
 }{m_9!
 (\Delta_{1234,9}-2)_{-m_9} \Gamma\left(\frac{\Delta_1+\Delta_2+\Delta_7}{2}-h\right)
 \Gamma\left(\frac{\Delta_3+\Delta_7+\Delta_8}{2}-h\right)
  \Gamma\left(\frac{\Delta_4+\Delta_8+\Delta_9}{2}-h\right)
 }\ ,
\end{align}
corresponding to the 2 leftmost bulk propagators in the left diagram in Fig. 
\ref{6ptFigure}.
By using this rule recursively, we can compute the sum in (\ref{FRnorma}) for some Witten diagrams. However, the general Witten diagram requires another type of basic sum, corresponding to a part of the Witten diagram connecting 4 external legs 
 ($\Delta_1,\Delta_2,\Delta_3,\Delta_4$)  to a bulk propagator ($\Delta_9,m_9$) like in
 the right diagram in Fig. \ref{6ptFigure}, 
\begin{align}
&\sum_{m_7,m_8=0}^\infty V_{\Delta_1 \Delta_2 \Delta_7}(0,0,m_7) 
S_{\Delta_7}(m_7) V_{\Delta_3 \Delta_4 \Delta_8}(0,0,m_8)
S_{\Delta_8}(m_8) V_{\Delta_7 \Delta_8 \Delta_9}(m_7,m_8,m_9)
\label{FRsum2}\\=&
\frac{  4\pi^{2h}\lambda_{127}\lambda_{348}\lambda_{789}
\Gamma^2(\Delta_7-h+1)\Gamma^2(\Delta_8-h+1)
 \Gamma\left(\frac{\Delta_1+\Delta_2+\Delta_3+\Delta_4+\Delta_9}{2}-h-2\right)
 }{m!
 (\Delta_{1234,9}-2)_{-m} \Gamma\left(\frac{\Delta_1+\Delta_2+\Delta_7}{2}-h\right)
 \Gamma\left(\frac{\Delta_3+\Delta_4+\Delta_8}{2}-h\right)
  \Gamma\left(\frac{\Delta_7+\Delta_8+\Delta_9}{2}-h\right)
 }\ .\nn
\end{align}

Using (\ref{FRsum1}) and (\ref{FRsum2}) recursively, it is easy to show that a general tree level $n$-point Witten diagram with $s$ internal propagators has
\be
\sum_{\{m_a\}} M(m_1,\dots,m_s)=
\Gamma\left(\frac{\sum_{i=1}^n\Delta_i}{2}-h-p\right)
(-\pi^h )^s
\prod_{a=1}^{s}  \Gamma^2(\Delta_a-h+1) 
\prod_{\rm vert } 
\frac{\lambda_{ijk}}{\Gamma\left(\frac{\Delta_i+\Delta_j+\Delta_k}{2}-h\right)}\nn
\ee
where
\be
\lambda_{123}=\frac{\pi^h}{2} \Gamma\left(\frac{\Delta_1+\Delta_2+\Delta_3}{2}-h\right)
\prod_{i=1}^3\frac{\mathcal{C}_{\Delta_i}}{\Gamma(\Delta_i)}
=\frac{\Gamma\left(\frac{\Delta_1+\Delta_2+\Delta_3}{2}-h\right)
}{16\pi^{2h}\prod_{i=1}^3 \Gamma(\Delta_i-h+1)}
\ee
is the 3-point Mellin amplitude. Finally, using the fact that a tree level $n$-point Witten diagram has $s=n-3$ bulk propagators and $n-2$ cubic vertices, it is straightforward to prove ($\ref{FRnorma}$).

\section{Discussion}

We have argued that Conformal Field Theory correlation functions have a natural home in Mellin space, and we have given dramatic evidence for this claim in the case of CFTs with a weakly coupled AdS dual.  

The easiest way to summarize our results is to list the profound similarities between Mellin space for AdS/CFT and momentum space for scattering amplitudes.  As with scattering amplitudes in momentum space, in Mellin space 
\begin{itemize}
\item CFT correlation functions have poles corresponding to the exchange of operators, which we have dubbed `OPE factorization', and on these poles the correlation functions factorize into lower point correlators;
\item the differential equations that define AdS/CFT correlators as Green's functions turn into simple, purely algebraic functional equations for the Mellin amplitude;
\item there are simple diagrammatic rules that enable a direct construction of the Mellin amplitude corresponding to any Witten diagram;
\item the Mellin space `momentum' flows through these Witten diagrams in such a way that it is conserved at all vertices.
\end{itemize}
Furthermore, the connection becomes totally explicit when we take the flat space limit of AdS/CFT, where Mellin variables turn into flat space kinematic invariants via $\delta_{ij} \to p_i \cdot p_j$ and the Mellin amplitudes themselves reduce to the flat space S-Matrix of the bulk theory.

In this paper we have only dealt in a precise and systematic way with CFTs dual to theories of scalar fields living in the bulk of AdS, and we have only computed the correlators at tree-level, \emph{i.e.} at leading order in $1/N$.  We expect that this is only the beginning.  It will be interesting to understand how our results generalize to theories with higher spin fields, at loop level, and to more general CFTs that do not have a weakly coupled AdS dual. Let us conclude with a few comments about these possibilities.

As discussed in appendix \ref{sect:TensorFields}, there exists a straightforward method for generalizing our results to vector and tensor fields.  However, more efficient methods might very well exist which more naturally incorporate the helicity structure of the fields dual to the conserved currents and the stress-energy tensor. In this respect, (standard) momentum space seems to have an advantage since the current conservation condition, which is a differential equation in position space, turns into the more tractable algebraic equation $p^\mu J_\mu = 0$ in momentum space. Indeed, recently some simple results have been obtained for tensor correlators in momentum space \cite{Maldacena:2011nz}.  Second, computing gravitational amplitudes using Witten diagrams is tedious because gravity has an infinite number of interaction vertices. To efficiently compute these amplitudes in Mellin space, we would require some version of the BCFW recursion relations \cite{Britto:2005fq, Britto:2004ap, ArkaniHamed:2008yf} analogous to the one developed in momentum space for AdS \cite{Raju:2010by,Raju:2011mp}. Unfortunately, conformal invariance is obscured in momentum space.  Moreover, in some cases (logarithmic) divergences in the Fourier transform of CFT correlation functions in momentum space lead to conformal anomalies which complicate the analysis for higher-point functions, see for example \cite{vanRees:2011ir}. These issues do not arise in Mellin space.

At loop level in the bulk of AdS, there may be two distinct but natural forms for the Mellin amplitude to take -- one generalizing our factorization formula, and the other generalizing our diagrammatic rules.  As we discussed in section \ref{sect:FactorizationofCFT}, we expect that at loop-level and even non-perturbatively, the Mellin amplitude will be meromorphic, with poles corresponding to the exchange of operators in the OPE.  So at loop-level it should still be possible to write the amplitude in a form that makes the poles corresponding to various multitrace operators manifest; a form with exactly these properties was found for the 4-pt amplitude in \cite{Penedones:2010ue}.  However, given the success of our diagrammatic rules, we also expect to find loop-level rules, where infinite sums replace the momentum space integrals familiar from flat spacetime.  In these diagrams, factorization may be obscured, but perhaps bulk unitarity will be manifest.  
It is not hard to guess a generalization of our rules from tree to loop level, but we will leave the exploration of their nature and their physical interpretation to future work.

We should note that the complete holographic computation of correlation functions in for example $\mathcal N=4$ SYM using supergravity is more involved than just the evaluation of the Witten diagrams themselves. In particular, in order to obtain for example the five-dimensional Lagrangian a Kaluza-Klein reduction has to be performed \cite{Kim:1985ez} which beyond the first order is rendered computationally difficult because of the non-linear relations between the Kaluza-Klein coefficients and the eventual five-dimensional bulk fields \cite{Lee:1998bxa}. Using superconformal invariance and inspired by the results obtained in this way, a general conjecture for certain scalar four-point functions of $\mathcal N = 4$ SYM in the supergravity limit was written down in \cite{Dolan:2006ec}.

Results like those of \cite{Dolan:2006ec} for AdS/CFT four-point functions were often conveniently expressed in terms of the so-called D-functions which were first introduced in \cite{D'Hoker}. Such D-functions correspond to contact Witten diagram in AdS and as we have seen in equation \eqref{ContactInteraction} their Mellin transforms are just constants. One may therefore attempt to use existing techniques in the literature \cite{D'Hoker} to convert bulk to bulk propagators to contact interactions and amplitudes to sums over D-functions, and then reformulate these results as Mellin amplitudes. We have verified that this leads to answers in a different form than we presented above and the two forms are related by resummation (as, for example, in the four point calculation of \cite{Penedones:2010ue}). 
The results of this paper indicate that the Mellin transform may be a more natural object to describe higher-point correlation functions. We hope that this can enable an extension of the current results for e.g. $\mathcal N=4$ correlation functions in the supergravity limit to more than four external points.
We also expect Mellin amplitudes to be useful in $\cn =4$ SYM beyond the supergravity limit. An important question is if higher point functions of single-trace operators can be constructed from the knowledge of 2-pt and 3-pt functions of single-trace operators.
The factorization of Mellin amplitudes provides a concrete way to do this in CFTs whose AdS dual can be constructed from cubic interaction vertices.
More generally, we can hope that a BCFW-like construction  is possible for type IIB string theory in AdS$_5\times S^5$, which recursively reduces $n$-pt  to 3-pt Mellin amplitudes.

Will our results extend usefully to general CFTs?  The present definition of the Mellin amplitude \cite{Mack} includes factors of $\Gamma(\delta_{ij})$ which are especially convenient for theories with a perturbative expansion, where the existence of operators with dimensions $\Delta_1$ and $\Delta_2$ implies the existence of an operator with dimension $\Delta_1 + \Delta_2$, up to perturbative corrections.  We have only studied theories with a small number of low-dimension operators, so it may be interesting to investigate Mellin amplitudes in say weakly coupled ${\cal N} = 4$ SYM (not necessarily at large $N$) , where the perturbation expansion involves a larger number of operators.  But the larger question is whether the Mellin representation can be useful at a non-perturbative level.  Our arguments based on the OPE from section \ref{sect:FactorizationofCFT} suggest that it will be, but we will need to make those arguments more precise in order to apply them quantitatively.  It would be interesting to investigate this question by studying simple, exactly solvable models such as minimal models.  One by-product of our analysis that could be immediately useful is the functional equation, since its homogeneous solutions are nothing other than the conformal blocks.  In the future we expect to present results using the functional equation to obtain explicit expressions for the conformal blocks.

Finally, we have extended the discussion of \cite{Penedones:2010ue}, showing that when we take the flat space limit of a classical scalar field theory in AdS, the Mellin amplitude of the dual CFT morphs into the S-Matrix of the bulk theory.  
This suggest that Mellin amplitudes can be used to give a holographic and non-perturbative definition of the gravitational S-matrix, through the flat space limit of AdS.
The main open question is how much progress can be made in the computation of Mellin amplitudes beyond the planar limit (i.e. tree level in AdS).
 It will be exciting to pursue this research avenue towards a holographic description of flat spacetime.

%%% Local Variables: 
%%% mode: latex
%%% TeX-master: "MellinFactorization"
%%% End: 

\section*{Acknowledgements}

We thank Nima Arkani-Hamed, Shamit Kachru, Ami Katz, and Pedro Vieira for helpful discussions.  
SR is supported by a Ramanujan Fellowship of the Department of Science and Technology of the Government of India. JK is supported by SLAC; SLAC is operated by Stanford University for the US Department of Energy under contract DE-AC02-76SF00515.  ALF is supported by DOE grant DE-FG02-01ER-40676 and NSF CAREER grant PHY-0645456. 
Research at the Perimeter Institute is supported in part by the Government of Canada through NSERC and by the Province of Ontario through the Ministry of Research \& Innovation. JP was partially funded by FCT project CERN/FP/116358/2010 and PTDC/FIS/099293/2008.  We would all like to thank the organizers of the ``Back to the Bootstrap'' conference, and the Perimeter Institute for Theoretical Physics, where this collaboration was initiated, for hospitality.

\appendix

\section{The Functional Equation and Large $\delta_{ij}$ Behavior}
\label{sect:MellinDeterminedByPoles}

We showed in section \ref{sec:FeynmanRules} that the diagrammatic rules are identical to the factorization formula, and also that the diagrammatic rules satisfy the functional equation of \ref{sect:FunctionalEquation}. Furthermore, we know that our formulas reproduce the poles of Witten diagrams from the discussion of the factorization formula in section \ref{sect:AdSCFTFactorization}, so if there is any difference between our formulas and the Witten diagrams then this difference must be an analytic function. However, by linearity of the functional equation we find that any possible difference should also satisfy the general functional equation \eqref{GeneralFunctionalEquation} with $M_0 = 0$.

Our formulas explicitly vanish as $\delta_{ij} \to \infty$, and on physical grounds we expect that Witten diagrams must be polynomially bounded in this limit.  Among other things, if this were not true then our Mellin amplitudes would not reproduce the flat space S-Matrix, as we know they do.  So to really prove that our formulas are identical to the Mellin amplitude of the corresponding Witten diagrams, it suffices to prove that no polynomial can satisfy the homogeneous functional equation.

Let us see why a pure polynomial cannot satisfy the homogeneous functional equation.  Roughly speaking, one would expect this to follow because the functional equation requires a certain periodicity, and polynomials clearly cannot be periodic, but we can easily make a more precise and direct argument.  The general homogeneous functional equation is
\be
(\delta_{LR} - \Delta)(d - \Delta - \delta_{LR})M + \sum_{ab \leq k <  ij}  \left( \delta_{ai} \delta_{bj} M - \delta_{aj} \delta_{bi} M_{ai,bj}^{aj,bi} + \delta_{ab} \delta_{ij} M_{ai,bj}^{ab,ij} \right) = 0
\ee
Let us use a basis for the $\delta_{ij}$ that is unconstrained by identities involving the dimensions $\Delta_i$.  Any polynomial in this restricted set of $\delta_{ij}$ variables will have a term of highest degree $D$ of the form
\be
H(\delta_{ij}) = \sum_{\sum d_{ij} = D} f(d_{ij}) \prod_{i<j}^n \delta_{ij}^{d_{ij}} 
\ee
Plugging this into the functional equation, we see that the term in that equation of greatest degree is unaffected by the shifts of the $\delta_{ij}$, so the first two terms in the summand cancel and we find that
\be
\left[ (\delta_{LR} - \Delta)(d - \Delta - \delta_{LR})  + \sum_{ab \leq k <  ij}  \delta_{ab} \delta_{ij} \right]  H(\delta_{ij}) = 0
\ee
The term in brackets of order $\delta_{LR}^2$ will automatically cancel, but the term of order $\delta_{LR}$ will not, at least for generic (physical) values of the $\Delta_i$. It also cannot be canceled by the next-to-leading degree terms in $M$ because the order $\delta_{ij}^2$ coefficient of those terms also cancels, so the only solution to this equation is $H = 0$, completing the proof.

Now, for completeness, we will present an argument that the functional equation can be directly used to argue that our factorization formula correctly computes Witten diagrams. (Notice that in the body of the text we proved this via a detour, through the equivalence of factorization and the diagrammatic rules.)  Namely, we will prove that our factorization formula satisfies the functional equation when it is evaluated on its poles in
$\delta_{LR}$.  We know that the poles and residues of the factorization formula are identical to the poles and residues of the Witten diagram in question, from the development of section \ref{sect:AdSCFTFactorization}, and so from the argument above we then obtain that our factorization formula is indeed correct.

We want to prove that our factorization formula is sufficient to determine the Mellin amplitude. More precisely, we claim that
\be
M =  \sum_{m=0}^\infty   \frac{Res(m)}{\delta_{LR} - \Delta - 2m}  
\label{factorizationansatz}
\ee
with  
\be
Res(m) = -\frac{ 4\pi^h \Gamma^2(\Delta-h+1)  m!}{\Gamma(\Delta-h+1+m)} \left[ \sum_{\sum n_{ij} = m} \prod_{i<j \leq k} \frac{(\delta_{ij})_{n_{ij}}}{n_{ij}!} M^L(\delta_{ij} + n_{ij}) \right]_{\Delta_{LR} = \Delta + 2m}  R_m
\ee
where of course $R_m$ takes the same form as the left piece, so we have not written it out explicitly.  The arguments from section \ref{sec:FeynmanRules} immediately apply, so
the functional equation reduces to
 \be
M_0= (\delta_{LR} - \Delta)(d - \Delta - \delta_{LR})M +
 \sum_{ab \leq k <  ij}  \delta_{ab} \delta_{ij} M_{ai,bj}^{ab,ij} \label{reducedfequ}
\ee
The expression on the right hand side appears to be of order $\delta_{LR}^2 M$ as $\delta_{LR} \to \infty$, but as before one can check explicitly that this dependence cancels between the first term and the summand.  Then the leading behavior for large $\delta_{LR}$ is of order $\delta_{LR} M$, which approaches a constant, independent of $\delta_{LR}$.  Thus up to a constant, we can evaluate the right hand side of this equation on its poles.
We will find a dramatic simplification.  Let us focus on a particular term in the summand and study
\begin{align}
\delta_{AB} \delta_{IJ} M_{AI,BJ}^{AB,IJ} =& \sum_m 
\frac{-1}{\delta_{LR} - \Delta - 2m - 2} 
\frac{ 4 \pi^h \Gamma^2(\Delta-h+1)  m!}{\Gamma(\Delta-h+1+m)}
\ \delta_{IJ} R_m(\delta_{IJ} \to \delta_{IJ} + 1)\\
& \sum_{\sum n_{ij} = m} \left( \prod_{i<j \leq k \atop ij \neq AB} 
\frac{(\delta_{ij})_{n_{ij}}}{n_{ij}!} \right) \frac{(\delta_{AB}+1)_{n_{AB}}}{n_{AB}!} \delta_{AB} M^L(\delta_{ij} + n_{ij}; \delta_{AB} + n_{AB} + 1) 
\nonumber 
\end{align}
where we have capitalized the $AB,IJ$ to differentiate them from the $ij$ that are being summed over, and again we are leaving right piece implicit because we will be manipulating it and the left piece in an identical way. Furthermore, since we are ignoring the constant piece and only consider the the poles and their residues, we are implicitly evaluating all the $\delta_{ij}$ in this expression at the pole $\delta_{LR} = \Delta + 2m$, in particular the $\delta_{AB}$ and $\delta_{IJ}$ are now \emph{also} assumed to be subject to this constraint. Focusing on the second line, notice that
\be
\frac{(\delta_{AB}+1)_{n_{AB}}}{n_{AB}!} \delta_{AB} = (n_{AB}+1) \frac{(\delta_{AB})_{n_{AB}+1}}{(n_{AB}+1)!}
\ee
Let us use this fact and switch the order of summation, so that we sum over the $AB$ labels inside the sum over the $n_{ij}$,
\begin{align}
\sum_{\sum n_{ij} = m} \sum_{A<B}^k (n_{AB}+1) \left( \prod_{i<j \leq k \atop ij \neq AB} 
\frac{(\delta_{ij})_{n_{ij}}}{n_{ij}!} \right) \frac{(\delta_{AB})_{n_{AB}+1}}{(n_{AB}+1)!}  M^L(\delta_{ij} + n_{ij}; \delta_{AB} + n_{AB} + 1) 
\end{align}
The result is that we obtain precisely the $(m+1)$th term of the series, except for $(n_{AB}+1)$ type factors. 
 However, when viewed from the perspective of the $(m+1)$th term, this just means that we must multiply by $\sum n_{ij} = m+1$.  
 This follows from the following identity regarding sums over partitions of integers,
\begin{align}
\sum_{\sum  n_{i} = m} \sum_{A=1}^k (n_{A}+1) 
F( n_{1},\dots,  n_{A} + 1,\dots,n_k) =
(m+1)\sum_{\sum  n_{i} = m+1} 
F( n_{1},\dots ,n_k)\ .
\end{align}
Thus we find that the entire summand in (\ref{reducedfequ}) has the following poles
\be
\sum_{m=0}^\infty 2m(2\Delta-d+2m) \frac{Res(m)}{\delta_{LR} - \Delta - 2m}
\ee
Evaluating also the first term on its poles gives
\be
(\delta_{LR} - \Delta)(d - \Delta - \delta_{LR})M \to -\sum_{m=0}^\infty 2m(2\Delta-d+2m) \frac{Res(m)}{\delta_{LR} - \Delta - 2m}\ ,
\ee
proving that the right hand side of the functional equation is analytic in $\delta_{LR}$.
To complete the proof that our ansatz solves the functional equation it is sufficient to show that the right hand side of (\ref{reducedfequ}) tends to $M_0$ as $\delta_{LR}\to \infty$.  This depends on the theory and the structure of the diagram, and proving this was precisely the subject of section \ref{sec:FeynmanRules}.

\section{Exchange of Vector Fields}
\label{sect:TensorFields}

In this appendix, we wish to examine how Mellin amplitudes factorize when a gauge
boson 
is exchanged in a bulk to bulk propagator. This analysis can be generalized to the exchange of a higher spin field in the 
bulk. We will only consider amplitudes with external scalars.

We need a formula that relates the bulk to bulk propagator to the bulk to 
boundary propagator
\begin{equation}
\label{bulkbulkdecompose}
G_{BB, \Delta}^{\mu \nu} (x, y) = \int \chi(c) d c \int d^d z G_{\partial B, (h + c)}^{\mu \rho} (x, z) G_{ \partial B, (h - c)}^{\nu \sigma} \eta_{\rho \sigma},
\end{equation}
where $G_{ \partial B, \Delta}$ indicates the bulk to boundary propagator
for a spin-1 field of dimension $\Delta$, and $G_{BB,\Delta}$ is the bulk to bulk
propagator for the same field. The existence of such a formula follows from 
the existence of the analogous formula for scalars. All we will assume about the function $\chi(c)$ here is that it has a pole
at $c = \Delta - h$.  (A formula of the kind that we need was developed in \cite{Balitsky:2011tw} but we will not need its detailed form.) 

By inserting this formula in a Witten diagram we get
\begin{equation}
\label{factgauge}
A(x_1, \ldots x_n) = \int \chi(c) dc \int d^d z A_L^{\mu}(x_1, \ldots x_m, z) A_R^{\nu}(z, x_{m+1}, \ldots x_n) \eta_{\mu \nu}.
\end{equation}
Let us lift the vector fields $A_L$ and $A_R$ to vector fields on the boundary
of AdS in the embedding space.
As explained in \cite{OurDIS,Weinberg:2010fx}, a spin-1 primary operator 
$A_\mu(x)$ is uplifted to a transverse spin-1 field $A_M(P)$ on the light-cone of $\mathbb{M}^{d+2}$, such that
\be
P^M \, A_M(P)=0\ ,\ \ \ \ \ \ \ \ \ \ \
A_\mu(x)= \frac{\partial P^M}{\partial x^\mu} A_M(P)\ ,
\ee 
where $P^M = (1,x^2, x^m)$ is the Poincar\'e section of the lightcone.

It follows that
\be
\eta_{\mu\nu} A_L^\mu A_R^\nu = \eta_{MN} A_L^M A_R^N\ ,
\ee
which uplifts equation (\ref{factgauge}) to the embedding space.

Conformal invariance tells us that $A_L$ must have the form
\begin{equation}
\label{lineardecompose}
A_L^M(P_1, \ldots P_m, P) = \sum_{p=0}^m A_L^p(P_1, \ldots P_m, P)P_p^M
\end{equation}
with $P_0 \equiv P$ and where the $A_L^p(P_1, \ldots P_m, P)$ are $m+1$
scalar functions.

 We will also need the fact that the function $A_L^p$ has a conformal weight at the point $P_q$
which is given by
\begin{equation}
\label{newdims}
\Delta^p_q = \left\{\begin{array}{ll} \Delta_q, &\text{if}~q \neq p \\  \Delta_q + 1, &\text{if}~ q = p  \end{array} \right.
\end{equation}
where $\Delta_q$ are the dimensions of the external operators in the 
original amplitude $A$.

Inserting this into \eqref{factgauge}, we get
\begin{eqnarray}
\label{factgauge2}
A(x_1, \ldots x_n) &=& \int \chi(c) dc \int d^d P  \sum_{p=1}^{m} \sum_{q=m+1}^n A_L^p(P_1, \ldots P_m, P) A_R^{q}(P, P_{m+1}, \ldots P_n) (P_p \cdot P_q) \nn\\
& \equiv & \sum_{p=1}^m \sum_{q=m+1}^n A_{pq}(P_1,\dots, P_n) (P_p \cdot P_q) .
\end{eqnarray}
Note that the sum over $p$ starts from $1$, because the terms involving $P \cdot P_1$ etc. have dropped out because of the transversality constraint \eqref{lineardecompose}. From now, the range of the sums will be kept implicit. Also in the term $\left(P_p \cdot P_q\right)$, one factor comes from the left, and the other 
factor comes from the right.  In general, the effect on the Mellin amplitude of multiplying the correlation function by the factor $P_i \cdot P_j$ is
to send $M(\delta_{ij} ) \rightarrow \delta_{ij} M(\delta_{ij} -1)$. We can therefore for the moment consider the Mellin amplitude in the absence of this factor, and reintroduce it later. But, this is exactly of the form we encountered in scalar theories, where we have seen how factorization works.  So, repeating the analysis for scalars, we now find that
\begin{equation}
\label{factorMellingauge}
M_{pq}(\delta_{i j}) = \int \chi(c) d c    L_p \times R_q,
\end{equation}
where $M_{pq}$ is the Mellin transform of $A_{pq}$.  
Here, the subscripts on $L$ and $R$ come from \eqref{factgauge2} and {\em do not}
indicate the number of particles inside $L$ and $R$. We have kept that information
implicit to avoid clutter. We have
\begin{equation}
L_p=\int  [d \tilde \delta]_L[dl]_L\,
M_{p}^L(\tilde{\delta}_{ij},l_i)
\prod_{i<j}^k \frac{\Gamma(\tilde{\delta}_{ij}) \Gamma(\delta_{ij}-\tilde{\delta}_{ij})}{\Gamma(\delta_{ij})}
\end{equation}
where the constraints on the $\tilde{\delta}$ are now
\begin{equation}
\sum_{j \neq i} \tilde{\delta}_{i j} = \Delta^{p}_{i}
\end{equation}
where $\Delta^{p}_i$ is given by \eqref{newdims}.

We are almost done now, since we can repeat the analysis for poles 
when a scalar is exchanged.
We find that $M_{pq}$ contains a pole when
\begin{equation}
\label{polegauge}
\sum_{i j} (\delta_{ij} + n_{ij}) - \frac{1}{2} (\sum_i \Delta_i^p - h - c) = 0 .
\end{equation}
However, since $\sum_i \Delta_i^p = \sum_i \Delta_i + 1$, we can rewrite 
\eqref{polegauge} as 
\begin{equation}
\delta_{L R} = \sum_i \Delta_i - 2 \sum_{i j} \delta_{i j} = h + c - 1 + 2 m .
\label{eq:polegaugeII}
\end{equation}
In order to obtain $M$ from the $M_{pq}$'s, we take
$M= \sum_{pq} \delta_{pq} M_{pq}(\delta_{pq}-1)$.  However, since $p$ and $q$ are always on opposite sides of the propagator, their shift has no effect on the position of the poles in $\delta_{LR}$.  Therefore, (\ref{eq:polegaugeII}) 
is the position of the poles in $M$ as well.
If we now assume that the function $\chi(c)$ has a pole at $c = \Delta - h$,
we find that for the exchange of a gauge boson, the Mellin amplitude has a pole
at 
\begin{equation}
\delta_{L R} = \Delta - 1 + 2 m,
\end{equation}
with a residue that can be read off from \eqref{factorMellingauge}. 

A very similar analysis can be performed for gravity. As we mentioned above, however, the primary complication in computing graviton amplitudes is that Witten diagrams involving the interaction of gravitons are inordinately complicated. It would be interesting to see if this can be ameliorated using BCFW recursion relations and to compare their form to the recursion relations developed for correlation functions of stress tensors and conserved currents in \cite{Raju:2010by,Raju:2011mp}.

\section{Some Technical Developments}

\subsection{Shadow Field Identities}
\label{app:shadow}

 We saw in section \ref{sect:FactorizationofCFT} that to implement factorization and unitarity, for each operator with dimension $\Delta$ we need to introduce a shadow operator with dimension $d - \Delta$.  Our goal is to understand factorization in the Mellin representation, so we should first understand the relationship between the Mellin amplitude of a product of operators $\CO_i(x_i)$ and the identical amplitude where one operator, say $\CO_1(x_1)$, is replaced by its shadow $\tilde \CO_1(x_1)$.  The result is
 \be
 \tilde{M}(\tilde{\delta}_{ij})=-\frac{1}{\Gamma(h-\Delta_1)}
 \int [d\delta] M(\delta_{ij}) \prod_{1<i<j}^n \frac{
 \Gamma(\delta_{ij})\Gamma(\tilde{\delta}_{ij}-\delta_{ij})}{\Gamma(\tilde{\delta}_{ij})}\ .
 \label{eq:shadowidentityI}
 \ee
 Now we will derive it.

 The starting point is an identity relating propagation with dimension $\Delta$ and $d-\Delta$:
 \be
 \frac{\mathcal{C}_{d-\Delta} }{
 (-2 P \cdot X)^{d-\Delta}}=-\frac{1}{\pi^h \Gamma(h-\Delta)} \int d^d P' 
 \frac{\Gamma(d-\Delta)}{(-2 P \cdot P')^{d-\Delta}}
 \frac{\mathcal{C}_\Delta}{(-2 P' \cdot X)^{\Delta}}
 \label{shadowidentity}
 \ee
 where $X$ is a point in the bulk of AdS, but $P$ and $P'$ are boundary points.  If we apply this identity to the computation of an AdS amplitude $A$ and then represent $A$ using a Mellin amplitude $M$, we find
 \begin{align}
 \tilde{A}(x_1,\dots,x_n) &=
 -\frac{1}{\pi^h \Gamma(h-\Delta_1)} \int dy \frac{\Gamma(d-\Delta_1)}{(x_1-y)^{2(d-\Delta_1)}}A(y, x_2,\dots,x_n)\\
  &=  -\frac{1}{ \Gamma(h-\Delta_1)} 
 \int [d\tilde{\delta}]\left( 
 \int [d\delta] M(\delta_{ij}) \prod_{1<i<j}^n \frac{
 \Gamma(\delta_{ij})\Gamma(\tilde{\delta}_{ij}-\delta_{ij})}{\Gamma(\tilde{\delta}_{ij})}
 \right)
   \prod_{i<j}^n \Gamma(\tilde{\delta}_{ij})
  (x_{ij}^2)^{-\tilde{\delta}_{ij}}\nn
 \end{align}
 Or in other words, we have derived the relation (\ref{eq:shadowidentityI}) between the amplitude $M$ and the equivalent amplitude with $\CO_1 \to \tilde \CO_1$.

Now, we would like to use this relation to evaluate the $R$ piece in eq. (\ref{eq:MellinLeftI}). 
Let us rewrite it with variables relabeled in a way that suits our present purposes:
\ba
M_{k+1}(\delta_{ij} ) = - \frac{1}{\Gamma(\Delta_{k+1}-h)} \int [d \tilde{\delta}] \tilde{M}(\tilde{\delta}_{ij}) \prod_{ i<j}^k \frac{\Gamma(\tilde{\delta}_{ij}) \Gamma(\delta_{ij} - \tilde{\delta}_{ij})}{\Gamma(\delta_{ij})} ,
\ea
where the integration variables are constrained by $\sum_{j \ne i} \tilde{\delta}_{ij} = \tilde{\Delta}_i$, and the amplitudes $M(\delta_{ij})$ and $\tilde{M}(\tilde{\delta}_{ij})$ differ
only through $\tilde{\Delta}_{k+1} = d-\Delta_{k+1}$ in the first external leg (for all other legs, $\tilde{\Delta}_i=\Delta_i$). In general, this applies
only in the case that the $\delta_{ij}$ variables also satisfy
similar constraints, $\sum_{j\ne i} \delta_{ij} = \Delta_i$, whereas in our factorization eq. (\ref{eq:MellinLeftI}) we need to be able to take $\delta_{LR} = \Delta_{k+1} + 2m$ when $m \ne 0$.  So this identity is not yet directly usable.  
 However,
for the special case of three-point functions ($k=2$), the integrations
are vacuous, and we can derive a shadow field identity for arbitrary
$\delta_{ij}$.  To do this, let us define new variables $\hat{\delta}_{ij}$
that {\em do} satisfy the constraints, and write
\ba
M_3(\delta_{ij}) &=&  -\frac{1}{\Gamma(\Delta_3-h)} \frac{\Gamma(\delta_{12} - \tilde{\delta}_{12}) \Gamma(\hat{\delta}_{12}) }
{\Gamma(\hat{\delta}_{12} - \tilde{\delta}_{12}) \Gamma(\delta_{12})}
\int [d \tilde{\delta}] \tilde{M}(\tilde{\delta}_{ij})
\prod_{i<j}^2 \frac{\Gamma(\tilde{\delta}_{ij})\Gamma(\hat{\delta}_{ij} - \tilde{\delta}_{ij})}{\Gamma(\hat{\delta}_{ij})} \nn\\
 &=&\frac{\Gamma(\delta_{12} - \tilde{\delta}_{12}) \Gamma(\hat{\delta}_{12}) }
{\Gamma(\hat{\delta}_{12} - \tilde{\delta}_{12}) \Gamma(\delta_{12})}
M_3(\hat{\delta}_{ij}).
\ea
One can now substitute this relation
 into eq. (\ref{eq:MellinLeftI}), with
$(1,2)\rightarrow (n-1,n)$ for continuity of index labels and
$\Delta_3 \rightarrow c+h, \hat{\delta}_{13} \rightarrow \hat{l}_{n-1}, \hat{\delta}_{23} \rightarrow \hat{l}_n $, and constraints that give
\ba
2\tilde{\delta}_{n-1,n} &=& \Delta_{n-1} + \Delta_n -h+c \nn\\
2\hat{\delta}_{n-1,n} &=& \Delta_{n-1} + \Delta_n -h -c\nn\\
2 \hat{l}_{n-1} &=& h+c + \Delta_{n-1} - \Delta_n \nn\\
2 \hat{l}_n &=& h+c + \Delta_n - \Delta_{n-1} \nn\\
2 \delta_{n-1,n} &=& \Delta_{n-1} + \Delta_n - \delta_{LR}.
\ea
Thus, we find that on $c=\Delta-h, \delta_{LR} = \Delta + 2m$, we have
\ba
R &=&- \Gamma(\Delta-h) \frac{(h-\Delta)_{-m}}{\left( \frac{\Delta_{n-1} + \Delta_n-\Delta}{2}\right)_{-m} }
M^R_3(\hat{\delta}_{n-1,n}, \hat{l}_i)
=
-\frac{\Gamma(\Delta-h) (-1)^m}{(\Delta-h+1)_m}
M_3(\delta_{n-1,n}+m) (\delta_{n-1,n})_m  \nn
\ea
This is sufficient to prove the factorization formula (\ref{MainFormula}) for all poles of $\delta_{LR}$ in $n$-point functions when adding on three-point functions:
\ba
M_n(\delta_{ij}) &\sim& \sum_{n_{ij}} 
 \frac{\pi^h \Gamma(\Delta-h+1)}{\delta_{LR} - \Delta - 2m} M^L_{k+1} (\delta_{ij} + n_{ij})\left(  \prod_{i < j}^k \frac{(-1)^{n_{ij}}}{n_{ij}!} \frac{\Gamma(\delta_{ij} + n_{ij})}{\Gamma(\delta_{ij})} \right)\nn\\ &&
\times  \frac{(h-\Delta)_{-m}}{\left( \frac{\Delta_{n-1} + \Delta_n-\Delta}{2}\right)_{-m} }
M^R_3(\hat{\delta}_{ij}, \hat{l}_i) \label{eq:recurse3point}
\ea
Now, the key to generalizing beyond three-point functions is that when we factorize the propagator to obtain eq. (\ref{eq:MellinLeftI}), either side of the diagram may be chosen to be the ``Left'' or the ``Right'', and the answer must be the same regardless.  Instead of adding a three-point function onto a $k+1$-point function, we could have made the $k+1$-point function the ``Right'' piece.

 Applying
eq. (\ref{eq:MellinLeftI}) in this way with $L \leftrightarrow R$, we obtain
\ba
M_n(\delta_{ij}) &\sim &
\sum_m \frac{\pi^h (h-\Delta)}{\delta_{LR} - \Delta-2m} M_3^R (-1)^m
 \frac{(\delta_{n-1,n})_m }{m!} \times L.
\label{eq:decomp2}
\ea
By the constraints, we have $\delta_{n-1,n}+m = \frac{\Delta_{n-1} + \Delta_n - \Delta}{2} $ in the above formula.  Matching residues of the poles in $\delta_{LR}$ in eqs. (\ref{eq:decomp2}) and (\ref{eq:recurse3point}), we therefore obtain
\ba
&&\sum_{\sum n_{ij}=m} \Gamma(\Delta-h+1) M_{k+1}^L(\delta_{ij} + n_{ij})
\left( \prod_{i<j}^k (-1)^{n_{ij}} \frac{ (\delta_{ij})_{n_{ij}} }{n_{ij}!}\right)
\frac{ (h-\Delta)_{-m}}{\left( \frac{\Delta_{n-1} + \Delta_n - \Delta}{2}
\right)_{-m}} \nn \\ && =\frac{(h-\Delta)(-1)^m}{m! \left( \frac{\Delta_{n-1} + \Delta_n - \Delta}{2}
\right)_{-m} } \times L.
\ea
After some simplifications, this reduces to
\ba
L &=& - \frac{\Gamma(\Delta-h)m! (-1)^m}{(\Delta-h+1)_m} \sum_{\sum n_{ij}=m}
 M_{k+1}^L(\delta_{ij} + n_{ij}) 
\left( \prod_{i<j}^k  \frac{ (\delta_{ij})_{n_{ij}} }{n_{ij}!}\right)
 ,
\ea
which proves the identity (\ref{eq:genshadow}).

\subsection{General Diagrammatic Identities and Vertices from the Functional Equation}
\label{sec:diagid}

We saw in section \ref{sec:FeynmanRules} that although the factorization formula
(\ref{MainFormula}) looks diagrammatically non-local, in that the residues of the poles in one $\delta_i$ appear naively to depend on the positions of all the other $\delta_j$'s, it in fact satisfies a set of diagrammatic rules for each vertex and propagator that depend only on adjacent pole positions.  A key technical aspect of this fact was an identity (\ref{eq:rulesidentity}) that allowed the rules to be proved recursively, so that larger diagrams could be effectively reduced to smaller ones with fewer external lines.  For the same result to occur in a general $\phi^n$ theory, there must exist a general identity of the form shown in equation (\ref{eq:GeneralDiagrammaticIdentity}). We expect this identity to be true in all $\phi^n$ theories, and we have explicitly computed the $\phi^4$ vertex 
\bea
&& V_{abcd} =  \!\!\!\!\sum_{n_a, n_b, n_c, \nt_a, \nt_b, \nb_a} 
   \left[ \frac{1}{n_a! n_b! n_c! (1-h+m_a + \Delta_a)_{-n_a} 
 (1-h+m_b + \Delta_b)_{-n_b} (1-h+m_c + \Delta_c)_{-n_c}}
\right. \nn\\
&& \frac{(\Delta_{abc,d} + m_a + m_b + m_c - m_d)_{m_d - n_a -n_b - n_c} 
(\Delta_{ab,c} + m_a + m_b - m_c - n_a - n_b + n_c)_{m_c - n_c - \nt_a - \nt_b}}
{ (m_c- n_a - n_b - n_c)! \nt_a! \nt_b! (m_c - n_c -\nt_a - \nt_b)!} \nn \\
&& \left. \frac{V_{cd a}(m_a - n_a - \nt_a - \nb_a)
(\Delta_{cda,b} + m_a - n_a - \nt_a - m_b + n_b + \nt_b)_{m_b - n_b - \nt_b - \nb_a}}
{(1-h+m_b- n_b + \Delta_b)_{-\nt_b} \nb_a! (1-h+m_a - n_a + \Delta_a)_{-\nb_a -\nt_a} (m_b - n_b - \nt_b-\nb_a)!} \right] \nn 
\eea
and checked our identity (\ref{eq:GeneralDiagrammaticIdentity}) numerically for the case of $\phi^3$ and $\phi^4$.

We can use the functional equation to generate more general vertices.  The idea is that if we construct the $(s+t)$-pt Mellin amplitude from $(s+1)$ and $(t+1)$-pt vertices and plug the result into the functional equation, then we will automatically generate an $(s+t)$-pt diagrammatic vertex.  This means we begin by taking (schematically)
\be
M_{s+t}(\delta) = \sum_m \frac{V_{\Delta_1...\Delta_s \Delta}(m_1,\ldots,m_s,m)  S_\Delta(m) V_{\Delta_1...\Delta_t \Delta}(m_{s+1},\ldots,m_{s+t},m) }{\delta - m}
\ee
where we are imagining that the external legs are themselves propagators coupling to further vertices, and $\delta$ is the single propagator variable.  Once we plug this formula into the functional equation, we need to deal appropriately with the fact that the external legs are `off-shell', which means that the $m_i$ can be shifted.  The resulting identities can be used as a definition or a check of the $n$-pt vertex; we also obtain very non-trivial identities from the symmetry properties of these vertices.

\subsection{Flat Space Limit Identities}
\label{sect:FlatSpaceLimitIdentities}

In this appendix we prove two identities that were used in the study of the flat space limit of the AdS factorization formula in section \ref{sect:FlatSpaceLimit}. Firstly, we want to prove that
\begin{align}
\sum_{m=0}^\infty \frac{1}{m!} 
\left. \left(\frac{\partial}{\partial t_L}\frac{\partial}{\partial t_R}\right)^m
\frac{(t_Lt_R)^{h -\Delta-1}e^{-\beta_L/t_L-\beta_R/t_R}}{\Gamma(\Delta-h+1+m)}
\right|_{t_L=t_R=1} 
=\beta_L^{h-\Delta} e^{-\beta_L} \, \delta(\beta_L-\beta_R)\ .
\end{align}
This can be proven by integrating both sides against
$\beta_L^{x-1} \beta_R^{y-1}$ over $\beta_L$ and $\beta_R$ from $0$ to $\infty$. 
After performing the integrals, the left hand side gives
\begin{align}
&
\sum_{m=0}^\infty \frac{\Gamma(x)\Gamma(y)}{m!\Gamma(\Delta-h+1+m)} 
\left. \left(\frac{\partial}{\partial t_L}\frac{\partial}{\partial t_R}\right)^m
(t_Lt_R)^{h -\Delta-1}t_L^x t_R^y
\right|_{t_L=t_R=1}  \nn \\
=& \sum_{m=0}^\infty \frac{\Gamma(x)\Gamma(y)}{m!\Gamma(\Delta-h+1+m)
(x+h-\Delta)_{-m}(y+h-\Delta)_{-m} }\\
=& 
\Gamma(x+y+h-\Delta-1) \nn
\end{align}
and the right hand side trivially gives the same.

Secondly, we want to prove that the limit
\begin{align}
\lim_{\Delta\to \infty} \sum_{m=0}^\infty \left. \left(\frac{\partial}{\partial t_L}\frac{\partial}{\partial t_R}\right)^m\frac{ (\beta_L\beta_R)^{\frac{\Delta}{2}} 
(t_Lt_R)^{h -\Delta-1}e^{-\beta_L/t_L-\beta_R/t_R}
}{(  u +\frac{\Delta+2m}{\Delta^2})\Gamma(\Delta-h+1+m)m! } 
\right|_{t_L=t_R=1}
\label{LHSidenFSL}
\end{align}
is given by
\be
\delta(\beta_L - \beta_R) e^{-\beta_L} \beta_L^h \frac{2\beta_L}{2\beta_L u + 1}\ .
\label{RHSidenFSL}
\ee
This was used in section \ref{subsect:MassiveFlatSpaceLimit} with the identification $u=-\delta_{LR}/\Delta^2$.

The strategy to prove that  (\ref{LHSidenFSL}) equals (\ref{RHSidenFSL})  is to compute their double Mellin transform by integrating them against
$\beta_L^{x-1} \beta_R^{y-1}$ over $\beta_L$ and $\beta_R$ from $0$ to $\infty$.
The Mellin transform of (\ref{RHSidenFSL}) is
\be
\int_0^\infty d\beta e^{-\beta} \beta^{x+y+h-1} \frac{2 }{2\beta u + 1}
=\int_0^\infty ds \frac{e^{-1/s}}{ s^{x+y+h}} \frac{1 }{ u + s/2}\ ,
\label{MPropidentity}
\ee
where we changed integration variable to $s=1/\beta$.
The Mellin transform of (\ref{LHSidenFSL}) is
\begin{align}
&\lim_{\Delta\to \infty} \sum_{m=0}^\infty \left. \left(\frac{\partial}{\partial t_L}\frac{\partial}{\partial t_R}\right)^m\frac{ 
\Gamma\left(x+\frac{\Delta}{2}\right)\Gamma\left(y+\frac{\Delta}{2}\right)
(t_Lt_R)^{h -\frac{\Delta}{2}-1}t_L^x t_R^y
}{(  u +\frac{\Delta+2m}{\Delta^2})\Gamma(\Delta-h+1+m)m! } 
\right|_{t_L=t_R=1}\\
=&
\lim_{\Delta\to \infty} \sum_{m=0}^\infty 
\frac{ 
\Gamma\left(x+\frac{\Delta}{2}\right)\Gamma\left(y+\frac{\Delta}{2}\right)
}{m!  \Gamma(\Delta-h+1+m)
\left(h+x-\frac{\Delta}{2}\right)_{-m}\left(h+y-\frac{\Delta}{2}\right)_{-m}}  
\frac{1}{  u +\frac{ 2m}{\Delta^2}}\nn
\end{align}
Comparing with (\ref{MPropidentity}), we identify the continuous integration variable $s$ with the limit of the discrete variable $4m/\Delta^2$. Then, the sum over $m$ turns into the integral over $s$
\be
\sum_{m=0}^\infty \frac{4}{\Delta^2}\dots  \to \int_0^\infty ds \dots
\ee
and, using the Stirling approximation,  the summand reduces to the correct integrand,
\begin{align}
\lim_{\Delta\to \infty}  \left.
\frac{\Delta^2
\Gamma\left(x+\frac{\Delta}{2}\right)\Gamma\left(y+\frac{\Delta}{2}\right)
}{4m! \Gamma(\Delta-h+1+m) 
\left(h+x-\frac{\Delta}{2}\right)_{-m}\left(h+y-\frac{\Delta}{2}\right)_{-m}}
\right|_{m=\frac{s\Delta^2}{4}} =
\frac{e^{-1/s}}{ s^{x+y+h}}\ .
\end{align}

\bibliographystyle{utphys}
\bibliography{Mellinbib}

\providecommand{\href}[2]{#2}\begingroup\raggedright\begin{thebibliography}{10}

\bibitem{Mack}
G.~Mack, ``{D-independent representation of Conformal Field Theories in D
  dimensions via transformation to auxiliary Dual Resonance Models. Scalar
  amplitudes},''
\href{http://arxiv.org/abs/0907.2407}{{\ttfamily arXiv:0907.2407 [hep-th]}}.
%%CITATION = 0907.2407;%%.

\bibitem{MackSummary}
G.~Mack, ``{D-dimensional Conformal Field Theories with anomalous dimensions as
  Dual Resonance Models},''
\href{http://arxiv.org/abs/0909.1024}{{\ttfamily arXiv:0909.1024 [hep-th]}}.
%%CITATION = 0909.1024;%%.

\bibitem{Penedones:2010ue}
J.~Penedones, ``{Writing CFT correlation functions as AdS scattering
  amplitudes},'' \href{http://dx.doi.org/10.1007/JHEP03(2011)025}{{\em JHEP}
  {\bfseries 03} (2011) 025},
\href{http://arxiv.org/abs/1011.1485}{{\ttfamily arXiv:1011.1485 [hep-th]}}.
%%CITATION = 1011.1485;%%.

\bibitem{Ferrara:1973vz}
S.~Ferrara, A.~F. Grillo, G.~Parisi, and R.~Gatto, ``{Covariant expansion of
  the conformal four-point function},''
\href{http://dx.doi.org/10.1016/0550-3213(72)90587-1}{{\em Nucl. Phys.}
  {\bfseries B49} (1972) 77--98}.
%%CITATION = NUPHA,B49,77;%%.

\bibitem{Ferrara:1974nf}
S.~Ferrara, A.~F. Grillo, R.~Gatto, and G.~Parisi, ``{Analyticity properties
  and asymptotic expansions of conformal covariant green's functions},''
\href{http://dx.doi.org/10.1007/BF02813413}{{\em Nuovo Cim.} {\bfseries A19}
  (1974) 667--695}.
%%CITATION = NUCIA,A19,667;%%.

\bibitem{Polyakov:1974gs}
A.~M. Polyakov, ``{Nonhamiltonian approach to conformal quantum field
  theory},''
{\em Zh. Eksp. Teor. Fiz.} {\bfseries 66} (1974) 23--42.
%%CITATION = ZETFA,66,23;%%.

\bibitem{Sofia}
V.~K. Dobrev, V.~B. Petkova, S.~G. Petrova, and I.~T. Todorov, ``{Dynamical
  Derivation of Vacuum Operator Product Expansion in Euclidean Conformal
  Quantum Field Theory},''
\href{http://dx.doi.org/10.1103/PhysRevD.13.887}{{\em Phys. Rev.} {\bfseries
  D13} (1976) 887}.
%%CITATION = PHRVA,D13,887;%%.

\bibitem{Maldacena:1997re}
J.~M. Maldacena, ``{The Large N limit of superconformal field theories and
  supergravity},'' \href{http://dx.doi.org/10.1023/A:1026654312961,
  10.1023/A:1026654312961}{{\em Adv.Theor.Math.Phys.} {\bfseries 2} (1998)
  231--252}, \href{http://arxiv.org/abs/hep-th/9711200}{{\ttfamily
  arXiv:hep-th/9711200 [hep-th]}}.

\bibitem{Witten:1998qj}
E.~Witten, ``{Anti-de Sitter space and holography},'' {\em
  Adv.Theor.Math.Phys.} {\bfseries 2} (1998) 253--291,
  \href{http://arxiv.org/abs/hep-th/9802150}{{\ttfamily arXiv:hep-th/9802150
  [hep-th]}}.

\bibitem{Gubser:1998bc}
S.~Gubser, I.~R. Klebanov, and A.~M. Polyakov, ``{Gauge theory correlators from
  noncritical string theory},''
  \href{http://dx.doi.org/10.1016/S0370-2693(98)00377-3}{{\em Phys.Lett.}
  {\bfseries B428} (1998) 105--114},
  \href{http://arxiv.org/abs/hep-th/9802109}{{\ttfamily arXiv:hep-th/9802109
  [hep-th]}}.

\bibitem{LiuTseytlin}
H.~Liu and A.~A. Tseytlin, ``{On four-point functions in the CFT/AdS
  correspondence},'' \href{http://dx.doi.org/10.1103/PhysRevD.59.086002}{{\em
  Phys. Rev.} {\bfseries D59} (1999) 086002},
\href{http://arxiv.org/abs/hep-th/9807097}{{\ttfamily arXiv:hep-th/9807097}}.
%%CITATION = HEP-TH/9807097;%%.

\bibitem{Liu}
H.~Liu, ``{Scattering in anti-de Sitter space and operator product
  expansion},'' \href{http://dx.doi.org/10.1103/PhysRevD.60.106005}{{\em Phys.
  Rev.} {\bfseries D60} (1999) 106005},
\href{http://arxiv.org/abs/hep-th/9811152}{{\ttfamily arXiv:hep-th/9811152}}.
%%CITATION = HEP-TH/9811152;%%.

\bibitem{D'Hoker:1998mz}
E.~D'Hoker and D.~Z. Freedman, ``{General scalar exchange in AdS(d+1)},''
  \href{http://dx.doi.org/10.1016/S0550-3213(99)00169-8}{{\em Nucl. Phys.}
  {\bfseries B550} (1999) 261--288},
\href{http://arxiv.org/abs/hep-th/9811257}{{\ttfamily arXiv:hep-th/9811257}}.
%%CITATION = HEP-TH/9811257;%%.

\bibitem{Freedman:1998bj}
D.~Z. Freedman, S.~D. Mathur, A.~Matusis, and L.~Rastelli, ``{Comments on
  4-point functions in the CFT/AdS correspondence},''
  \href{http://dx.doi.org/10.1016/S0370-2693(99)00229-4}{{\em Phys. Lett.}
  {\bfseries B452} (1999) 61--68},
\href{http://arxiv.org/abs/hep-th/9808006}{{\ttfamily arXiv:hep-th/9808006}}.
%%CITATION = HEP-TH/9808006;%%.

\bibitem{Freedman:1998tz}
D.~Z. Freedman, S.~D. Mathur, A.~Matusis, and L.~Rastelli, ``{Correlation
  functions in the CFT($d$)/AdS($d+1$) correspondence},''
  \href{http://dx.doi.org/10.1016/S0550-3213(99)00053-X}{{\em Nucl. Phys.}
  {\bfseries B546} (1999) 96--118},
\href{http://arxiv.org/abs/hep-th/9804058}{{\ttfamily arXiv:hep-th/9804058}}.
%%CITATION = HEP-TH/9804058;%%.

\bibitem{D'Hoker}
E.~D'Hoker, D.~Z. Freedman, S.~D. Mathur, A.~Matusis, and L.~Rastelli,
  ``{Graviton exchange and complete 4-point functions in the AdS/CFT
  correspondence},''
  \href{http://dx.doi.org/10.1016/S0550-3213(99)00525-8}{{\em Nucl. Phys.}
  {\bfseries B562} (1999) 353--394},
\href{http://arxiv.org/abs/hep-th/9903196}{{\ttfamily arXiv:hep-th/9903196}}.
%%CITATION = HEP-TH/9903196;%%.

\bibitem{Arutyunov:2000py}
G.~Arutyunov and S.~Frolov, ``{Four-point functions of lowest weight CPOs in N
  = 4 SYM(4) in supergravity approximation},''
  \href{http://dx.doi.org/10.1103/PhysRevD.62.064016}{{\em Phys. Rev.}
  {\bfseries D62} (2000) 064016},
\href{http://arxiv.org/abs/hep-th/0002170}{{\ttfamily arXiv:hep-th/0002170}}.
%%CITATION = HEP-TH/0002170;%%.

\bibitem{Arutyunov:2002fh}
G.~Arutyunov, F.~A. Dolan, H.~Osborn, and E.~Sokatchev, ``{Correlation
  functions and massive Kaluza-Klein modes in the AdS/CFT correspondence},''
  \href{http://dx.doi.org/10.1016/S0550-3213(03)00448-6}{{\em Nucl. Phys.}
  {\bfseries B665} (2003) 273--324},
\href{http://arxiv.org/abs/hep-th/0212116}{{\ttfamily arXiv:hep-th/0212116}}.
%%CITATION = HEP-TH/0212116;%%.

\bibitem{Arutyunov:2003ae}
G.~Arutyunov and E.~Sokatchev, ``{On a large N degeneracy in N = 4 SYM and the
  AdS/CFT correspondence},''
  \href{http://dx.doi.org/10.1016/S0550-3213(03)00353-5}{{\em Nucl. Phys.}
  {\bfseries B663} (2003) 163--196},
\href{http://arxiv.org/abs/hep-th/0301058}{{\ttfamily arXiv:hep-th/0301058}}.
%%CITATION = HEP-TH/0301058;%%.

\bibitem{Berdichevsky:2007xd}
L.~Berdichevsky and P.~Naaijkens, ``{Four-point functions of different-weight
  operators in the AdS/CFT correspondence},''
  \href{http://dx.doi.org/10.1088/1126-6708/2008/01/071}{{\em JHEP} {\bfseries
  01} (2008) 071},
\href{http://arxiv.org/abs/0709.1365}{{\ttfamily arXiv:0709.1365 [hep-th]}}.
%%CITATION = 0709.1365;%%.

\bibitem{Uruchurtu:2008kp}
L.~I. Uruchurtu, ``{Four-point correlators with higher weight superconformal
  primaries in the AdS/CFT Correspondence},''
  \href{http://dx.doi.org/10.1088/1126-6708/2009/03/133}{{\em JHEP} {\bfseries
  03} (2009) 133},
\href{http://arxiv.org/abs/0811.2320}{{\ttfamily arXiv:0811.2320 [hep-th]}}.
%%CITATION = 0811.2320;%%.

\bibitem{Buchbinder:2010vw}
E.~I. Buchbinder and A.~A. Tseytlin, ``{On semiclassical approximation for
  correlators of closed string vertex operators in AdS/CFT},''
  \href{http://dx.doi.org/10.1007/JHEP08(2010)057}{{\em JHEP} {\bfseries 08}
  (2010) 057},
\href{http://arxiv.org/abs/1005.4516}{{\ttfamily arXiv:1005.4516 [hep-th]}}.
%%CITATION = 1005.4516;%%.

\bibitem{Uruchurtu:2011wh}
L.~I. Uruchurtu, ``{Next-next-to-extremal Four Point Functions of N=4 1/2 BPS
  Operators in the AdS/CFT Correspondence},''
\href{http://arxiv.org/abs/1106.0630}{{\ttfamily arXiv:1106.0630 [hep-th]}}.
%%CITATION = 1106.0630;%%.

\bibitem{Dolan:2006ec}
F.~A. Dolan, M.~Nirschl, and H.~Osborn, ``{Conjectures for large N N = 4
  superconformal chiral primary four point functions},''
  \href{http://dx.doi.org/10.1016/j.nuclphysb.2006.05.009}{{\em Nucl. Phys.}
  {\bfseries B749} (2006) 109--152},
\href{http://arxiv.org/abs/hep-th/0601148}{{\ttfamily arXiv:hep-th/0601148}}.
%%CITATION = HEP-TH/0601148;%%.

\bibitem{Howtozintegrals}
E.~D'Hoker, D.~Z. Freedman, and L.~Rastelli, ``{AdS/CFT 4-point functions: How
  to succeed at z-integrals without really trying},''
  \href{http://dx.doi.org/10.1016/S0550-3213(99)00526-X}{{\em Nucl. Phys.}
  {\bfseries B562} (1999) 395--411},
\href{http://arxiv.org/abs/hep-th/9905049}{{\ttfamily arXiv:hep-th/9905049}}.
%%CITATION = HEP-TH/9905049;%%.

\bibitem{Dolan:2003hv}
F.~A. Dolan and H.~Osborn, ``{Conformal partial waves and the operator product
  expansion},'' \href{http://dx.doi.org/10.1016/j.nuclphysb.2003.11.016}{{\em
  Nucl. Phys.} {\bfseries B678} (2004) 491--507},
\href{http://arxiv.org/abs/hep-th/0309180}{{\ttfamily arXiv:hep-th/0309180}}.
%%CITATION = HEP-TH/0309180;%%.

\bibitem{susskind}
L.~Susskind, ``{Holography in the flat space limit},''
\href{http://arxiv.org/abs/hep-th/9901079}{{\ttfamily arXiv:hep-th/9901079}}.
%%CITATION = HEP-TH/9901079;%%.

\bibitem{polchinski}
J.~Polchinski, ``{S-matrices from AdS spacetime},''
\href{http://arxiv.org/abs/hep-th/9901076}{{\ttfamily arXiv:hep-th/9901076}}.
%%CITATION = HEP-TH/9901076;%%.

\bibitem{GGP}
M.~Gary, S.~B. Giddings, and J.~Penedones, ``{Local bulk S-matrix elements and
  CFT singularities},''
  \href{http://dx.doi.org/10.1103/PhysRevD.80.085005}{{\em Phys. Rev.}
  {\bfseries D80} (2009) 085005},
\href{http://arxiv.org/abs/0903.4437}{{\ttfamily arXiv:0903.4437 [hep-th]}}.
%%CITATION = 0903.4437;%%.

\bibitem{JP}
I.~Heemskerk, J.~Penedones, J.~Polchinski, and J.~Sully, ``{Holography from
  Conformal Field Theory},''
  \href{http://dx.doi.org/10.1088/1126-6708/2009/10/079}{{\em JHEP} {\bfseries
  10} (2009) 079},
\href{http://arxiv.org/abs/0907.0151}{{\ttfamily arXiv:0907.0151 [hep-th]}}.
%%CITATION = 0907.0151;%%.

\bibitem{Katz}
A.~L. Fitzpatrick, E.~Katz, D.~Poland, and D.~Simmons-Duffin, ``{Effective
  Conformal Theory and the Flat-Space Limit of AdS},''
\href{http://arxiv.org/abs/1007.2412}{{\ttfamily arXiv:1007.2412 [hep-th]}}.
%%CITATION = 1007.2412;%%.

\bibitem{TakuyaFSL}
T.~Okuda and J.~Penedones, ``{String scattering in flat space and a scaling
  limit of Yang-Mills correlators},''
\href{http://arxiv.org/abs/1002.2641}{{\ttfamily arXiv:1002.2641 [hep-th]}}.
%%CITATION = 1002.2641;%%.

\bibitem{Fitzpatrick:2011jn}
A.~Fitzpatrick and J.~Kaplan, ``{Scattering States in AdS/CFT},''
  \href{http://arxiv.org/abs/1104.2597}{{\ttfamily arXiv:1104.2597 [hep-th]}}.

\bibitem{Gary:2009mi}
M.~Gary and S.~B. Giddings, ``{The flat space S-matrix from the AdS/CFT
  correspondence?},'' \href{http://dx.doi.org/10.1103/PhysRevD.80.046008}{{\em
  Phys. Rev.} {\bfseries D80} (2009) 046008},
\href{http://arxiv.org/abs/0904.3544}{{\ttfamily arXiv:0904.3544 [hep-th]}}.
%%CITATION = 0904.3544;%%.

\bibitem{GiddingsBulkLoc}
S.~B. Giddings, ``{Flat-space scattering and bulk locality in the AdS/CFT
  correspondence},'' \href{http://dx.doi.org/10.1103/PhysRevD.61.106008}{{\em
  Phys. Rev.} {\bfseries D61} (2000) 106008},
\href{http://arxiv.org/abs/hep-th/9907129}{{\ttfamily arXiv:hep-th/9907129}}.
%%CITATION = HEP-TH/9907129;%%.

\bibitem{Gary:2011kk}
M.~Gary and S.~B. Giddings, ``{Constraints on a fine-grained AdS/CFT
  correspondence},'' \href{http://arxiv.org/abs/1106.3553}{{\ttfamily
  arXiv:1106.3553 [hep-th]}}. 

\bibitem{MPaulos}
M.~F. Paulos, ``{Towards Feynman rules for Mellin amplitudes in AdS/CFT},''.

\bibitem{Rattazzi:2008pe}
R.~Rattazzi, V.~S. Rychkov, E.~Tonni, and A.~Vichi, ``{Bounding scalar operator
  dimensions in 4D CFT},''
  \href{http://dx.doi.org/10.1088/1126-6708/2008/12/031}{{\em JHEP} {\bfseries
  12} (2008) 031},
\href{http://arxiv.org/abs/0807.0004}{{\ttfamily arXiv:0807.0004 [hep-th]}}.
%%CITATION = 0807.0004;%%.

\bibitem{Rychkov:2009ij}
V.~S. Rychkov and A.~Vichi, ``{Universal Constraints on Conformal Operator
  Dimensions},'' \href{http://dx.doi.org/10.1103/PhysRevD.80.045006}{{\em Phys.
  Rev.} {\bfseries D80} (2009) 045006},
\href{http://arxiv.org/abs/0905.2211}{{\ttfamily arXiv:0905.2211 [hep-th]}}.
%%CITATION = 0905.2211;%%.

\bibitem{Dirac}
P.~A.~M. Dirac, ``{Wave equations in conformal space},''
{\em Annals Math.} {\bfseries 37} (1936) 429--442.
%%CITATION = ANMAA,37,429;%%.

\bibitem{Weinberg:2010fx}
S.~Weinberg, ``{Six-dimensional Methods for Four-dimensional Conformal Field
  Theories},'' \href{http://dx.doi.org/10.1103/PhysRevD.82.045031}{{\em
  Phys.Rev.} {\bfseries D82} (2010) 045031},
  \href{http://arxiv.org/abs/1006.3480}{{\ttfamily arXiv:1006.3480 [hep-th]}}.

\bibitem{Dolan:2000ut}
F.~A. Dolan and H.~Osborn, ``{Conformal four point functions and the operator
  product expansion},''
  \href{http://dx.doi.org/10.1016/S0550-3213(01)00013-X}{{\em Nucl. Phys.}
  {\bfseries B599} (2001) 459--496},
\href{http://arxiv.org/abs/hep-th/0011040}{{\ttfamily arXiv:hep-th/0011040}}.
%%CITATION = HEP-TH/0011040;%%.

\bibitem{Symanzik}
K.~Symanzik, ``{On Calculations in conformal invariant field theories},''
{\em Lett. Nuovo Cim.} {\bfseries 3} (1972) 734--738.
%%CITATION = NCLTA,3,734;%%.

\bibitem{Maldacena:2011nz}
J.~M. Maldacena and G.~L. Pimentel, ``{On graviton non-Gaussianities during
  inflation},'' \href{http://arxiv.org/abs/1104.2846}{{\ttfamily
  arXiv:1104.2846 [hep-th]}}.

\bibitem{Britto:2005fq}
R.~Britto, F.~Cachazo, B.~Feng, and E.~Witten, ``{Direct proof of tree-level
  recursion relation in Yang-Mills theory},''
  \href{http://dx.doi.org/10.1103/PhysRevLett.94.181602}{{\em Phys.Rev.Lett.}
  {\bfseries 94} (2005) 181602},
  \href{http://arxiv.org/abs/hep-th/0501052}{{\ttfamily arXiv:hep-th/0501052
  [hep-th]}}.

\bibitem{Britto:2004ap}
R.~Britto, F.~Cachazo, and B.~Feng, ``{New recursion relations for tree
  amplitudes of gluons},''
  \href{http://dx.doi.org/10.1016/j.nuclphysb.2005.02.030}{{\em Nucl.Phys.}
  {\bfseries B715} (2005) 499--522},
  \href{http://arxiv.org/abs/hep-th/0412308}{{\ttfamily arXiv:hep-th/0412308
  [hep-th]}}.

\bibitem{ArkaniHamed:2008yf}
N.~Arkani-Hamed and J.~Kaplan, ``{On Tree Amplitudes in Gauge Theory and
  Gravity},'' \href{http://dx.doi.org/10.1088/1126-6708/2008/04/076}{{\em JHEP}
  {\bfseries 0804} (2008) 076},
  \href{http://arxiv.org/abs/0801.2385}{{\ttfamily arXiv:0801.2385 [hep-th]}}.

\bibitem{Raju:2010by}
S.~Raju, ``{BCFW for Witten Diagrams},''
  \href{http://dx.doi.org/10.1103/PhysRevLett.106.091601}{{\em Phys.Rev.Lett.}
  {\bfseries 106} (2011) 091601},
  \href{http://arxiv.org/abs/1011.0780}{{\ttfamily arXiv:1011.0780 [hep-th]}}.
%%CITATION = 1011.0780;%%.

\bibitem{Raju:2011mp}
S.~Raju, ``{Recursion Relations for AdS/CFT Correlators},''
  \href{http://dx.doi.org/10.1103/PhysRevD.83.126002}{{\em Phys. Rev.}
  {\bfseries D83} (2011) 126002},
\href{http://arxiv.org/abs/1102.4724}{{\ttfamily arXiv:1102.4724 [hep-th]}}.
%%CITATION = 1102.4724;%%.

\bibitem{vanRees:2011ir}
B.~C. van Rees, ``{Irrelevant deformations and the holographic Callan- Symanzik
  equation},''
\href{http://arxiv.org/abs/1105.5396}{{\ttfamily arXiv:1105.5396 [hep-th]}}.
%%CITATION = 1105.5396;%%.

\bibitem{Kim:1985ez}
H.~Kim, L.~Romans, and P.~van Nieuwenhuizen, ``{The Mass Spectrum of Chiral N=2
  D=10 Supergravity on S**5},''
  \href{http://dx.doi.org/10.1103/PhysRevD.32.389}{{\em Phys.Rev.} {\bfseries
  D32} (1985) 389}.

\bibitem{Lee:1998bxa}
S.~Lee, S.~Minwalla, M.~Rangamani, and N.~Seiberg, ``{Three-point functions of
  chiral operators in D = 4, N = 4 SYM at large N},'' {\em Adv. Theor. Math.
  Phys.} {\bfseries 2} (1998) 697--718,
\href{http://arxiv.org/abs/hep-th/9806074}{{\ttfamily arXiv:hep-th/9806074}}.
%%CITATION = HEP-TH/9806074;%%.

\bibitem{Balitsky:2011tw}
I.~Balitsky, ``{Mellin representation of the graviton bulk-to-bulk propagator
  in AdS},'' \href{http://dx.doi.org/10.1103/PhysRevD.83.087901}{{\em
  Phys.Rev.} {\bfseries D83} (2011) 087901},
  \href{http://arxiv.org/abs/1102.0577}{{\ttfamily arXiv:1102.0577 [hep-th]}}.

\bibitem{OurDIS}
L.~Cornalba, M.~S. Costa, and J.~Penedones, ``{Deep Inelastic Scattering in
  Conformal QCD},'' \href{http://dx.doi.org/10.1007/JHEP03(2010)133}{{\em JHEP}
  {\bfseries 03} (2010) 133},
\href{http://arxiv.org/abs/0911.0043}{{\ttfamily arXiv:0911.0043 [hep-th]}}.
%%CITATION = 0911.0043;%%.

\end{thebibliography}\endgroup

\end{document}